\documentclass[aps,prx,twocolumn,floatfix,longbibliography,nofootinbib]{revtex4-2}
\usepackage[utf8]{inputenc}
\usepackage{natbib}
\usepackage{graphicx}
\usepackage{xcolor}
\usepackage[tbtags]{amsmath}
\usepackage[colorlinks=true, linkcolor=blue]{hyperref}
\usepackage{amssymb}
\usepackage{gensymb}
\usepackage{comment}
\usepackage{float}
\usepackage{amsmath}
\usepackage{tabularx,graphicx}
\usepackage{epstopdf}
\usepackage{latexsym}
\usepackage{color, colortbl}
\usepackage{psfrag}
\usepackage{bbm}
\usepackage{bm}
\usepackage{titlesec}
\usepackage{dsfont}
\usepackage{feynmp}
\usepackage{slashed}
\usepackage{subcaption}
\usepackage{multirow}
\usepackage{subcaption}
\usepackage[newcommands]{ragged2e}
\usepackage[normalem]{ulem}
\renewcommand{\vec}[1]{{\boldsymbol{#1}}}
\usepackage{braket}
\usepackage{subcaption}

\def \v {{\bf v}}

\def \x {{\vec x}}

\def \beq {\begin{eqnarray}}
\def \eeq {\end{eqnarray}}

\def \vp{\varphi}

\def \nn {\nonumber}
\def \omd {{\omega_{\rm D}}}

\newcommand{\change}[1]{{ #1}}
\newcommand{\rd}{{\rm d}}

\newcommand{\vn}[1]{{\left|\vec{#1}\right|}}
\newcommand{\calN}{{\mathcal N}}

\newcommand{\calL}{{\mathcal L}}

\usepackage{tikz-feynman}
\usetikzlibrary{arrows}
\usetikzlibrary{arrows.meta, positioning,shapes.geometric}
\tikzset{
  mid arrow/.style={postaction={decorate,decoration={
        markings,
        mark=at position .575 with {\arrow[#1]{stealth}}
      }}},
  near arrow/.style={postaction={decorate,decoration={
        markings,
        mark=at position .275 with {\arrow[#1]{stealth}}
      }}},
   far arrow/.style={postaction={decorate,decoration={
        markings,
        mark=at position .800 with {\arrow[#1]{stealth}}
      }}},
   boson/.style={decorate, draw=black,
    decoration={snake,amplitude=1pt, segment length=5pt},
      },
   mid triangle/.style={postaction={decorate,decoration={
        markings,
        mark=at position .575 with {\arrow[#1]{triangle 45}}
      }}}
}

\usepackage{pdfpages}
\makeatletter
\AtBeginDocument{\let\LS@rot\@undefined}
\makeatother

\begin{document}

\title{Phonon Induced Energy Relaxation in Quantum Critical Metals}
\author{Haoyu Guo}
\author{Debanjan Chowdhury}
\affiliation{Department of Physics, Cornell University, Ithaca, New York 14853, USA.}
%\email{debanjanchowdhury@cornell.edu}
\begin{abstract}
Metals at the brink of electronic quantum phase transitions display high-temperature superconductivity, competing orders, and unconventional charge transport, revealing strong departures from conventional Fermi liquid behavior. Investigation of these fascinating intertwined phenomena has been at the center of research across a variety of correlated materials over the past many decades. A ubiquitous experimental observation is the emergence of a universal timescale that governs electrical transport and momentum relaxation. In this work, we analyze an equally important theoretical question of how the energy contained in the electronic degrees of freedom near a quantum phase transition relaxes to the environment via their coupling to acoustic phonons. Assuming that the bottleneck for energy dissipation is controlled by the coupling between electronic degrees of freedom and acoustic phonons, we present a universal theory of the temperature dependence of the energy relaxation rate in a marginal Fermi liquid. We find that the energy relaxation rate exhibits a complex set of temperature-dependent crossovers controlled by emergent energy scales in the problem. We place these results in the context of recent measurements of the energy relaxation rate via non-linear optical spectroscopy in the normal state of hole-doped cuprates.  
\end{abstract}

\maketitle

{\it Introduction.-} In recent years, the problem of {\it strange} metals has drawn significant attention due to their unconventional transport behavior \cite{JANBruin2013,SAHartnoll2022,DChowdhury2022a}. An experimental hallmark of these correlated metallic systems is a linear-in-temperature ($T$) electrical resistivity governed by a so-called {\it Planckian} relaxation rate of order $k_B T/\hbar$ down to low temperatures ---in stark contrast to an umklapp-dominated $T^2$-scaling predicted by Fermi liquid theory, or a $T^5$-Bloch-Gr\"uneissen regime due to scattering off acoustic phonons  \cite{AAAbrikosov1963,ziman2001electrons}. The electrical transport timescale reflects the rate at which momentum is lost from the electronic system to external degrees of freedom, such as the lattice or impurities. Given the scale-invariant nature of the Planckian scattering rate \cite{SSachdev1999b} and experimental evidence for electronic quantum criticality across these materials \cite{LTaillefer2010,matsuda,greeneSC,Paschen2021}, a vast majority of theoretical efforts have focused on a scenario involving a Fermi surface coupled to the long-wavelength fluctuations of a bosonic collective mode of electronic origin \cite{SSLee2018,EBerg2019,varmaRMP}, e.g. an electronic nematic \cite{nematicreview}. The strong quantum critical fluctuations destroy the long-lived electronic quasiparticles near the Fermi surface, giving rise to non-Fermi liquid behavior. {Early important theoretical progress  \cite{CMVarma1989,PWAnderson1991a, PWAnderson1992b} already captured many experimental signatures of the strange metals in cuprates at a phenomenological level, especially the linear-in-$T$ single-particle scattering rate.} However, this does not automatically yield a $T-$linear momentum-relaxation rate, which has remained one of the central mysteries in the field. A number of theoretical works over the past few years have suggested additional microscopic ingredients that could potentially resolve this mystery \cite{AAPatel2018b,DChowdhury2018c,DVElse2021,AAPatel2023,HGuo2022a,ZDShi2023,NBashan2024,ETulipman2024,NBashan2025,PALee2021,PALee2024,CHMousatov2020,AAPatel2024,AAPatel2024a,AHardy2025, CLi2024}. 

\setlength{\abovecaptionskip}{0pt}  % space above caption
\setlength{\belowcaptionskip}{-0.5cm} 

\begin{figure}
    \centering
    \begin{subfigure}[t]{1.0\linewidth}
    \includegraphics[width=\linewidth]{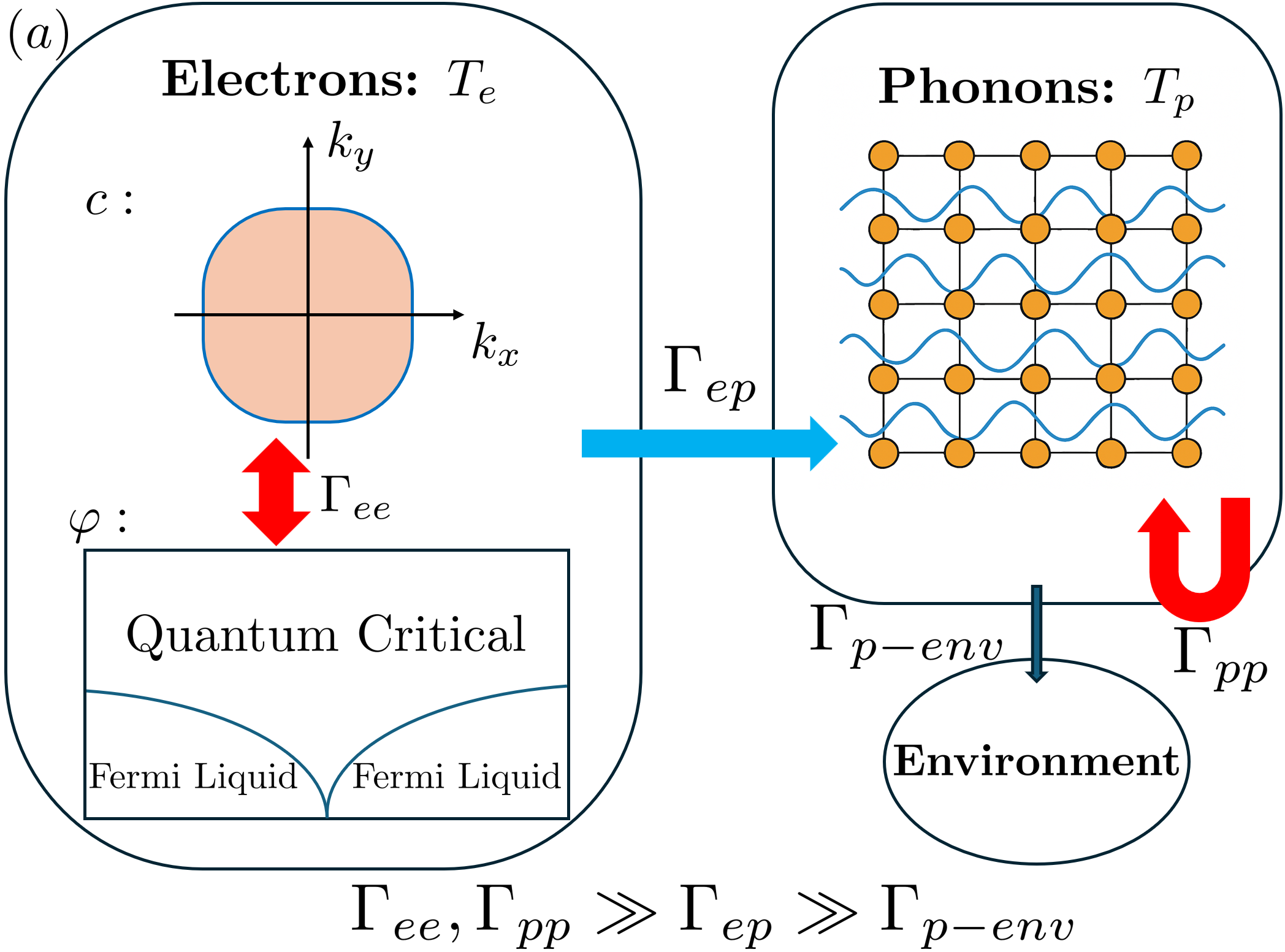}
    \end{subfigure}
    \begin{subfigure}[t]{\linewidth}
    \includegraphics[width=0.9\linewidth]{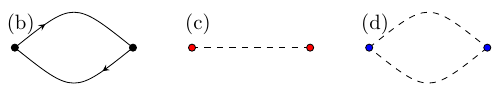}
    \end{subfigure}
    \caption{(a) The bottleneck for energy relaxation from the electronic d.o.f. at local temperature $T_e$ to the phonons at local temperature $T_p$ is governed by $\Gamma_{ep}$, which is controlled by the processes illustrated in (b)-(d). The lowest-order Feynman diagrams for the phonon self-energy due to the couplings: (b) $\calL_{ep}^{(1)}$ Eq.~\eqref{eq:coupling1}, (c) $\calL_{ep}^{(2)}$ Eq.~\eqref{eq:coupling2}, and (d) $\calL_{ep}^{(3)}$ Eq.~\eqref{eq:coupling3}, respectively. Solid (dashed) lines denote renormalized electron (bosonic collective mode, $\vp$) Green's function.} \label{fig:fig1}
\end{figure}

In this work, we focus on a related but even more poorly understood question: How does a strongly correlated metal relax its {\it energy} to the environment? {Previous studies have focused a lot on how an electron becomes incoherent through inelastic scattering, but most of the transferred energy is assumed to remain in the electronic system. In this work, we query how this energy ultimately dissipates into the environment through various symmetry-allowed couplings to phonons.} While it is unlikely that phonons by themselves are responsible for the anomalous low-temperature momentum-relaxation in the strange-metal in the vicinity of quantum criticality \cite{CHMousatov2021}, they are centrally important for energy relaxation (Fig.~\ref{fig:fig1}a). Phonons certainly act as a {\it bath} for the combined system consisting of electrons coupled to the bosonic collective mode, but have not received much theoretical attention in the context of non-Fermi liquids. The classic theoretical treatment \cite{PBAllen1987} of energy relaxation rates, $\Gamma_E$, in a three-dimensional Fermi liquid coupled to acoustic phonons predicts a crossover from $\Gamma_E \propto T^3$ at low temperatures to $\Gamma_E \propto 1/T$ at high temperatures. In general, the momentum relaxation rate and $\Gamma_E$ are {\it not} automatically identical. How this behavior is modified in a non-Fermi liquid without long-lived electronic quasiparticles is presently unclear. \change{While our results are broadly applicable to a variety of correlated electron systems in the vicinity of quantum phase transitions, it is partly inspired by the quantum critical scenario in the cuprates, which remains an active area of investigation \cite{rams15,BMichon2019,SBadoux2016,BMichon2023a,SDChen2019b,AShekhter2013d,KFujita2014a, AGourgout2022e}; see also alternative points of view \cite{JBobroff2002,KKGomes2007b,DPelc2022,NDoiron-Leyraud2007b,NBarisic2019}.} We note that the theoretical framework developed below can be easily adapted to correlated insulating states that host Fermi surfaces of emergent neutral fermions (e.g. spinons) coupled to dynamical gauge-fields \cite{LNW,QSL}, assuming that these degrees of freedom (d.o.f.) also relax their energy via coupling to acoustic phonons \cite{PAL_sound}.

{\it Model of thermal relaxation.-} We propose a minimal low-energy model consisting of three d.o.f.: electrons, a bosonic collective mode of electronic origin tuned to the vicinity of quantum criticality, and acoustic phonons (Fig.~\ref{fig:fig1}a). Our goal is to understand the temperature dependence of energy transfer from the combined electronic degrees of freedom to the phonons. The new ingredient relative to previous considerations is the presence of the bosonic collective mode, which we expect to couple to phonons via symmetry-allowed channels. The direct coupling between the bosonic collective mode and the phonons in the low-energy effective theory are associated with a distinct kinematic regime, compared to any usual local electron-phonon coupling. We investigate the effects of three distinct couplings: (i) a conventional electron–phonon deformation potential, and (ii) linear and (iii) quadratic couplings between the bosonic mode and phonons, each motivated by symmetry and microscopic considerations.  We thus treat all couplings as weak perturbations compared to dominant electronic interactions.

To model the thermal relaxation process, we use the standard two-temperature model (illustrated in Fig.~\ref{fig:fig1}) which ignores the coupling between the phonons and the environment $\Gamma_\text{p-env}$, and assumes that the electrons and the phonons scattering rapidly within themselves and separately equilibrate first (with temperatures $T_e$ and $T_p$ respectively) and the relaxation bottleneck is set by the electron-phonon coupling. This is summarized by the hierarchy of scattering rates $\Gamma_{ee},\Gamma_{pp}\gg \Gamma_{ep}\gg \Gamma_{\text{p-env}}$ {(see \cite{Supp} for further discussion)}. The energy-flux between the electron and the phonon sector is given by $C_p\rd T_p/\rd t=-C_e \rd T_e/\rd t=\kappa(T_e-T_p)$, and the energy relaxation rate is given by 
\begin{equation}\label{eq:GammaE}
     \Gamma_E=\kappa\left(\frac{1}{C_e}+\frac{1}{C_p}\right)\,,
   \end{equation}  where $C_{e(p)}$ is the specific heat of the eletrons (phonons). Note that the parametrically different dependence of $C_e,~C_p$ on temperature inevitably accounts for some of the crossovers in $\Gamma_E$. To disentangle the thermodynamic $T-$dependent contributions to $C_e,~C_p$ from the intrinsic $T-$dependence of the energy flux ($\kappa$) from electrons to phonons, in the remainder of this manuscript we focus primarily on $\kappa$ due to the distinct couplings described above. In this work, we compute $\kappa$ using the Keldysh kinetic equation \cite{Supp,GStefanucci2013a,AKamenev2023}, and an alternative approach based on the Kubo formula is developed in \cite{PGlorioso2022a}. We will return to the full $T-$dependence of $\Gamma_E$ when we place our  results in the context of the recent experiments \cite{DChaudhuri2025}.

{\it Low-energy theory.-} Consider a low-energy Lagrangian for electrons in the vicinity of quantum criticality, coupled to acoustic phonons via $\calL=\calL_e+\calL_p+\calL_{ep}$, where \vspace{-0.3cm}
\begin{subequations}
\beq\label{eq:Le}
   \calL_e & =& \int \rd\tau \Big[\sum_{\vec{k}} c_{\vec{k}}^\dagger \left(\partial_\tau+\varepsilon_{\vec{k}}\right)c_{\vec{k}} \nn\\
   &&+\sum_{\vec{q}} \varphi_{-\vec{q}}\left(\partial_\tau^2+v_{\vp}^2\vec{q}^2+ r\right)\varphi_{\vec{q}} \nn \\
     && + g  \sum_{\vec{k},\vec{q}} f_{\vec{k},\vec{{q}}} c_{\vec{k+q/2}}^\dagger c_{\vec{k-q/2}}\varphi_{\vec{q}}\Big]+ ..., \,\\
     \label{eq:Lp}
  \calL_{p} &=& \int\rd \tau \sum_i\sum_{\vn{q}<\omd/c}X_{i,-\vec{q}} \left(\partial_{\tau}^2+c^2\vec{q}^2\right) X_{i,\vec{q}}\,.
\eeq
\end{subequations} 
Here, $c_{\vec{k}}^\dagger$ creates a fermion with momentum $\vec{k}$ and dispersion $\varepsilon_{\vec{k}}$, and $\varphi_{\vec{q}} = \varphi_{-\vec{q}}^\dagger$ denotes a critical bosonic mode (e.g., Ising-nematic order \cite{nematicreview}) near a quantum critical point (QCP) tuned by $r$. \change{Microscopically, the $\varphi$ field arises from Hubbard-Stratonovich decoupling of the electron interaction.} The boson velocity $v_\varphi$ is comparable to the Fermi velocity $v_F$, and it couples to fermions via a Yukawa coupling $g$ and form factor $f_{\vec{k},\vec{q}}$ (e.g., $B_{1g}$ irreducible representation for the Ising-nematic). $X_i$ represents the $i$-th component of lattice displacement, with sound speed $c$, and we adopt a Debye phonon model with cutoff $\omega_D/c$. For simplicity, all phonon modes are taken to have the same velocity. The $\dots$ in Eq.~\eqref{eq:Le} includes generic perturbations such as disorder. 

In the clean limit, the theory yields a Landau-damped boson with dynamical exponent $z_\varphi = 3$ and a non-Fermi liquid with $z_c = 3/2$ \cite{SSLee2018,PALee1989,AJMillis1993,JPolchinski1994, BIHalperin1993,YBKim1994a,CNayak1994,SSLee2009,MAMetlitski2010,
    DFMross2010,SSur2014,MAMetlitski2015,SAHartnoll2014,
    AEberlein2017,THolder2015,THolder2015a,ALFitzpatrick2014,
    JADamia2019,JADamia2020,JADamia2021,SPRidgway2015,AAbanov2020,YMWu2020,
    XWang2019,AKlein2020,OGrossman2021,DChowdhury2020,
    VOganesyan2001,AVChubukov2017,DLMaslov2017,
    SLi2023,EEAldape2022,IEsterlis2021,HGuo2022a,YBKim1995a,LVDelacretaz2022a,
SEHan2023,DVElse2021a,DVElse2021,ZDShi2022,ZDShi2023}. Motivated by extensive experimental evidence \cite{varmaRMP}, we focus on a regime where electrons display marginal Fermi liquid (MFL) behavior \cite{CMVarma1989}, with self-energy $\Sigma_c(i\omega) \sim -\omega\ln\omega$. This can arise from various mechanisms including disorder \cite{IEsterlis2021, HGuo2022a, AAPatel2023}, which reduce the boson exponent to $z_\varphi = 2$. %{However, Eq. \eqref{eq:Le} by itself does not relax energy outside of the electronic system, due to the time translational symmetry and absence of coupling to the phonon sector.} 
Importantly, our analysis of energy relaxation via phonons is agnostic to the microscopic origin of MFL, assuming only that electron–phonon couplings are weak.

Motivated by experiments in cuprates and other quasi-two-dimensional strange metals, we assume that the electronic degrees of freedom are effectively two-dimensional, with negligible interlayer hopping, while phonons remain fully three-dimensional. We consider three symmetry-allowed electron-phonon couplings,
$\calL_{ep} = \calL_{ep}^{(1)} + \calL_{ep}^{(2)} + \calL_{ep}^{(3)}$, where:
\begin{eqnarray}\label{eq:coupling1}
  \calL_{ep}^{(1)} &=& \int\rd \tau \sum_{\vec{k},\vec{q}} M^i_{\vec{k},\vec{q}} c^\dagger_{\vec{k+q/2}}c_{\vec{k-q/2}} X_{i,\vec{q}}\,,\\
  \label{eq:coupling2}
  \calL_{ep}^{(2)} &=&\int \rd \tau \sum_{\vec{q}} N_{\vec{q}}^i X_{i,-\vec{q}} \varphi_{\vec{q}}\,,\\
  \label{eq:coupling3}
  \calL_{ep}^{(3)} &=& \frac{1}{2}\int \rd \tau \sum_{\vec{k},\vec{{q}}} L^{i}_{\vec{k},\vec{q}} \varphi_{-\vec{k-q/2}} \varphi_{\vec{k+q/2}} X_{i,\vec{q}}\,.
\end{eqnarray}
Here, $\calL_{ep}^{(1)}$ is the standard deformation-potential coupling, where $M^i_{\vec{k},\vec{q}}$ encodes strain coupling to the electron density. $\calL_{ep}^{(2)}$ describes a linear coupling between strain and the bosonic collective mode $\varphi$, allowed by symmetry (e.g., nematic modes and $B_{1g}$ strain \cite{QiXu}). $\calL_{ep}^{(3)}$ captures inelastic scattering between phonons and pairs of $\varphi$ bosons \cite{Halperin76}. By Goldstone’s theorem, the matrix elements $M^i_{\vec{k},\vec{q}}, N^i_{\vec{q}}, L^i_{\vec{k},\vec{q}}$ scale linearly with the phonon momentum $\vec{q}$. {The theory introduced above can be formally controlled using an extension of the large-$N$ Yukawa-SYK model \cite{DChowdhury2022a,IEsterlis2021,HGuo2022a,AAPatel2023,HGuo2024,ZDShi2022,ZDShi2023,EEAldape2022,CLi2024}, which we review in the End Matter.}

For the electron coupling $M^i_{\vec{k},\vec{q}}$, only the $\vec{q}$-scaling is relevant; the details of the Fermi surface geometry are largely unimportant. In contrast, the quasi-2D nature of $\varphi$ strongly affects the boson-phonon couplings. Summing over polarizations, we model the matrix elements as $\sum_i |N^i_{\vec{q}}|^2, \sum_i |L^i_{\vec{k},\vec{q}}|^2 \sim A_{N,L} (\vn{q_\text{2D}}^2 + \lambda q_z^2)$, where $A_N$ and $A_L$ have dimensions $[\text{energy}]^2$ and $[\text{energy}]^0$, respectively. We assume $\lambda \ll 1$ to reflect weak coupling to out-of-plane strain, and comment later on its effects.

Throughout, we assume the couplings in $\calL_{ep}$ are weak compared to electronic or phononic self-interactions, justifying a perturbative treatment of the energy flux $\kappa$. While $\calL_{ep}^{(2)}$ could in principle be eliminated via a change of basis for $\varphi$, we retain it explicitly, since we treat phonons as the primary channel for energy dissipation into the environment. Removing it would spuriously couple $\varphi$ directly to the bath.

\begin{figure}
  \centering
  \includegraphics[width=\linewidth]{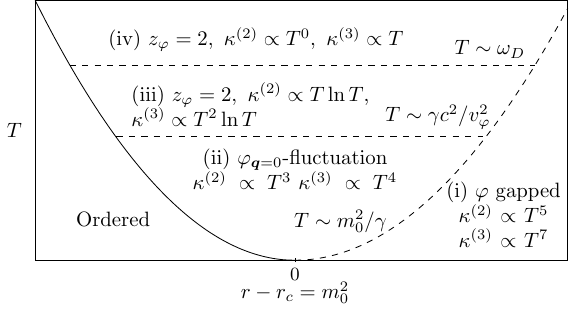}
  \vspace{-0.2cm}
  \caption{
  %(a) Numerical plot of electronic contribution $\kappa^{(1)}$ to the energy flux.  (b)
  Different regimes of the energy flux due to phonon-boson coupling (Eqs.~\eqref{eq:coupling2},\eqref{eq:coupling3}) near the QCP. $\omega_D$ is the Debye frequency of the phonons; $c$ is the speed of sound; $v_\varphi$ is the velocity of the $\varphi$ boson; $\gamma$ is the Landau damping coefficient (see Eq.~\eqref{eq:Aphi}); $m_0$ is the renormalized boson mass which vanishes at the QCP.}\label{fig:results}
\end{figure}

The contribution $\kappa^{(1)}$ from the deformation-potential electron-phonon coupling reproduces Allen's well-known result, i.e. a crossover from $T^4$ to constant near the Debye frequency $\omega_D$ \cite{PBAllen1987}. We  review this in the End Matter and also demonstrate that the MFL self-energy does not change this picture. Fig.~\ref{fig:results} summarizes the contributions $\kappa^{(2,3)}$ from the new boson-phonon couplings proposed here, which shows a richer $T$-dependent structure that we analyze below.

{\it Energy-relaxation due to linear coupling between phonon and collective mode.-} Let us now turn to the contribution due to $\calL_{ep}^{(2)}$ in Eq.~\eqref{eq:coupling2}, with a phonon self-energy shown in Fig.~\ref{fig:fig1}(c). As noted previously, the renormalized propagator for $\vp$ in the MFL is overdamped with $z_\vp=2$, with a spectral function,
 \begin{equation}\label{eq:Aphi}
  A_\varphi(\omega,\vec{q})=\frac{- 2\gamma\omega}{\gamma^2\omega^2+[\vn{q_\text{2D}}^2 v_{\vp}^2+m^2(T)]^2}\,.
\end{equation} 
Here, $\gamma=\calN g^2/\Gamma$ is the Landau-damping coefficient with dimension of energy ($\calN$: fermion density of states,  $\Gamma$: fermion elastic scattering rate), expected to be an electronic energy scale. The boson mass term, $m^2(T)$  includes the corrections due to a finite temperature, that are potentially important in the quantum critical regime. The QCP corresponds to $m(T=0)=0$, and the thermal fluctuations induce a $m^2(T)\sim \gamma T\ln(\gamma/T)$ \cite{IEsterlis2021,AAPatel2014}.

Assuming the $\varphi$ boson is equilibrated with the electrons at $T_e \approx T_p \equiv T$, the energy transfer rate takes the form $\partial_t E = \kappa^{(2)}(T_e - T_p)$, where 
\begin{equation}\label{eq:kappa2}
\kappa^{(2)} = \frac{1}{4} \sum_i \int \frac{d^3 \vec{q}}{(2\pi)^3} |N^i_{\vec{q}}|^2\, [-A_\varphi(c|\vec{q}|, \vec{q})]\, \frac{c|\vec{q}|}{2T^2 \sinh^2(c|\vec{q}|/2T)}.
\end{equation}
 For $T \ll \omega_D$, we approximate $|\vec{q}| \sim T/c$ and, using Goldstone’s theorem, take $\sum_i |N^i_{\vec{q}}|^2 \approx A_N \vn{q_\text{2D}}^2$. This yields
\begin{equation}\label{eq:kappa21}
\kappa^{(2)} \approx \frac{A_N}{c^5} \int_0^\pi d\theta\, \frac{\gamma T^5 \sin^3 \theta}{\gamma^2 T^2 + \left[m^2(T) + T^2 v_\varphi^2 \sin^2\theta / c^2\right]^2},
\end{equation}
where $\theta$ is the angle between $\vec{q}$ and the $z$-axis.

\begin{figure}[htb]
   \centering
   \includegraphics[width=\linewidth]{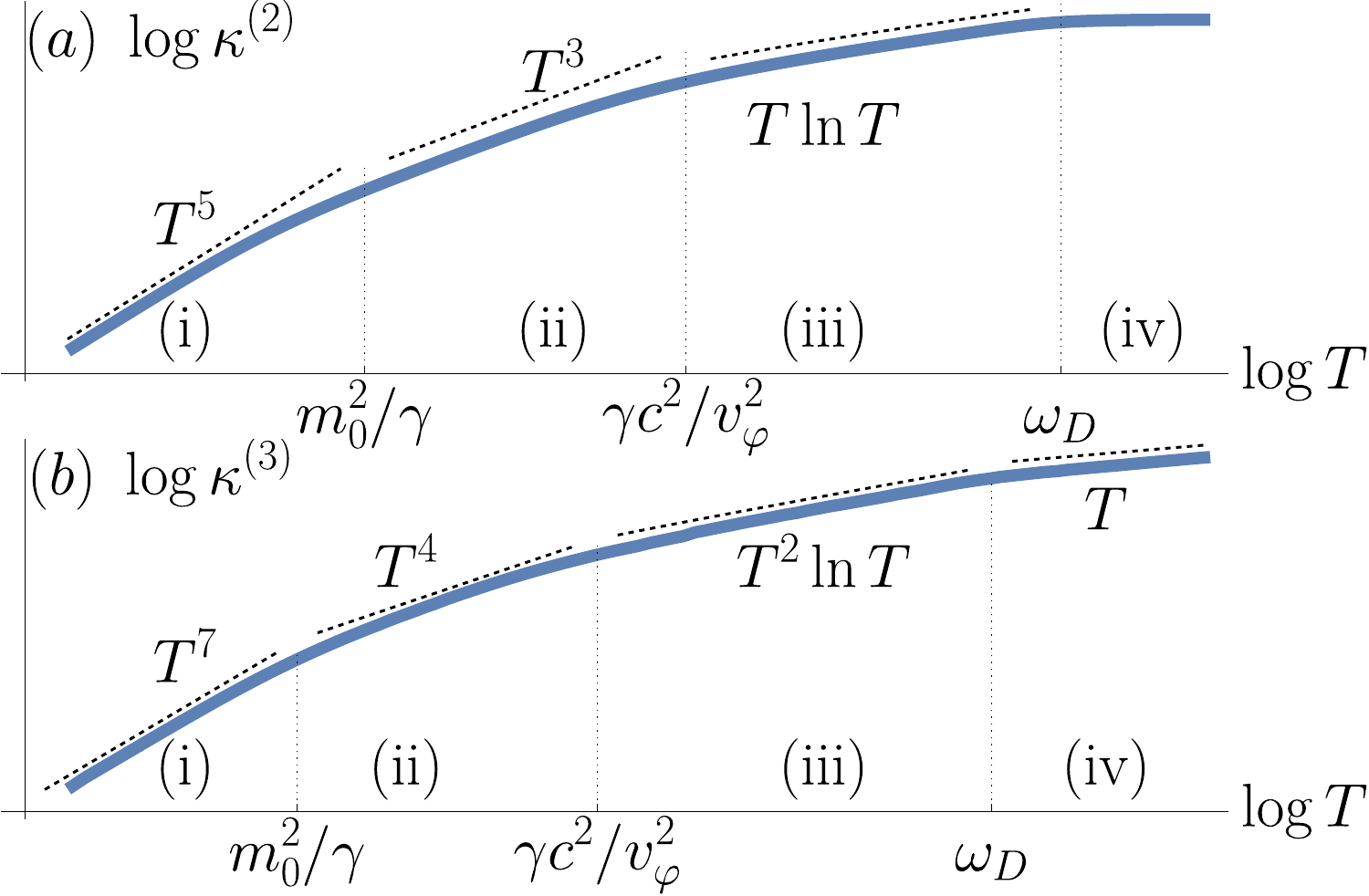}
   \caption{Numerical plot of (a) $\kappa^{(2)}(T)$ and (b) $\kappa^{(3)}(T)$ on log-log scale. The blue curves are numerical evaluation of Eqs.~\eqref{eq:kappa2} and \eqref{eq:kappa3}. The dashed lines denote the approximate scaling forms deduced analytically. } \label{fig:kappa23}
 \end{figure}

 $\kappa^{(2)}$  exhibits a complex sequence of crossovers marked by different power-laws as a function of temperature, as illustrated in Fig.~\ref{fig:results}  and 
Fig.~\ref{fig:kappa23} (a). They are as follows: 

(i) Let us start slightly away from the QCP, $m(T=0)=m_0>0$, which is assumed to be smaller than all other energy scales, including $\omd$. At the lowest temperature, the denominator of Eq.~\eqref{eq:kappa21} is dominated by $m_0$, so we obtain $\kappa^{(2)}\sim (A_N/c^5) (\gamma/m_0^4) T^5$.  In this regime, $\varphi$  mediates \emph{local} interaction between the electrons, and the effects of damping and dispersion are suppressed relative to $m_0$. 
Alternatively, the result can be derived from $\kappa^{(2)}\sim A_N T^4\times\braket{\varphi\varphi}$, where the $T^4$ prefactor is the result of phonon phase counting and Gold-stone coupling.
%Alternatively, by counting the phonon phase-space, energy transfer and expansion around the thermal equilibrium, we can obtain $\kappa^{(2)}\sim A_N T^4\times\braket{\varphi\varphi}$.
In the low-energy and long-wavelength limit, the boson $\varphi$ is expected to behave as an ohmic bath, which yields $\braket{\varphi\varphi}\propto \gamma T/m_0^4$.
We also remark in passing that similar bosonic collective modes are generically present in interacting Fermi liquids, and $\kappa^{(2)}$ can be interpreted as an interaction correction to energy relaxation on top of the more dominant free fermion result $\kappa^{(1)}$. When the system is deep in the Fermi liquid phase, the boson mass $m_0$ is expected to be an electronic energy scale that overwhelms other crossovers, i.e. $\kappa^{(2)}$ grows as $T^5$ until being cutoff at $T\sim \omd$.

(ii) With increasing $T>m_0^2/\gamma$, the $\gamma^2 T^2$ term in the denominator of Eq.~\eqref{eq:kappa21} dominates, so $\kappa^{(2)}\sim (A_N/c^5) T^3/\gamma$, which becomes comparable to $\kappa^{(1)}$, albeit with a different coefficient. In this regime, the $z_{\varphi}=2$ dynamics first rears its head; however, the typical phonon momentum is small compared to the typical $\varphi$ frequencies. Therefore, the phonons are more sensitive to the homogeneous fluctuations of $\varphi$, as opposed to their full diffusive character. Mathematically, regimes (i) and (ii) can be described in a unified fashion by defining $\Delta(T)=\max(m(T),\sqrt{\gamma T})$ and writing $\kappa^{(2)}\sim (A_N/c^5)\gamma T (T/\Delta(T))^4$; this applies to the quantum critical fan ($m_0=0$) as well as for $\Delta(T)\to m(T)$. 

(iii) When $T\gtrsim \gamma(c^2/v_\varphi^2)$, the $T^2v_\varphi^2/c^2$ term in the denominator becomes relevant. The $\theta$-integral in Eq.~\eqref{eq:kappa21} should be cutoff at $\theta_\text{min}\sim (c/v_\varphi)\Delta(T)/T$, which yields $\kappa^{(2)}\sim A_N/(cv_\varphi^4)\gamma T\ln (v_\varphi^2T/(c^2 \gamma))$ (see End Matter), which signals an enhanced contribution due to phonons moving in the $z$ direction. In this regime, the typical phonon momentum is large compared to the typical $\varphi$ frequencies, and unlike the previous lower-temperature regime, the phonons are now sensitive to diffusive character of $\varphi$. %This regime is unique to phonons for the following reason. Suppose if we were to replace the phonons by a distinct bosonic mode of electronic origin, but now with $c\approx v_\varphi$. Then this regime would onset at $T\sim\gamma c^2/v_\varphi^2\approx \gamma$, which is a high electronic energy scale. In the present setting associated with phonons, the onset of the $\kappa^{(2)}\sim T\ln(T)$ regime is tied to the parametric suppression in $c\ll v_\varphi$. 
 Such a regime is unique to phonons: for a bosonic mode of electronic origin with $c \sim v_\varphi$, it would only emerge at $T \sim \gamma$, an electronic scale. Here, it appears at much lower temperatures due to $c \ll v_\varphi$.

(iv) With a further increase $T\gtrsim\omd$, the integral saturates, and we obtain a constant $\kappa^{(2)}\sim A_N/(cv_\varphi^4)\gamma \omega_D \ln (v_\varphi^2 \omega_D/(c^2 \gamma))$. 

It is useful to make a few remarks about the underlying dimensionality of the problem. The lowest temperature regimes (i), (ii)  exhibiting $\kappa^{(2)}\sim T^5$ and $\kappa^{(2)}\sim T^3$ depend on the phonon dimensionality in a straightforward fashion via their density of states. However, the $\kappa^{(2)}\sim T\ln(T)$ regime (iii) involves a subtle kinematic interplay between the phonon and $\varphi$. In particular, the phonon dispersion along the $z-$direction contributes more significantly compared to the other directions, leading to the $\ln$ enhancement; the latter would be absent if the phonons were purely two-dimensional. Furthermore, $\kappa^{(2)}$ in this regime also depends crucially on the assumption of completely decoupled nature of the two-dimensional electronic layers, via the simplification, $N^i_{\vec{q}}\propto\vn{q_\text{2D}}$, instead of $\vn{q}$. Finally, let us also consider the additional perturbation due to coupling the $\varphi$ boson to out-of-plane strain, which leads to a correction, $\lambda q_z^2$ term in $|N^i_{\vec{q}}|^2$ with $\lambda\ll 1$. This does not lead to qualitatively different results in regimes (i) and (ii), but it can \emph{enhance} the $\ln$ factor in regime (iii) to a linear factor of $v_\varphi^2 T/(c^2 \gamma)$ (see End Matter). Since this regime is upper bounded by $T\sim \omega_D$, this term remains a small perturbation if $\lambda\ll \gamma c^2/(\omega_D v_\varphi^2)$.

\setlength{\abovecaptionskip}{0.1cm}  % space above caption
\setlength{\belowcaptionskip}{-0.4cm} 

 {\it Energy-relaxation due to non-linear coupling between phonon and collective mode.-}  Finally, we consider the contribution due to $\calL_{ep}^{(3)}$ in Eq.~\eqref{eq:coupling3}, with a phonon self-energy shown in Fig.~\ref{fig:fig1}(d), which involves scattering between $\varphi^2$ and the acoustic phonons. %The renormalized spectral function for $\varphi$ is still given by Eq.~\eqref{eq:Aphi}. 
 The general expression for the flux $\kappa^{(3)}$ reads 
 \begin{equation}\label{eq:kappa3}
 \begin{split}
  &\kappa^{(3)}=-\frac{1}{16} \int_{\vec{q}}\frac{\rd^3\vec{q}}{(2\pi)^3}   \frac{c\vn{q}}{2T^2\sinh^2\frac{c\vn{q}}{2T}}  \int\frac{\rd \nu}{2\pi}\int\frac{\rd^3\vec{k}}{(2\pi)^3} \\
  &\times \sum_{i} |L^i_{\vec{k}+\vec{q}/2,\vec{q}}|^2A_{\varphi}(\nu,\vec{k}) A_{\varphi}(\nu+c\vn{q},\vec{k}+\vec{q}) \\
  &\times \left[\coth\bigg(\frac{\nu+c\vn{q}}{2T}\bigg)-\coth\bigg(\frac{\nu}{2T}\bigg)\right]\,.
 \end{split}
\end{equation}
  Once again, there is a complex sequence of temperature dependent crossovers that can be obtained with the following approximations: replace $\sum_i |L^i_{\vec{k}+\vec{q}/2,\vec{q}}|^2$ by $A_L \vn{q_\text{2D}}^2$, and for $T\lesssim\omd$, include contributions from phase-space where $c\vn{q}\sim\nu\sim T$. %Then $\kappa^{(3)}$ is schematically given by,
  We obtain
\begin{equation}\label{eq:kappa32}
\begin{split}
  \kappa^{(3)} &\approx\frac{A_L}{a_zc^5} T^5 \int_0^\pi \rd \theta \sin^3 \theta \int \rd^2\vec{k}_\text{2D} \\
  &\times A_\varphi(T,\vec{k}_\text{2D}) A_\varphi(T,\vec{k}_\text{2D}+\vec{q}_\text{2D}(\theta))\,,
\end{split}
\end{equation} 
where the vector $\vec{q}_\text{2D}(\theta)$ has norm $(T/c)\sin \theta$, and $\theta$ is the angle between $\vec{q}$ and the z-axis.

The regimes where $\kappa^{(3)}$ exhibits the distinct $T-$dependent crossovers are identical to the ones that arise  in the discussion for $\kappa^{(2)}$ above. (i) For $T\lesssim m_0^2/\gamma$, the $\vp$-spectral functions are dominated by $m_0$ and the $\vec{k}$-integral can be estimated as $A_\varphi^2(T,k_\text{typ}) k_\text{typ}^2 $ treating the $\vec{q}_\text{2D}(\theta)$ term as being unimportant. Here $k_\text{typ}$ is the typical magnitude of $\vec{k}$, estimated to be $k_\text{typ}\sim m_0/v_\varphi$. Therefore, we obtain $\kappa^{(3)}\sim (A_L/a_z c^5 v_\varphi^2) \gamma^2 T^7/m_0^6$, which is strongly suppressed compared to $\kappa^{(2)}$, as one might expect based on scaling arguments. Alternatively, a simple counting argument leads to $\kappa^{(3)}\sim A_L T^4 \times \braket{\varphi^4}$. Here $\braket{\varphi^4}\propto T^3/m_0^6$ where the $T^2/m_0^8$ comes from two ohmic baths and there are $N\propto Tm_0^2$ different ways to distribute the energy between and the momentum these two baths. (ii) As before, when $m_0^2/\gamma<T<\gamma(c^2/v_\varphi^2)$, we can replace $m_0^2$ by $\gamma T$, and obtain $\kappa^{(3)}\sim (A_L/a_z c^5 v_\varphi^2) T^4/\gamma$. If we take $m_0\rightarrow0$ in the quantum critical region, these regimes crossover smoothly into each other with $\kappa^{(3)}\sim (A_L/a_z c^5 v_\varphi^2)\gamma^2 T^7/m^6(T)$.
(iii) When $\gamma(c^2/v_\varphi^2)<T<\omd$, the $\vec{q}_\text{2D}(\theta)$ term in the second $A_\varphi(...)$ factor becomes more important than $\vec{k}_\text{2D}$, leading to $\kappa^{(3)}\sim (A_L/a_zc v_\varphi^6) \gamma T^2 \ln (v_\varphi^2T/(c^2 \gamma))$. The logarithmic factor has the same origin as that in $\kappa^{(2)}$. In the quantum critical regime, this becomes $\kappa^{(3)} \sim (A_L/a_z cv_\varphi^6) (\gamma^2 T^3/m^2(T))\ln (v_\varphi^2T/(c^2 \gamma))$. (iv) Finally, when $T>\omd$, $\kappa^{(3)}\sim (A_L/a_zc v_\varphi^6) \gamma T \omd \ln (v_\varphi^2\omd/(c^2 \gamma))$. Here, the high-$T$ limit of $\kappa^{(3)}$ is different from $\kappa^{(1)}$ and $\kappa^{(2)}$ in that it does not saturate but grows linearly in $T$. This is expected as $\kappa^{(3)}$ involves an additional boson compared to $\kappa^{(2)}$, and therefore the result should be proportional to the density of that additional boson, which is proportional to $T$, and this will be ultimately cut off at an electronic energy scale where $\varphi$ reveals its electronic nature.  
A comparison between analytical estimate and numerical result is shown in Fig.~\ref{fig:kappa23}.

{\it Connection to Experiments.-} Recent optical experiments involving the technique of two-dimensional coherent spectroscopy \cite{keithnelson,2DCS,Mahmood2021} have studied the nonlinear response of correlated metals and its relationship to energy relaxation \cite{DBarbalas2023,DChaudhuri2025}, providing a further impetus to a careful theoretical examination of the questions raised above. 
Over a broad range of hole-dopings in a specific cuprate family \cite{IBozovic2016,FMahmood2018}, recent pump-probe measurements \cite{DChaudhuri2025} have identified energy relaxation as the dominant nonlinear process that governs the experimental phenomenology. At low temperatures across most of the doping phase diagram, $\Gamma_E\sim k_BT/\hbar$, with a rate that is significantly slower (approximately by an order of magnitude) compared to the Planckian momentum-relaxation rate \cite{JANBruin2013,PGiraldo-Gallo2018}. With increasing temperature, there is a tendency for $\Gamma_E$ to saturate across a doping dependent crossover scale. Both of these regimes are in stark contrast to the results expected in a conventional Fermi liquid coupled to acoustic phonons \cite{PBAllen1987}. In THz experiments, it is the second rate $\Gamma_{ep}$ that is measured, which leads to a fast initial decay of the pump-probe signal to a plateau at the order of pico seconds, whereas the phonon-environment relaxation $\Gamma_\text{p-env}$ leads to a slower decay on longer timescales. 

To make contact with experiments, we convert $\kappa = [\kappa^{(1)}+\kappa^{(2)}+\kappa^{(3)}]$ to the energy relaxation rate, $\Gamma_E$, via Eq.~\eqref{eq:GammaE}. Here, due to the comparison between the heat capacities $C_p \sim (T/\omd)^3$ and $C_e \sim (T/E_F) \ln(\gamma/T)$, an additional crossover is present, whose temperature scale $T_*\sim \sqrt{\omd^2/E_F}$ is set by equating $C_p$ and $C_e$. Eq.~\eqref{eq:GammaE} states that the sector with the smaller heat capacity dominates as its temperature changes more rapidly. When $T\ll T_*$, we have $\Gamma_E\approx \kappa / C_p$ and when $T\gg T_*$, $\Gamma_E \approx \kappa/ C_e$. From thermodynamic measurements in a representative family of the cuprates \cite{BMichon2019}, $T_*$ is at the order of $1$-$10$ K. For the experimental range of $T\sim O(100 \textrm{ K})$, the latter is more relevant, so we would then divide the scaling of $\kappa$ by a power of $T$ to get scaling of $\Gamma_E$. Our theoretical results for $\kappa^{(3)}$ provide an appealing scenario for the experimental observations. In the experimental temperature range, the energy relaxation rate $\Gamma_E$ is related to $\kappa^{(3)}$ by $\Gamma_E\approx \kappa^{(3)}/C_e$. Therefore, in regime (iii) and regime (iv) we would obtain a crossover from $T\ln T$ to $T$-independent constant in $\Gamma_E$, which qualitatively agrees with the observation in \cite{DChaudhuri2025} for samples near optimal doping (see Fig. 6c in \cite{DChaudhuri2025}).

{\it Discussion and Outlook.-} We have shown that near an electronic QCP, symmetry-allowed couplings to acoustic phonons generate rich, temperature-dependent crossovers in the energy relaxation rate. When momentum relaxation in a marginal Fermi liquid is governed by electronic processes and disorder, while energy relaxation is limited by phonon coupling, there is no {\it a priori} reason for the two relaxation rates to exhibit a similar behavior. Our analysis reveals that energy relaxation follows distinct scaling behaviors across emergent low-energy regimes, independent of momentum relaxation. \change{We also note that (a) while the electron-phonon coupling, can in principle, modify the structure of the underlying QCP \cite{QiXu,Halperin76,MZacharias2015,IPaul2017}, the interesting regime (iii) obtained above will not be affected, and (b) the energy relaxation process we address is not affected by potential boson drag effects \cite{YWerman2017,HGuo2019,DVElse2021,HGuo2022a}; see  \cite{Supp} for  details.}

Recent experiments \cite{JZhang2017a,JZhang2019,JZhang2019a} have motivated proposals that phonons in strongly correlated metals may lose their quasiparticle character, exhibiting Planckian lifetimes $O(\hbar/k_B T)$ \cite{YWerman2017,YWerman2017a,ETulipman2020a,ETulipman2021,HGuo2019}. Despite this broadening, our results for energy flux should still apply, as the relevant phonon frequencies remain $O(k_B T)$. A full theoretical treatment of a non-Fermi liquid entangled with incoherent phonons is left for future work. {Additionally, it would be interesting to investigate how the phonon dynamics is modified by the non-Fermi liquid through numerical approaches, such as quantum Monte-Carlo \cite{AAPatel2024a,EBerg2019,AKlein2020,XYXu2020,XYXu2017}.}

In this work we have focused on the normal state, but pairing fluctuations above $T_c$ and Bogoliubov quasiparticles below $T_c$ may also influence energy relaxation through their coupling to phonons \cite{ARothwarf1967}, an interesting direction for future study. Moreover, the complex crossovers we identify—arising from mismatched velocities and dimensionality of phonons and collective modes—suggest similar behavior may appear in sound attenuation near quantum criticality \cite{mason,morse,Pippard,Blount,tsuneto,Allen_US,PAL_sound,ABBhatia1961}, which we leave for future investigation.

{\it Acknowledgments.-}  We thank N.P. Armitage, D. Chaudhuri, and B. Ramshaw for stimulating discussions. H.G. is supported by a Wilkins postdoctoral fellowship at Cornell University. D.~C. is supported in part by a NSF CAREER grant (DMR-2237522), and a Sloan Research Fellowship. 

\bibliography{Refs1}

\begin{thebibliography}{126}%
\makeatletter
\providecommand \@ifxundefined [1]{%
 \@ifx{#1\undefined}
}%
\providecommand \@ifnum [1]{%
 \ifnum #1\expandafter \@firstoftwo
 \else \expandafter \@secondoftwo
 \fi
}%
\providecommand \@ifx [1]{%
 \ifx #1\expandafter \@firstoftwo
 \else \expandafter \@secondoftwo
 \fi
}%
\providecommand \natexlab [1]{#1}%
\providecommand \enquote  [1]{``#1''}%
\providecommand \bibnamefont  [1]{#1}%
\providecommand \bibfnamefont [1]{#1}%
\providecommand \citenamefont [1]{#1}%
\providecommand \href@noop [0]{\@secondoftwo}%
\providecommand \href [0]{\begingroup \@sanitize@url \@href}%
\providecommand \@href[1]{\@@startlink{#1}\@@href}%
\providecommand \@@href[1]{\endgroup#1\@@endlink}%
\providecommand \@sanitize@url [0]{\catcode `\\12\catcode `\$12\catcode
  `\&12\catcode `\#12\catcode `\^12\catcode `\_12\catcode `\%12\relax}%
\providecommand \@@startlink[1]{}%
\providecommand \@@endlink[0]{}%
\providecommand \url  [0]{\begingroup\@sanitize@url \@url }%
\providecommand \@url [1]{\endgroup\@href {#1}{\urlprefix }}%
\providecommand \urlprefix  [0]{URL }%
\providecommand \Eprint [0]{\href }%
\providecommand \doibase [0]{https://doi.org/}%
\providecommand \selectlanguage [0]{\@gobble}%
\providecommand \bibinfo  [0]{\@secondoftwo}%
\providecommand \bibfield  [0]{\@secondoftwo}%
\providecommand \translation [1]{[#1]}%
\providecommand \BibitemOpen [0]{}%
\providecommand \bibitemStop [0]{}%
\providecommand \bibitemNoStop [0]{.\EOS\space}%
\providecommand \EOS [0]{\spacefactor3000\relax}%
\providecommand \BibitemShut  [1]{\csname bibitem#1\endcsname}%
\let\auto@bib@innerbib\@empty
%</preamble>
\bibitem [{\citenamefont {Bruin}\ \emph {et~al.}(2013)\citenamefont {Bruin},
  \citenamefont {Sakai}, \citenamefont {Perry},\ and\ \citenamefont
  {Mackenzie}}]{JANBruin2013}%
  \BibitemOpen
  \bibfield  {author} {\bibinfo {author} {\bibfnamefont {J.~A.~N.}\
  \bibnamefont {Bruin}}, \bibinfo {author} {\bibfnamefont {H.}~\bibnamefont
  {Sakai}}, \bibinfo {author} {\bibfnamefont {R.~S.}\ \bibnamefont {Perry}},\
  and\ \bibinfo {author} {\bibfnamefont {A.~P.}\ \bibnamefont {Mackenzie}},\
  }\bibfield  {title} {\bibinfo {title} {Similarity of scattering rates in
  metals showing {{T-Linear}} resistivity},\ }\href
  {https://doi.org/10.1126/science.1227612} {\bibfield  {journal} {\bibinfo
  {journal} {Science}\ }\textbf {\bibinfo {volume} {339}},\ \bibinfo {pages}
  {804} (\bibinfo {year} {2013})}\BibitemShut {NoStop}%
\bibitem [{\citenamefont {Hartnoll}\ and\ \citenamefont
  {Mackenzie}(2022)}]{SAHartnoll2022}%
  \BibitemOpen
  \bibfield  {author} {\bibinfo {author} {\bibfnamefont {S.~A.}\ \bibnamefont
  {Hartnoll}}\ and\ \bibinfo {author} {\bibfnamefont {A.~P.}\ \bibnamefont
  {Mackenzie}},\ }\bibfield  {title} {\bibinfo {title} {Colloquium:
  {{Planckian}} dissipation in metals},\ }\href
  {https://doi.org/10.1103/RevModPhys.94.041002} {\bibfield  {journal}
  {\bibinfo  {journal} {Rev. Mod. Phys.}\ }\textbf {\bibinfo {volume} {94}},\
  \bibinfo {pages} {041002} (\bibinfo {year} {2022})}\BibitemShut {NoStop}%
\bibitem [{\citenamefont {Chowdhury}\ \emph {et~al.}(2022)\citenamefont
  {Chowdhury}, \citenamefont {Georges}, \citenamefont {Parcollet},\ and\
  \citenamefont {Sachdev}}]{DChowdhury2022a}%
  \BibitemOpen
  \bibfield  {author} {\bibinfo {author} {\bibfnamefont {D.}~\bibnamefont
  {Chowdhury}}, \bibinfo {author} {\bibfnamefont {A.}~\bibnamefont {Georges}},
  \bibinfo {author} {\bibfnamefont {O.}~\bibnamefont {Parcollet}},\ and\
  \bibinfo {author} {\bibfnamefont {S.}~\bibnamefont {Sachdev}},\ }\bibfield
  {title} {\bibinfo {title} {Sachdev-{{Ye-Kitaev}} models and beyond:
  {{Window}} into non-{{Fermi}} liquids},\ }\href
  {https://doi.org/10.1103/RevModPhys.94.035004} {\bibfield  {journal}
  {\bibinfo  {journal} {Rev. Mod. Phys.}\ }\textbf {\bibinfo {volume} {94}},\
  \bibinfo {pages} {035004} (\bibinfo {year} {2022})}\BibitemShut {NoStop}%
\bibitem [{\citenamefont {Abrikosov}(1963)}]{AAAbrikosov1963}%
  \BibitemOpen
  \bibfield  {author} {\bibinfo {author} {\bibfnamefont {A.~A.}\ \bibnamefont
  {Abrikosov}},\ }\href@noop {} {\emph {\bibinfo {title} {Methods of Quantum
  Field Theory in Statistical Physics}}},\ \bibinfo {edition} {rev. english ed.
  translated and edited by richard a. silverman.}\ ed.,\ Selected {{Russian}}
  Publications in the Mathematical Sciences\ (\bibinfo  {publisher}
  {Prentice-Hall},\ \bibinfo {address} {Englewood Cliffs, N.J.},\ \bibinfo
  {year} {1963})\BibitemShut {NoStop}%
\bibitem [{\citenamefont {Ziman}(2001)}]{ziman2001electrons}%
  \BibitemOpen
  \bibfield  {author} {\bibinfo {author} {\bibfnamefont {J.~M.}\ \bibnamefont
  {Ziman}},\ }\href@noop {} {\emph {\bibinfo {title} {Electrons and phonons:
  the theory of transport phenomena in solids}}}\ (\bibinfo  {publisher}
  {Oxford university press},\ \bibinfo {year} {2001})\BibitemShut {NoStop}%
\bibitem [{\citenamefont {Sachdev}(1999)}]{SSachdev1999b}%
  \BibitemOpen
  \bibfield  {author} {\bibinfo {author} {\bibfnamefont {S.}~\bibnamefont
  {Sachdev}},\ }\href
  {http://www.cambridge.org/us/academic/subjects/physics/condensed-matter-physics-nanoscience-and-mesoscopic-physics/quantum-phase-transitions-2nd-edition?format=HB&isbn=9780521514682}
  {\emph {\bibinfo {title} {Quantum Phase Transitions}}},\ \bibinfo {edition}
  {1st}\ ed.\ (\bibinfo  {publisher} {Cambridge University Press},\ \bibinfo
  {address} {Cambridge, UK},\ \bibinfo {year} {1999})\BibitemShut {NoStop}%
\bibitem [{\citenamefont {Taillefer}(2010)}]{LTaillefer2010}%
  \BibitemOpen
  \bibfield  {author} {\bibinfo {author} {\bibfnamefont {L.}~\bibnamefont
  {Taillefer}},\ }\bibfield  {title} {\bibinfo {title} {Scattering and pairing
  in cuprate superconductors},\ }\href
  {https://doi.org/10.1146/annurev-conmatphys-070909-104117} {\bibfield
  {journal} {\bibinfo  {journal} {Annual Review of Condensed Matter Physics}\
  }\textbf {\bibinfo {volume} {1}},\ \bibinfo {pages} {51} (\bibinfo {year}
  {2010})},\ \Eprint {https://arxiv.org/abs/1003.2972} {arXiv:1003.2972
  [cond-mat.supr-con]} \BibitemShut {NoStop}%
\bibitem [{\citenamefont {Shibauchi}\ \emph {et~al.}(2014)\citenamefont
  {Shibauchi}, \citenamefont {Carrington},\ and\ \citenamefont
  {Matsuda}}]{matsuda}%
  \BibitemOpen
  \bibfield  {author} {\bibinfo {author} {\bibfnamefont {T.}~\bibnamefont
  {Shibauchi}}, \bibinfo {author} {\bibfnamefont {A.}~\bibnamefont
  {Carrington}},\ and\ \bibinfo {author} {\bibfnamefont {Y.}~\bibnamefont
  {Matsuda}},\ }\bibfield  {title} {\bibinfo {title} {A quantum critical point
  lying beneath the superconducting dome in iron pnictides},\ }\href
  {https://doi.org/https://doi.org/10.1146/annurev-conmatphys-031113-133921}
  {\bibfield  {journal} {\bibinfo  {journal} {Annual Review of Condensed Matter
  Physics}\ }\textbf {\bibinfo {volume} {5}},\ \bibinfo {pages} {113} (\bibinfo
  {year} {2014})}\BibitemShut {NoStop}%
\bibitem [{\citenamefont {Greene}\ \emph {et~al.}(2020)\citenamefont {Greene},
  \citenamefont {Mandal}, \citenamefont {Poniatowski},\ and\ \citenamefont
  {Sarkar}}]{greeneSC}%
  \BibitemOpen
  \bibfield  {author} {\bibinfo {author} {\bibfnamefont {R.~L.}\ \bibnamefont
  {Greene}}, \bibinfo {author} {\bibfnamefont {P.~R.}\ \bibnamefont {Mandal}},
  \bibinfo {author} {\bibfnamefont {N.~R.}\ \bibnamefont {Poniatowski}},\ and\
  \bibinfo {author} {\bibfnamefont {T.}~\bibnamefont {Sarkar}},\ }\bibfield
  {title} {\bibinfo {title} {The strange metal state of the electron-doped
  cuprates},\ }\href
  {https://doi.org/https://doi.org/10.1146/annurev-conmatphys-031119-050558}
  {\bibfield  {journal} {\bibinfo  {journal} {Annual Review of Condensed Matter
  Physics}\ }\textbf {\bibinfo {volume} {11}},\ \bibinfo {pages} {213}
  (\bibinfo {year} {2020})}\BibitemShut {NoStop}%
\bibitem [{\citenamefont {Paschen}\ and\ \citenamefont
  {Si}(2021)}]{Paschen2021}%
  \BibitemOpen
  \bibfield  {author} {\bibinfo {author} {\bibfnamefont {S.}~\bibnamefont
  {Paschen}}\ and\ \bibinfo {author} {\bibfnamefont {Q.}~\bibnamefont {Si}},\
  }\bibfield  {title} {\bibinfo {title} {Quantum phases driven by strong
  correlations},\ }\href {https://doi.org/10.1038/s42254-020-00262-6}
  {\bibfield  {journal} {\bibinfo  {journal} {Nature Reviews Physics}\ }\textbf
  {\bibinfo {volume} {3}},\ \bibinfo {pages} {9} (\bibinfo {year}
  {2021})}\BibitemShut {NoStop}%
\bibitem [{\citenamefont {Lee}(2018)}]{SSLee2018}%
  \BibitemOpen
  \bibfield  {author} {\bibinfo {author} {\bibfnamefont {S.-S.}\ \bibnamefont
  {Lee}},\ }\bibfield  {title} {\bibinfo {title} {Recent {{Developments}} in
  {{Non-Fermi Liquid Theory}}},\ }\href
  {https://doi.org/10.1146/annurev-conmatphys-031016-025531} {\bibfield
  {journal} {\bibinfo  {journal} {Annual Review of Condensed Matter Physics}\
  }\textbf {\bibinfo {volume} {9}},\ \bibinfo {pages} {227} (\bibinfo {year}
  {2018})}\BibitemShut {NoStop}%
\bibitem [{\citenamefont {Berg}\ \emph {et~al.}(2019)\citenamefont {Berg},
  \citenamefont {Lederer}, \citenamefont {Schattner},\ and\ \citenamefont
  {Trebst}}]{EBerg2019}%
  \BibitemOpen
  \bibfield  {author} {\bibinfo {author} {\bibfnamefont {E.}~\bibnamefont
  {Berg}}, \bibinfo {author} {\bibfnamefont {S.}~\bibnamefont {Lederer}},
  \bibinfo {author} {\bibfnamefont {Y.}~\bibnamefont {Schattner}},\ and\
  \bibinfo {author} {\bibfnamefont {S.}~\bibnamefont {Trebst}},\ }\bibfield
  {title} {\bibinfo {title} {Monte carlo studies of quantum critical metals},\
  }\href {https://doi.org/10.1146/annurev-conmatphys-031218-013339} {\bibfield
  {journal} {\bibinfo  {journal} {Annual Review of Condensed Matter Physics}\
  }\textbf {\bibinfo {volume} {10}},\ \bibinfo {pages} {63} (\bibinfo {year}
  {2019})},\ \Eprint {https://arxiv.org/abs/1804.01988} {arXiv:1804.01988
  [cond-mat.str-el]} \BibitemShut {NoStop}%
\bibitem [{\citenamefont {Varma}(2020)}]{varmaRMP}%
  \BibitemOpen
  \bibfield  {author} {\bibinfo {author} {\bibfnamefont {C.~M.}\ \bibnamefont
  {Varma}},\ }\bibfield  {title} {\bibinfo {title} {Colloquium: Linear in
  temperature resistivity and associated mysteries including high temperature
  superconductivity},\ }\href {https://doi.org/10.1103/RevModPhys.92.031001}
  {\bibfield  {journal} {\bibinfo  {journal} {Rev. Mod. Phys.}\ }\textbf
  {\bibinfo {volume} {92}},\ \bibinfo {pages} {031001} (\bibinfo {year}
  {2020})}\BibitemShut {NoStop}%
\bibitem [{\citenamefont {Fradkin}\ \emph {et~al.}(2010)\citenamefont
  {Fradkin}, \citenamefont {Kivelson}, \citenamefont {Lawler}, \citenamefont
  {Eisenstein},\ and\ \citenamefont {Mackenzie}}]{nematicreview}%
  \BibitemOpen
  \bibfield  {author} {\bibinfo {author} {\bibfnamefont {E.}~\bibnamefont
  {Fradkin}}, \bibinfo {author} {\bibfnamefont {S.~A.}\ \bibnamefont
  {Kivelson}}, \bibinfo {author} {\bibfnamefont {M.~J.}\ \bibnamefont
  {Lawler}}, \bibinfo {author} {\bibfnamefont {J.~P.}\ \bibnamefont
  {Eisenstein}},\ and\ \bibinfo {author} {\bibfnamefont {A.~P.}\ \bibnamefont
  {Mackenzie}},\ }\bibfield  {title} {\bibinfo {title} {Nematic fermi fluids in
  condensed matter physics},\ }\href
  {https://doi.org/https://doi.org/10.1146/annurev-conmatphys-070909-103925}
  {\bibfield  {journal} {\bibinfo  {journal} {Annual Review of Condensed Matter
  Physics}\ }\textbf {\bibinfo {volume} {1}},\ \bibinfo {pages} {153} (\bibinfo
  {year} {2010})}\BibitemShut {NoStop}%
\bibitem [{\citenamefont {Varma}\ \emph {et~al.}(1989)\citenamefont {Varma},
  \citenamefont {Littlewood}, \citenamefont {{Schmitt-Rink}}, \citenamefont
  {Abrahams},\ and\ \citenamefont {Ruckenstein}}]{CMVarma1989}%
  \BibitemOpen
  \bibfield  {author} {\bibinfo {author} {\bibfnamefont {C.~M.}\ \bibnamefont
  {Varma}}, \bibinfo {author} {\bibfnamefont {P.~B.}\ \bibnamefont
  {Littlewood}}, \bibinfo {author} {\bibfnamefont {S.}~\bibnamefont
  {{Schmitt-Rink}}}, \bibinfo {author} {\bibfnamefont {E.}~\bibnamefont
  {Abrahams}},\ and\ \bibinfo {author} {\bibfnamefont {A.~E.}\ \bibnamefont
  {Ruckenstein}},\ }\bibfield  {title} {\bibinfo {title} {Phenomenology of the
  normal state of {{Cu-O}} high-temperature superconductors},\ }\href
  {https://doi.org/10.1103/PhysRevLett.63.1996} {\bibfield  {journal} {\bibinfo
   {journal} {Phys. Rev. Lett.}\ }\textbf {\bibinfo {volume} {63}},\ \bibinfo
  {pages} {1996} (\bibinfo {year} {1989})}\BibitemShut {NoStop}%
\bibitem [{\citenamefont {Anderson}(1991)}]{PWAnderson1991a}%
  \BibitemOpen
  \bibfield  {author} {\bibinfo {author} {\bibfnamefont {P.~W.}\ \bibnamefont
  {Anderson}},\ }\bibfield  {title} {\bibinfo {title} {Hall effect in the
  two-dimensional {Luttinger} liquid},\ }\href
  {https://doi.org/10.1103/PhysRevLett.67.2092} {\bibfield  {journal} {\bibinfo
   {journal} {Physical Review Letters}\ }\textbf {\bibinfo {volume} {67}},\
  \bibinfo {pages} {2092} (\bibinfo {year} {1991})},\ \bibinfo {note}
  {publisher: American Physical Society}\BibitemShut {NoStop}%
\bibitem [{\citenamefont {Anderson}(1992)}]{PWAnderson1992b}%
  \BibitemOpen
  \bibfield  {author} {\bibinfo {author} {\bibfnamefont {P.~W.}\ \bibnamefont
  {Anderson}},\ }\bibfield  {title} {\bibinfo {title} {Experimental
  {Constraints} on the {Theory} of {High}-{Tc} {Superconductivity}},\ }\href
  {https://doi.org/10.1126/science.256.5063.1526} {\bibfield  {journal}
  {\bibinfo  {journal} {Science}\ }\textbf {\bibinfo {volume} {256}},\ \bibinfo
  {pages} {1526} (\bibinfo {year} {1992})},\ \bibinfo {note} {publisher:
  American Association for the Advancement of Science}\BibitemShut {NoStop}%
\bibitem [{\citenamefont {Patel}\ \emph {et~al.}(2018)\citenamefont {Patel},
  \citenamefont {McGreevy}, \citenamefont {Arovas},\ and\ \citenamefont
  {Sachdev}}]{AAPatel2018b}%
  \BibitemOpen
  \bibfield  {author} {\bibinfo {author} {\bibfnamefont {A.~A.}\ \bibnamefont
  {Patel}}, \bibinfo {author} {\bibfnamefont {J.}~\bibnamefont {McGreevy}},
  \bibinfo {author} {\bibfnamefont {D.~P.}\ \bibnamefont {Arovas}},\ and\
  \bibinfo {author} {\bibfnamefont {S.}~\bibnamefont {Sachdev}},\ }\bibfield
  {title} {\bibinfo {title} {Magnetotransport in a {{Model}} of a {{Disordered
  Strange Metal}}},\ }\href {https://doi.org/10.1103/PhysRevX.8.021049}
  {\bibfield  {journal} {\bibinfo  {journal} {Phys. Rev. X}\ }\textbf {\bibinfo
  {volume} {8}},\ \bibinfo {pages} {021049} (\bibinfo {year}
  {2018})}\BibitemShut {NoStop}%
\bibitem [{\citenamefont {Chowdhury}\ \emph {et~al.}(2018)\citenamefont
  {Chowdhury}, \citenamefont {Werman}, \citenamefont {Berg},\ and\
  \citenamefont {Senthil}}]{DChowdhury2018c}%
  \BibitemOpen
  \bibfield  {author} {\bibinfo {author} {\bibfnamefont {D.}~\bibnamefont
  {Chowdhury}}, \bibinfo {author} {\bibfnamefont {Y.}~\bibnamefont {Werman}},
  \bibinfo {author} {\bibfnamefont {E.}~\bibnamefont {Berg}},\ and\ \bibinfo
  {author} {\bibfnamefont {T.}~\bibnamefont {Senthil}},\ }\bibfield  {title}
  {\bibinfo {title} {Translationally invariant non-{{Fermi}} liquid metals with
  critical {{Fermi-surfaces}}: {{Solvable}} models},\ }\href
  {https://doi.org/10.1103/PhysRevX.8.031024} {\bibfield  {journal} {\bibinfo
  {journal} {Phys. Rev. X}\ }\textbf {\bibinfo {volume} {8}},\ \bibinfo {pages}
  {031024} (\bibinfo {year} {2018})},\ \Eprint
  {https://arxiv.org/abs/1801.06178} {arXiv:1801.06178 [cond-mat.str-el]}
  \BibitemShut {NoStop}%
\bibitem [{\citenamefont {Else}\ and\ \citenamefont
  {Senthil}(2021)}]{DVElse2021}%
  \BibitemOpen
  \bibfield  {author} {\bibinfo {author} {\bibfnamefont {D.~V.}\ \bibnamefont
  {Else}}\ and\ \bibinfo {author} {\bibfnamefont {T.}~\bibnamefont {Senthil}},\
  }\bibfield  {title} {\bibinfo {title} {Critical drag as a mechanism for
  resistivity},\ }\href {https://doi.org/10.1103/PhysRevB.104.205132}
  {\bibfield  {journal} {\bibinfo  {journal} {Phys. Rev. B}\ }\textbf {\bibinfo
  {volume} {104}},\ \bibinfo {eid} {205132} (\bibinfo {year} {2021})},\ \Eprint
  {https://arxiv.org/abs/2106.15623} {arXiv:2106.15623 [cond-mat.str-el]}
  \BibitemShut {NoStop}%
\bibitem [{\citenamefont {Patel}\ \emph {et~al.}(2023)\citenamefont {Patel},
  \citenamefont {Guo}, \citenamefont {Esterlis},\ and\ \citenamefont
  {Sachdev}}]{AAPatel2023}%
  \BibitemOpen
  \bibfield  {author} {\bibinfo {author} {\bibfnamefont {A.~A.}\ \bibnamefont
  {Patel}}, \bibinfo {author} {\bibfnamefont {H.}~\bibnamefont {Guo}}, \bibinfo
  {author} {\bibfnamefont {I.}~\bibnamefont {Esterlis}},\ and\ \bibinfo
  {author} {\bibfnamefont {S.}~\bibnamefont {Sachdev}},\ }\bibfield  {title}
  {\bibinfo {title} {Universal theory of strange metals from spatially random
  interactions},\ }\href {https://doi.org/10.1126/science.abq6011} {\bibfield
  {journal} {\bibinfo  {journal} {Science}\ }\textbf {\bibinfo {volume}
  {381}},\ \bibinfo {pages} {abq6011} (\bibinfo {year} {2023})},\ \Eprint
  {https://arxiv.org/abs/2203.04990} {arXiv:2203.04990 [cond-mat.str-el]}
  \BibitemShut {NoStop}%
\bibitem [{\citenamefont {Guo}\ \emph {et~al.}(2022)\citenamefont {Guo},
  \citenamefont {Patel}, \citenamefont {Esterlis},\ and\ \citenamefont
  {Sachdev}}]{HGuo2022a}%
  \BibitemOpen
  \bibfield  {author} {\bibinfo {author} {\bibfnamefont {H.}~\bibnamefont
  {Guo}}, \bibinfo {author} {\bibfnamefont {A.~A.}\ \bibnamefont {Patel}},
  \bibinfo {author} {\bibfnamefont {I.}~\bibnamefont {Esterlis}},\ and\
  \bibinfo {author} {\bibfnamefont {S.}~\bibnamefont {Sachdev}},\ }\bibfield
  {title} {\bibinfo {title} {Large {$N$} theory of critical {{Fermi}} surfaces
  {{II}}: Conductivity},\ }\href {https://doi.org/10.1103/PhysRevB.106.115151}
  {\bibfield  {journal} {\bibinfo  {journal} {Phys. Rev. B}\ }\textbf {\bibinfo
  {volume} {106}},\ \bibinfo {pages} {115151} (\bibinfo {year} {2022})},\
  \Eprint {https://arxiv.org/abs/2207.08841} {arXiv:2207.08841 [cond-mat,
  physics:hep-th]} \BibitemShut {NoStop}%
\bibitem [{\citenamefont {Shi}\ \emph {et~al.}(2023)\citenamefont {Shi},
  \citenamefont {Else}, \citenamefont {Goldman},\ and\ \citenamefont
  {Senthil}}]{ZDShi2023}%
  \BibitemOpen
  \bibfield  {author} {\bibinfo {author} {\bibfnamefont {Z.~D.}\ \bibnamefont
  {Shi}}, \bibinfo {author} {\bibfnamefont {D.~V.}\ \bibnamefont {Else}},
  \bibinfo {author} {\bibfnamefont {H.}~\bibnamefont {Goldman}},\ and\ \bibinfo
  {author} {\bibfnamefont {T.}~\bibnamefont {Senthil}},\ }\bibfield  {title}
  {\bibinfo {title} {Loop current fluctuations and quantum critical
  transport},\ }\href {https://doi.org/10.21468/SciPostPhys.14.5.113}
  {\bibfield  {journal} {\bibinfo  {journal} {SciPost Physics}\ }\textbf
  {\bibinfo {volume} {14}},\ \bibinfo {pages} {113} (\bibinfo {year}
  {2023})}\BibitemShut {NoStop}%
\bibitem [{\citenamefont {Bashan}\ \emph {et~al.}(2024)\citenamefont {Bashan},
  \citenamefont {Tulipman}, \citenamefont {Schmalian},\ and\ \citenamefont
  {Berg}}]{NBashan2024}%
  \BibitemOpen
  \bibfield  {author} {\bibinfo {author} {\bibfnamefont {N.}~\bibnamefont
  {Bashan}}, \bibinfo {author} {\bibfnamefont {E.}~\bibnamefont {Tulipman}},
  \bibinfo {author} {\bibfnamefont {J.}~\bibnamefont {Schmalian}},\ and\
  \bibinfo {author} {\bibfnamefont {E.}~\bibnamefont {Berg}},\ }\bibfield
  {title} {\bibinfo {title} {Tunable {{Non-Fermi Liquid Phase}} from
  {{Coupling}} to {{Two-Level Systems}}},\ }\href
  {https://doi.org/10.1103/PhysRevLett.132.236501} {\bibfield  {journal}
  {\bibinfo  {journal} {Physical Review Letters}\ }\textbf {\bibinfo {volume}
  {132}},\ \bibinfo {pages} {236501} (\bibinfo {year} {2024})}\BibitemShut
  {NoStop}%
\bibitem [{\citenamefont {Tulipman}\ \emph {et~al.}(2024)\citenamefont
  {Tulipman}, \citenamefont {Bashan}, \citenamefont {Schmalian},\ and\
  \citenamefont {Berg}}]{ETulipman2024}%
  \BibitemOpen
  \bibfield  {author} {\bibinfo {author} {\bibfnamefont {E.}~\bibnamefont
  {Tulipman}}, \bibinfo {author} {\bibfnamefont {N.}~\bibnamefont {Bashan}},
  \bibinfo {author} {\bibfnamefont {J.}~\bibnamefont {Schmalian}},\ and\
  \bibinfo {author} {\bibfnamefont {E.}~\bibnamefont {Berg}},\ }\bibfield
  {title} {\bibinfo {title} {Solvable models of two-level systems coupled to
  itinerant electrons: {{Robust}} non-{{Fermi}} liquid and quantum critical
  pairing},\ }\href {https://doi.org/10.1103/PhysRevB.110.155118} {\bibfield
  {journal} {\bibinfo  {journal} {Physical Review B}\ }\textbf {\bibinfo
  {volume} {110}},\ \bibinfo {pages} {155118} (\bibinfo {year}
  {2024})}\BibitemShut {NoStop}%
\bibitem [{\citenamefont {Bashan}\ \emph {et~al.}(2025)\citenamefont {Bashan},
  \citenamefont {Tulipman}, \citenamefont {Kivelson}, \citenamefont
  {Schmalian},\ and\ \citenamefont {Berg}}]{NBashan2025}%
  \BibitemOpen
  \bibfield  {author} {\bibinfo {author} {\bibfnamefont {N.}~\bibnamefont
  {Bashan}}, \bibinfo {author} {\bibfnamefont {E.}~\bibnamefont {Tulipman}},
  \bibinfo {author} {\bibfnamefont {S.~A.}\ \bibnamefont {Kivelson}}, \bibinfo
  {author} {\bibfnamefont {J.}~\bibnamefont {Schmalian}},\ and\ \bibinfo
  {author} {\bibfnamefont {E.}~\bibnamefont {Berg}},\ }\href
  {https://doi.org/10.48550/arXiv.2502.08699} {\bibinfo {title} {Extended
  strange metal regime from superconducting puddles}} (\bibinfo {year}
  {2025}),\ \Eprint {https://arxiv.org/abs/2502.08699} {arXiv:2502.08699
  [cond-mat]} \BibitemShut {NoStop}%
\bibitem [{\citenamefont {Lee}(2021)}]{PALee2021}%
  \BibitemOpen
  \bibfield  {author} {\bibinfo {author} {\bibfnamefont {P.~A.}\ \bibnamefont
  {Lee}},\ }\bibfield  {title} {\bibinfo {title} {Low-temperature {$T$}-linear
  resistivity due to umklapp scattering from a critical mode},\ }\href
  {https://doi.org/10.1103/PhysRevB.104.035140} {\bibfield  {journal} {\bibinfo
   {journal} {Phys. Rev. B}\ }\textbf {\bibinfo {volume} {104}},\ \bibinfo
  {pages} {035140} (\bibinfo {year} {2021})}\BibitemShut {NoStop}%
\bibitem [{\citenamefont {Lee}(2024)}]{PALee2024}%
  \BibitemOpen
  \bibfield  {author} {\bibinfo {author} {\bibfnamefont {P.~A.}\ \bibnamefont
  {Lee}},\ }\href {http://arxiv.org/abs/2402.10878} {\bibinfo {title} {A model
  of non-{{Fermi}} liquid with power law resistivity: Strange metal with a
  not-so-strange origin}} (\bibinfo {year} {2024}),\ \Eprint
  {https://arxiv.org/abs/2402.10878} {arXiv:2402.10878 [cond-mat]} \BibitemShut
  {NoStop}%
\bibitem [{\citenamefont {Mousatov}\ \emph {et~al.}(2020)\citenamefont
  {Mousatov}, \citenamefont {Berg},\ and\ \citenamefont
  {Hartnoll}}]{CHMousatov2020}%
  \BibitemOpen
  \bibfield  {author} {\bibinfo {author} {\bibfnamefont {C.~H.}\ \bibnamefont
  {Mousatov}}, \bibinfo {author} {\bibfnamefont {E.}~\bibnamefont {Berg}},\
  and\ \bibinfo {author} {\bibfnamefont {S.~A.}\ \bibnamefont {Hartnoll}},\
  }\bibfield  {title} {\bibinfo {title} {Theory of the strange metal
  {{Sr3Ru2O7}}},\ }\href {https://doi.org/10.1073/pnas.1915224117} {\bibfield
  {journal} {\bibinfo  {journal} {Proceedings of the National Academy of
  Sciences}\ }\textbf {\bibinfo {volume} {117}},\ \bibinfo {pages} {2852}
  (\bibinfo {year} {2020})}\BibitemShut {NoStop}%
\bibitem [{\citenamefont {Patel}\ \emph
  {et~al.}(2024{\natexlab{a}})\citenamefont {Patel}, \citenamefont {Lunts},\
  and\ \citenamefont {Sachdev}}]{AAPatel2024}%
  \BibitemOpen
  \bibfield  {author} {\bibinfo {author} {\bibfnamefont {A.~A.}\ \bibnamefont
  {Patel}}, \bibinfo {author} {\bibfnamefont {P.}~\bibnamefont {Lunts}},\ and\
  \bibinfo {author} {\bibfnamefont {S.}~\bibnamefont {Sachdev}},\ }\bibfield
  {title} {\bibinfo {title} {Localization of overdamped bosonic modes and
  transport in strange metals},\ }\href
  {https://doi.org/10.1073/pnas.2402052121} {\bibfield  {journal} {\bibinfo
  {journal} {Proceedings of the National Academy of Sciences}\ }\textbf
  {\bibinfo {volume} {121}},\ \bibinfo {pages} {e2402052121} (\bibinfo {year}
  {2024}{\natexlab{a}})}\BibitemShut {NoStop}%
\bibitem [{\citenamefont {Patel}\ \emph
  {et~al.}(2024{\natexlab{b}})\citenamefont {Patel}, \citenamefont {Lunts},\
  and\ \citenamefont {Albergo}}]{AAPatel2024a}%
  \BibitemOpen
  \bibfield  {author} {\bibinfo {author} {\bibfnamefont {A.~A.}\ \bibnamefont
  {Patel}}, \bibinfo {author} {\bibfnamefont {P.}~\bibnamefont {Lunts}},\ and\
  \bibinfo {author} {\bibfnamefont {M.~S.}\ \bibnamefont {Albergo}},\ }\href
  {https://doi.org/10.48550/arXiv.2410.05365} {\bibinfo {title} {Strange metals
  and planckian transport in a gapless phase from spatially random
  interactions}} (\bibinfo {year} {2024}{\natexlab{b}}),\ \Eprint
  {https://arxiv.org/abs/2410.05365} {arXiv:2410.05365 [cond-mat]} \BibitemShut
  {NoStop}%
\bibitem [{\citenamefont {Hardy}\ \emph {et~al.}(2025)\citenamefont {Hardy},
  \citenamefont {Parcollet}, \citenamefont {Georges},\ and\ \citenamefont
  {Patel}}]{AHardy2025}%
  \BibitemOpen
  \bibfield  {author} {\bibinfo {author} {\bibfnamefont {A.}~\bibnamefont
  {Hardy}}, \bibinfo {author} {\bibfnamefont {O.}~\bibnamefont {Parcollet}},
  \bibinfo {author} {\bibfnamefont {A.}~\bibnamefont {Georges}},\ and\ \bibinfo
  {author} {\bibfnamefont {A.~A.}\ \bibnamefont {Patel}},\ }\bibfield  {title}
  {\bibinfo {title} {Enhanced {{Strange Metallicity}} due to {{Hubbard-}}{$U$}
  {{Coulomb Repulsion}}},\ }\href
  {https://doi.org/10.1103/PhysRevLett.134.036502} {\bibfield  {journal}
  {\bibinfo  {journal} {Physical Review Letters}\ }\textbf {\bibinfo {volume}
  {134}},\ \bibinfo {pages} {036502} (\bibinfo {year} {2025})}\BibitemShut
  {NoStop}%
\bibitem [{\citenamefont {Li}\ \emph {et~al.}(2024)\citenamefont {Li},
  \citenamefont {Valentinis}, \citenamefont {Patel}, \citenamefont {Guo},
  \citenamefont {Schmalian}, \citenamefont {Sachdev},\ and\ \citenamefont
  {Esterlis}}]{CLi2024}%
  \BibitemOpen
  \bibfield  {author} {\bibinfo {author} {\bibfnamefont {C.}~\bibnamefont
  {Li}}, \bibinfo {author} {\bibfnamefont {D.}~\bibnamefont {Valentinis}},
  \bibinfo {author} {\bibfnamefont {A.~A.}\ \bibnamefont {Patel}}, \bibinfo
  {author} {\bibfnamefont {H.}~\bibnamefont {Guo}}, \bibinfo {author}
  {\bibfnamefont {J.}~\bibnamefont {Schmalian}}, \bibinfo {author}
  {\bibfnamefont {S.}~\bibnamefont {Sachdev}},\ and\ \bibinfo {author}
  {\bibfnamefont {I.}~\bibnamefont {Esterlis}},\ }\bibfield  {title} {\bibinfo
  {title} {Strange {{Metal}} and {{Superconductor}} in the {{Two-Dimensional
  Yukawa-Sachdev-Ye-Kitaev Model}}},\ }\href
  {https://doi.org/10.1103/PhysRevLett.133.186502} {\bibfield  {journal}
  {\bibinfo  {journal} {Phys. Rev. Lett.}\ }\textbf {\bibinfo {volume} {133}},\
  \bibinfo {pages} {186502} (\bibinfo {year} {2024})}\BibitemShut {NoStop}%
\bibitem [{\citenamefont {Mousatov}\ and\ \citenamefont
  {Hartnoll}(2021)}]{CHMousatov2021}%
  \BibitemOpen
  \bibfield  {author} {\bibinfo {author} {\bibfnamefont {C.~H.}\ \bibnamefont
  {Mousatov}}\ and\ \bibinfo {author} {\bibfnamefont {S.~A.}\ \bibnamefont
  {Hartnoll}},\ }\bibfield  {title} {\bibinfo {title} {Phonons, electrons and
  thermal transport in {{Planckian}} high {{Tc}} materials},\ }\href
  {https://doi.org/10.1038/s41535-021-00383-w} {\bibfield  {journal} {\bibinfo
  {journal} {npj Quantum Mater.}\ }\textbf {\bibinfo {volume} {6}},\ \bibinfo
  {pages} {1} (\bibinfo {year} {2021})}\BibitemShut {NoStop}%
\bibitem [{\citenamefont {Allen}(1987)}]{PBAllen1987}%
  \BibitemOpen
  \bibfield  {author} {\bibinfo {author} {\bibfnamefont {P.~B.}\ \bibnamefont
  {Allen}},\ }\bibfield  {title} {\bibinfo {title} {Theory of thermal
  relaxation of electrons in metals},\ }\href
  {https://doi.org/10.1103/PhysRevLett.59.1460} {\bibfield  {journal} {\bibinfo
   {journal} {Phys. Rev. Lett.}\ }\textbf {\bibinfo {volume} {59}},\ \bibinfo
  {pages} {1460} (\bibinfo {year} {1987})}\BibitemShut {NoStop}%
\bibitem [{\citenamefont {Ramshaw}\ \emph {et~al.}(2015)\citenamefont
  {Ramshaw}, \citenamefont {Sebastian}, \citenamefont {McDonald}, \citenamefont
  {Day}, \citenamefont {Tan}, \citenamefont {Zhu}, \citenamefont {Betts},
  \citenamefont {Liang}, \citenamefont {Bonn}, \citenamefont {Hardy},\ and\
  \citenamefont {Harrison}}]{rams15}%
  \BibitemOpen
  \bibfield  {author} {\bibinfo {author} {\bibfnamefont {B.~J.}\ \bibnamefont
  {Ramshaw}}, \bibinfo {author} {\bibfnamefont {S.~E.}\ \bibnamefont
  {Sebastian}}, \bibinfo {author} {\bibfnamefont {R.~D.}\ \bibnamefont
  {McDonald}}, \bibinfo {author} {\bibfnamefont {J.}~\bibnamefont {Day}},
  \bibinfo {author} {\bibfnamefont {B.~S.}\ \bibnamefont {Tan}}, \bibinfo
  {author} {\bibfnamefont {Z.}~\bibnamefont {Zhu}}, \bibinfo {author}
  {\bibfnamefont {J.~B.}\ \bibnamefont {Betts}}, \bibinfo {author}
  {\bibfnamefont {R.}~\bibnamefont {Liang}}, \bibinfo {author} {\bibfnamefont
  {D.~A.}\ \bibnamefont {Bonn}}, \bibinfo {author} {\bibfnamefont {W.~N.}\
  \bibnamefont {Hardy}},\ and\ \bibinfo {author} {\bibfnamefont
  {N.}~\bibnamefont {Harrison}},\ }\bibfield  {title} {\bibinfo {title}
  {Quasiparticle mass enhancement approaching optimal doping in a
  high-{{T}}{\textsubscript{c}} superconductor},\ }\href
  {https://doi.org/10.1126/science.aaa4990} {\bibfield  {journal} {\bibinfo
  {journal} {Science}\ }\textbf {\bibinfo {volume} {348}},\ \bibinfo {pages}
  {317} (\bibinfo {year} {2015})}\BibitemShut {NoStop}%
\bibitem [{\citenamefont {Michon}\ \emph {et~al.}(2019)\citenamefont {Michon},
  \citenamefont {Girod}, \citenamefont {Badoux}, \citenamefont {Ka{\v c}mar{\v
  c}{\'i}k}, \citenamefont {Ma}, \citenamefont {Dragomir}, \citenamefont
  {Dabkowska}, \citenamefont {Gaulin}, \citenamefont {Zhou}, \citenamefont
  {Pyon}, \citenamefont {Takayama}, \citenamefont {Takagi}, \citenamefont
  {Verret}, \citenamefont {{Doiron-Leyraud}}, \citenamefont {Marcenat},
  \citenamefont {Taillefer},\ and\ \citenamefont {Klein}}]{BMichon2019}%
  \BibitemOpen
  \bibfield  {author} {\bibinfo {author} {\bibfnamefont {B.}~\bibnamefont
  {Michon}}, \bibinfo {author} {\bibfnamefont {C.}~\bibnamefont {Girod}},
  \bibinfo {author} {\bibfnamefont {S.}~\bibnamefont {Badoux}}, \bibinfo
  {author} {\bibfnamefont {J.}~\bibnamefont {Ka{\v c}mar{\v c}{\'i}k}},
  \bibinfo {author} {\bibfnamefont {Q.}~\bibnamefont {Ma}}, \bibinfo {author}
  {\bibfnamefont {M.}~\bibnamefont {Dragomir}}, \bibinfo {author}
  {\bibfnamefont {H.~A.}\ \bibnamefont {Dabkowska}}, \bibinfo {author}
  {\bibfnamefont {B.~D.}\ \bibnamefont {Gaulin}}, \bibinfo {author}
  {\bibfnamefont {J.-S.}\ \bibnamefont {Zhou}}, \bibinfo {author}
  {\bibfnamefont {S.}~\bibnamefont {Pyon}}, \bibinfo {author} {\bibfnamefont
  {T.}~\bibnamefont {Takayama}}, \bibinfo {author} {\bibfnamefont
  {H.}~\bibnamefont {Takagi}}, \bibinfo {author} {\bibfnamefont
  {S.}~\bibnamefont {Verret}}, \bibinfo {author} {\bibfnamefont
  {N.}~\bibnamefont {{Doiron-Leyraud}}}, \bibinfo {author} {\bibfnamefont
  {C.}~\bibnamefont {Marcenat}}, \bibinfo {author} {\bibfnamefont
  {L.}~\bibnamefont {Taillefer}},\ and\ \bibinfo {author} {\bibfnamefont
  {T.}~\bibnamefont {Klein}},\ }\bibfield  {title} {\bibinfo {title}
  {Thermodynamic signatures of quantum criticality in cuprate
  superconductors},\ }\href {https://doi.org/10.1038/s41586-019-0932-x}
  {\bibfield  {journal} {\bibinfo  {journal} {Nature}\ }\textbf {\bibinfo
  {volume} {567}},\ \bibinfo {pages} {218} (\bibinfo {year}
  {2019})}\BibitemShut {NoStop}%
\bibitem [{\citenamefont {Badoux}\ \emph {et~al.}(2016)\citenamefont {Badoux},
  \citenamefont {Tabis}, \citenamefont {Lalibert{\'e}}, \citenamefont
  {Grissonnanche}, \citenamefont {Vignolle}, \citenamefont {Vignolles},
  \citenamefont {B{\'e}ard}, \citenamefont {Bonn}, \citenamefont {Hardy},
  \citenamefont {Liang}, \citenamefont {{Doiron-Leyraud}}, \citenamefont
  {Taillefer},\ and\ \citenamefont {Proust}}]{SBadoux2016}%
  \BibitemOpen
  \bibfield  {author} {\bibinfo {author} {\bibfnamefont {S.}~\bibnamefont
  {Badoux}}, \bibinfo {author} {\bibfnamefont {W.}~\bibnamefont {Tabis}},
  \bibinfo {author} {\bibfnamefont {F.}~\bibnamefont {Lalibert{\'e}}}, \bibinfo
  {author} {\bibfnamefont {G.}~\bibnamefont {Grissonnanche}}, \bibinfo {author}
  {\bibfnamefont {B.}~\bibnamefont {Vignolle}}, \bibinfo {author}
  {\bibfnamefont {D.}~\bibnamefont {Vignolles}}, \bibinfo {author}
  {\bibfnamefont {J.}~\bibnamefont {B{\'e}ard}}, \bibinfo {author}
  {\bibfnamefont {D.~A.}\ \bibnamefont {Bonn}}, \bibinfo {author}
  {\bibfnamefont {W.~N.}\ \bibnamefont {Hardy}}, \bibinfo {author}
  {\bibfnamefont {R.}~\bibnamefont {Liang}}, \bibinfo {author} {\bibfnamefont
  {N.}~\bibnamefont {{Doiron-Leyraud}}}, \bibinfo {author} {\bibfnamefont
  {L.}~\bibnamefont {Taillefer}},\ and\ \bibinfo {author} {\bibfnamefont
  {C.}~\bibnamefont {Proust}},\ }\bibfield  {title} {\bibinfo {title} {Change
  of carrier density at the pseudogap critical point of a cuprate
  superconductor},\ }\href {https://doi.org/10.1038/nature16983} {\bibfield
  {journal} {\bibinfo  {journal} {Nature}\ }\textbf {\bibinfo {volume} {531}},\
  \bibinfo {pages} {210} (\bibinfo {year} {2016})}\BibitemShut {NoStop}%
\bibitem [{\citenamefont {Michon}\ \emph {et~al.}(2023)\citenamefont {Michon},
  \citenamefont {Berthod}, \citenamefont {Rischau}, \citenamefont {Ataei},
  \citenamefont {Chen}, \citenamefont {Komiya}, \citenamefont {Ono},
  \citenamefont {Taillefer}, \citenamefont {{van der Marel}},\ and\
  \citenamefont {Georges}}]{BMichon2023a}%
  \BibitemOpen
  \bibfield  {author} {\bibinfo {author} {\bibfnamefont {B.}~\bibnamefont
  {Michon}}, \bibinfo {author} {\bibfnamefont {C.}~\bibnamefont {Berthod}},
  \bibinfo {author} {\bibfnamefont {C.~W.}\ \bibnamefont {Rischau}}, \bibinfo
  {author} {\bibfnamefont {A.}~\bibnamefont {Ataei}}, \bibinfo {author}
  {\bibfnamefont {L.}~\bibnamefont {Chen}}, \bibinfo {author} {\bibfnamefont
  {S.}~\bibnamefont {Komiya}}, \bibinfo {author} {\bibfnamefont
  {S.}~\bibnamefont {Ono}}, \bibinfo {author} {\bibfnamefont {L.}~\bibnamefont
  {Taillefer}}, \bibinfo {author} {\bibfnamefont {D.}~\bibnamefont {{van der
  Marel}}},\ and\ \bibinfo {author} {\bibfnamefont {A.}~\bibnamefont
  {Georges}},\ }\bibfield  {title} {\bibinfo {title} {Reconciling scaling of
  the optical conductivity of cuprate superconductors with {{Planckian}}
  resistivity and specific heat},\ }\href
  {https://doi.org/10.1038/s41467-023-38762-5} {\bibfield  {journal} {\bibinfo
  {journal} {Nat Commun}\ }\textbf {\bibinfo {volume} {14}},\ \bibinfo {pages}
  {3033} (\bibinfo {year} {2023})}\BibitemShut {NoStop}%
\bibitem [{\citenamefont {Chen}\ \emph {et~al.}(2019)\citenamefont {Chen},
  \citenamefont {Hashimoto}, \citenamefont {He}, \citenamefont {Song},
  \citenamefont {Xu}, \citenamefont {He}, \citenamefont {Devereaux},
  \citenamefont {Eisaki}, \citenamefont {Lu}, \citenamefont {Zaanen},\ and\
  \citenamefont {Shen}}]{SDChen2019b}%
  \BibitemOpen
  \bibfield  {author} {\bibinfo {author} {\bibfnamefont {S.-D.}\ \bibnamefont
  {Chen}}, \bibinfo {author} {\bibfnamefont {M.}~\bibnamefont {Hashimoto}},
  \bibinfo {author} {\bibfnamefont {Y.}~\bibnamefont {He}}, \bibinfo {author}
  {\bibfnamefont {D.}~\bibnamefont {Song}}, \bibinfo {author} {\bibfnamefont
  {K.-J.}\ \bibnamefont {Xu}}, \bibinfo {author} {\bibfnamefont {J.-F.}\
  \bibnamefont {He}}, \bibinfo {author} {\bibfnamefont {T.~P.}\ \bibnamefont
  {Devereaux}}, \bibinfo {author} {\bibfnamefont {H.}~\bibnamefont {Eisaki}},
  \bibinfo {author} {\bibfnamefont {D.-H.}\ \bibnamefont {Lu}}, \bibinfo
  {author} {\bibfnamefont {J.}~\bibnamefont {Zaanen}},\ and\ \bibinfo {author}
  {\bibfnamefont {Z.-X.}\ \bibnamefont {Shen}},\ }\bibfield  {title} {\bibinfo
  {title} {Incoherent strange metal sharply bounded by a critical doping in
  {{Bi2212}}},\ }\href {https://doi.org/10.1126/science.aaw8850} {\bibfield
  {journal} {\bibinfo  {journal} {Science}\ }\textbf {\bibinfo {volume}
  {366}},\ \bibinfo {pages} {1099} (\bibinfo {year} {2019})}\BibitemShut
  {NoStop}%
\bibitem [{\citenamefont {Shekhter}\ \emph {et~al.}(2013)\citenamefont
  {Shekhter}, \citenamefont {Ramshaw}, \citenamefont {Liang}, \citenamefont
  {Hardy}, \citenamefont {Bonn}, \citenamefont {Balakirev}, \citenamefont
  {McDonald}, \citenamefont {Betts}, \citenamefont {Riggs},\ and\ \citenamefont
  {Migliori}}]{AShekhter2013d}%
  \BibitemOpen
  \bibfield  {author} {\bibinfo {author} {\bibfnamefont {A.}~\bibnamefont
  {Shekhter}}, \bibinfo {author} {\bibfnamefont {B.~J.}\ \bibnamefont
  {Ramshaw}}, \bibinfo {author} {\bibfnamefont {R.}~\bibnamefont {Liang}},
  \bibinfo {author} {\bibfnamefont {W.~N.}\ \bibnamefont {Hardy}}, \bibinfo
  {author} {\bibfnamefont {D.~A.}\ \bibnamefont {Bonn}}, \bibinfo {author}
  {\bibfnamefont {F.~F.}\ \bibnamefont {Balakirev}}, \bibinfo {author}
  {\bibfnamefont {R.~D.}\ \bibnamefont {McDonald}}, \bibinfo {author}
  {\bibfnamefont {J.~B.}\ \bibnamefont {Betts}}, \bibinfo {author}
  {\bibfnamefont {S.~C.}\ \bibnamefont {Riggs}},\ and\ \bibinfo {author}
  {\bibfnamefont {A.}~\bibnamefont {Migliori}},\ }\bibfield  {title} {\bibinfo
  {title} {Bounding the pseudogap with a line of phase transitions in
  {YBa2Cu3O6}+${\delta}$},\ }\href {https://doi.org/10.1038/nature12165}
  {\bibfield  {journal} {\bibinfo  {journal} {Nature}\ }\textbf {\bibinfo
  {volume} {498}},\ \bibinfo {pages} {75} (\bibinfo {year} {2013})}\BibitemShut
  {NoStop}%
\bibitem [{\citenamefont {Fujita}\ \emph {et~al.}(2014)\citenamefont {Fujita},
  \citenamefont {Kim}, \citenamefont {Lee}, \citenamefont {Lee}, \citenamefont
  {Hamidian}, \citenamefont {Firmo}, \citenamefont {Mukhopadhyay},
  \citenamefont {Eisaki}, \citenamefont {Uchida}, \citenamefont {Lawler},
  \citenamefont {Kim},\ and\ \citenamefont {Davis}}]{KFujita2014a}%
  \BibitemOpen
  \bibfield  {author} {\bibinfo {author} {\bibfnamefont {K.}~\bibnamefont
  {Fujita}}, \bibinfo {author} {\bibfnamefont {C.~K.}\ \bibnamefont {Kim}},
  \bibinfo {author} {\bibfnamefont {I.}~\bibnamefont {Lee}}, \bibinfo {author}
  {\bibfnamefont {J.}~\bibnamefont {Lee}}, \bibinfo {author} {\bibfnamefont
  {M.~H.}\ \bibnamefont {Hamidian}}, \bibinfo {author} {\bibfnamefont {I.~A.}\
  \bibnamefont {Firmo}}, \bibinfo {author} {\bibfnamefont {S.}~\bibnamefont
  {Mukhopadhyay}}, \bibinfo {author} {\bibfnamefont {H.}~\bibnamefont
  {Eisaki}}, \bibinfo {author} {\bibfnamefont {S.}~\bibnamefont {Uchida}},
  \bibinfo {author} {\bibfnamefont {M.~J.}\ \bibnamefont {Lawler}}, \bibinfo
  {author} {\bibfnamefont {E.-A.}\ \bibnamefont {Kim}},\ and\ \bibinfo {author}
  {\bibfnamefont {J.~C.}\ \bibnamefont {Davis}},\ }\bibfield  {title} {\bibinfo
  {title} {Simultaneous transitions in cuprate momentum-space topology and
  electronic symmetry breaking},\ }\href
  {https://doi.org/10.1126/science.1248783} {\bibfield  {journal} {\bibinfo
  {journal} {Science}\ }\textbf {\bibinfo {volume} {344}},\ \bibinfo {pages}
  {612} (\bibinfo {year} {2014})}\BibitemShut {NoStop}%
\bibitem [{\citenamefont {Gourgout}\ \emph {et~al.}(2022)\citenamefont
  {Gourgout}, \citenamefont {Grissonnanche}, \citenamefont {Lalibert{\'e}},
  \citenamefont {Ataei}, \citenamefont {Chen}, \citenamefont {Verret},
  \citenamefont {Zhou}, \citenamefont {Mravlje}, \citenamefont {Georges},
  \citenamefont {{Doiron-Leyraud}},\ and\ \citenamefont
  {Taillefer}}]{AGourgout2022e}%
  \BibitemOpen
  \bibfield  {author} {\bibinfo {author} {\bibfnamefont {A.}~\bibnamefont
  {Gourgout}}, \bibinfo {author} {\bibfnamefont {G.}~\bibnamefont
  {Grissonnanche}}, \bibinfo {author} {\bibfnamefont {F.}~\bibnamefont
  {Lalibert{\'e}}}, \bibinfo {author} {\bibfnamefont {A.}~\bibnamefont
  {Ataei}}, \bibinfo {author} {\bibfnamefont {L.}~\bibnamefont {Chen}},
  \bibinfo {author} {\bibfnamefont {S.}~\bibnamefont {Verret}}, \bibinfo
  {author} {\bibfnamefont {J.-S.}\ \bibnamefont {Zhou}}, \bibinfo {author}
  {\bibfnamefont {J.}~\bibnamefont {Mravlje}}, \bibinfo {author} {\bibfnamefont
  {A.}~\bibnamefont {Georges}}, \bibinfo {author} {\bibfnamefont
  {N.}~\bibnamefont {{Doiron-Leyraud}}},\ and\ \bibinfo {author} {\bibfnamefont
  {L.}~\bibnamefont {Taillefer}},\ }\bibfield  {title} {\bibinfo {title}
  {Seebeck {{Coefficient}} in a {{Cuprate Superconductor}}: {{Particle-Hole
  Asymmetry}} in the {{Strange Metal Phase}} and {{Fermi Surface
  Transformation}} in the {{Pseudogap Phase}}},\ }\href
  {https://doi.org/10.1103/PhysRevX.12.011037} {\bibfield  {journal} {\bibinfo
  {journal} {Physical Review X}\ }\textbf {\bibinfo {volume} {12}},\ \bibinfo
  {pages} {011037} (\bibinfo {year} {2022})}\BibitemShut {NoStop}%
\bibitem [{\citenamefont {Bobroff}\ \emph {et~al.}(2002)\citenamefont
  {Bobroff}, \citenamefont {Alloul}, \citenamefont {Ouazi}, \citenamefont
  {Mendels}, \citenamefont {Mahajan}, \citenamefont {Blanchard}, \citenamefont
  {Collin}, \citenamefont {Guillen},\ and\ \citenamefont
  {Marucco}}]{JBobroff2002}%
  \BibitemOpen
  \bibfield  {author} {\bibinfo {author} {\bibfnamefont {J.}~\bibnamefont
  {Bobroff}}, \bibinfo {author} {\bibfnamefont {H.}~\bibnamefont {Alloul}},
  \bibinfo {author} {\bibfnamefont {S.}~\bibnamefont {Ouazi}}, \bibinfo
  {author} {\bibfnamefont {P.}~\bibnamefont {Mendels}}, \bibinfo {author}
  {\bibfnamefont {A.}~\bibnamefont {Mahajan}}, \bibinfo {author} {\bibfnamefont
  {N.}~\bibnamefont {Blanchard}}, \bibinfo {author} {\bibfnamefont
  {G.}~\bibnamefont {Collin}}, \bibinfo {author} {\bibfnamefont
  {V.}~\bibnamefont {Guillen}},\ and\ \bibinfo {author} {\bibfnamefont {J.-F.}\
  \bibnamefont {Marucco}},\ }\bibfield  {title} {\bibinfo {title} {Absence of
  {{Static Phase Separation}} in the {{High}} {${T}_{c}$} {{Cuprate}}
  {${\mathrm{Y}\mathrm{B}\mathrm{a}}_{2}{\mathrm{C}\mathrm{u}}_{3}{\mathrm{O}}_{6+y}$}},\
  }\href {https://doi.org/10.1103/PhysRevLett.89.157002} {\bibfield  {journal}
  {\bibinfo  {journal} {Physical Review Letters}\ }\textbf {\bibinfo {volume}
  {89}},\ \bibinfo {pages} {157002} (\bibinfo {year} {2002})}\BibitemShut
  {NoStop}%
\bibitem [{\citenamefont {Gomes}\ \emph {et~al.}(2007)\citenamefont {Gomes},
  \citenamefont {Pasupathy}, \citenamefont {Pushp}, \citenamefont {Ono},
  \citenamefont {Ando},\ and\ \citenamefont {Yazdani}}]{KKGomes2007b}%
  \BibitemOpen
  \bibfield  {author} {\bibinfo {author} {\bibfnamefont {K.~K.}\ \bibnamefont
  {Gomes}}, \bibinfo {author} {\bibfnamefont {A.~N.}\ \bibnamefont
  {Pasupathy}}, \bibinfo {author} {\bibfnamefont {A.}~\bibnamefont {Pushp}},
  \bibinfo {author} {\bibfnamefont {S.}~\bibnamefont {Ono}}, \bibinfo {author}
  {\bibfnamefont {Y.}~\bibnamefont {Ando}},\ and\ \bibinfo {author}
  {\bibfnamefont {A.}~\bibnamefont {Yazdani}},\ }\bibfield  {title} {\bibinfo
  {title} {Visualizing pair formation on the atomic scale in the high-{{Tc}}
  superconductor {{Bi2Sr2CaCu2O8}}+{$\delta$}},\ }\href
  {https://doi.org/10.1038/nature05881} {\bibfield  {journal} {\bibinfo
  {journal} {Nature}\ }\textbf {\bibinfo {volume} {447}},\ \bibinfo {pages}
  {569} (\bibinfo {year} {2007})}\BibitemShut {NoStop}%
\bibitem [{\citenamefont {Pelc}\ \emph {et~al.}(2022)\citenamefont {Pelc},
  \citenamefont {Spieker}, \citenamefont {Anderson}, \citenamefont {Krogstad},
  \citenamefont {Biniskos}, \citenamefont {Bielinski}, \citenamefont {Yu},
  \citenamefont {Sasagawa}, \citenamefont {Chauviere}, \citenamefont {Dosanjh},
  \citenamefont {Liang}, \citenamefont {Bonn}, \citenamefont {Damascelli},
  \citenamefont {Chi}, \citenamefont {Liu}, \citenamefont {Osborn},\ and\
  \citenamefont {Greven}}]{DPelc2022}%
  \BibitemOpen
  \bibfield  {author} {\bibinfo {author} {\bibfnamefont {D.}~\bibnamefont
  {Pelc}}, \bibinfo {author} {\bibfnamefont {R.~J.}\ \bibnamefont {Spieker}},
  \bibinfo {author} {\bibfnamefont {Z.~W.}\ \bibnamefont {Anderson}}, \bibinfo
  {author} {\bibfnamefont {M.~J.}\ \bibnamefont {Krogstad}}, \bibinfo {author}
  {\bibfnamefont {N.}~\bibnamefont {Biniskos}}, \bibinfo {author}
  {\bibfnamefont {N.~G.}\ \bibnamefont {Bielinski}}, \bibinfo {author}
  {\bibfnamefont {B.}~\bibnamefont {Yu}}, \bibinfo {author} {\bibfnamefont
  {T.}~\bibnamefont {Sasagawa}}, \bibinfo {author} {\bibfnamefont
  {L.}~\bibnamefont {Chauviere}}, \bibinfo {author} {\bibfnamefont
  {P.}~\bibnamefont {Dosanjh}}, \bibinfo {author} {\bibfnamefont
  {R.}~\bibnamefont {Liang}}, \bibinfo {author} {\bibfnamefont {D.~A.}\
  \bibnamefont {Bonn}}, \bibinfo {author} {\bibfnamefont {A.}~\bibnamefont
  {Damascelli}}, \bibinfo {author} {\bibfnamefont {S.}~\bibnamefont {Chi}},
  \bibinfo {author} {\bibfnamefont {Y.}~\bibnamefont {Liu}}, \bibinfo {author}
  {\bibfnamefont {R.}~\bibnamefont {Osborn}},\ and\ \bibinfo {author}
  {\bibfnamefont {M.}~\bibnamefont {Greven}},\ }\bibfield  {title} {\bibinfo
  {title} {Unconventional short-range structural fluctuations in cuprate
  superconductors},\ }\href {https://doi.org/10.1038/s41598-022-22150-y}
  {\bibfield  {journal} {\bibinfo  {journal} {Scientific Reports}\ }\textbf
  {\bibinfo {volume} {12}},\ \bibinfo {pages} {20483} (\bibinfo {year}
  {2022})}\BibitemShut {NoStop}%
\bibitem [{\citenamefont {{Doiron-Leyraud}}\ \emph {et~al.}(2007)\citenamefont
  {{Doiron-Leyraud}}, \citenamefont {Proust}, \citenamefont {LeBoeuf},
  \citenamefont {Levallois}, \citenamefont {Bonnemaison}, \citenamefont
  {Liang}, \citenamefont {Bonn}, \citenamefont {Hardy},\ and\ \citenamefont
  {Taillefer}}]{NDoiron-Leyraud2007b}%
  \BibitemOpen
  \bibfield  {author} {\bibinfo {author} {\bibfnamefont {N.}~\bibnamefont
  {{Doiron-Leyraud}}}, \bibinfo {author} {\bibfnamefont {C.}~\bibnamefont
  {Proust}}, \bibinfo {author} {\bibfnamefont {D.}~\bibnamefont {LeBoeuf}},
  \bibinfo {author} {\bibfnamefont {J.}~\bibnamefont {Levallois}}, \bibinfo
  {author} {\bibfnamefont {J.-B.}\ \bibnamefont {Bonnemaison}}, \bibinfo
  {author} {\bibfnamefont {R.}~\bibnamefont {Liang}}, \bibinfo {author}
  {\bibfnamefont {D.~A.}\ \bibnamefont {Bonn}}, \bibinfo {author}
  {\bibfnamefont {W.~N.}\ \bibnamefont {Hardy}},\ and\ \bibinfo {author}
  {\bibfnamefont {L.}~\bibnamefont {Taillefer}},\ }\bibfield  {title} {\bibinfo
  {title} {Quantum oscillations and the {{Fermi}} surface in an underdoped
  high-{{Tc}} superconductor},\ }\href {https://doi.org/10.1038/nature05872}
  {\bibfield  {journal} {\bibinfo  {journal} {Nature}\ }\textbf {\bibinfo
  {volume} {447}},\ \bibinfo {pages} {565} (\bibinfo {year}
  {2007})}\BibitemShut {NoStop}%
\bibitem [{\citenamefont {Bari{\v s}i{\'c}}\ \emph {et~al.}(2019)\citenamefont
  {Bari{\v s}i{\'c}}, \citenamefont {Chan}, \citenamefont {Veit}, \citenamefont
  {Dorow}, \citenamefont {Ge}, \citenamefont {Li}, \citenamefont {Tabis},
  \citenamefont {Tang}, \citenamefont {Yu}, \citenamefont {Zhao},\ and\
  \citenamefont {Greven}}]{NBarisic2019}%
  \BibitemOpen
  \bibfield  {author} {\bibinfo {author} {\bibfnamefont {N.}~\bibnamefont
  {Bari{\v s}i{\'c}}}, \bibinfo {author} {\bibfnamefont {M.~K.}\ \bibnamefont
  {Chan}}, \bibinfo {author} {\bibfnamefont {M.~J.}\ \bibnamefont {Veit}},
  \bibinfo {author} {\bibfnamefont {C.~J.}\ \bibnamefont {Dorow}}, \bibinfo
  {author} {\bibfnamefont {Y.}~\bibnamefont {Ge}}, \bibinfo {author}
  {\bibfnamefont {Y.}~\bibnamefont {Li}}, \bibinfo {author} {\bibfnamefont
  {W.}~\bibnamefont {Tabis}}, \bibinfo {author} {\bibfnamefont
  {Y.}~\bibnamefont {Tang}}, \bibinfo {author} {\bibfnamefont {G.}~\bibnamefont
  {Yu}}, \bibinfo {author} {\bibfnamefont {X.}~\bibnamefont {Zhao}},\ and\
  \bibinfo {author} {\bibfnamefont {M.}~\bibnamefont {Greven}},\ }\bibfield
  {title} {\bibinfo {title} {Evidence for a universal {{Fermi-liquid}}
  scattering rate throughout the phase diagram of the copper-oxide
  superconductors},\ }\href {https://doi.org/10.1088/1367-2630/ab4d0f}
  {\bibfield  {journal} {\bibinfo  {journal} {New Journal of Physics}\ }\textbf
  {\bibinfo {volume} {21}},\ \bibinfo {pages} {113007} (\bibinfo {year}
  {2019})}\BibitemShut {NoStop}%
\bibitem [{\citenamefont {Lee}\ \emph {et~al.}(2006)\citenamefont {Lee},
  \citenamefont {Nagaosa},\ and\ \citenamefont {Wen}}]{LNW}%
  \BibitemOpen
  \bibfield  {author} {\bibinfo {author} {\bibfnamefont {P.~A.}\ \bibnamefont
  {Lee}}, \bibinfo {author} {\bibfnamefont {N.}~\bibnamefont {Nagaosa}},\ and\
  \bibinfo {author} {\bibfnamefont {X.-G.}\ \bibnamefont {Wen}},\ }\bibfield
  {title} {\bibinfo {title} {Doping a mott insulator: Physics of
  high-temperature superconductivity},\ }\href
  {https://doi.org/10.1103/RevModPhys.78.17} {\bibfield  {journal} {\bibinfo
  {journal} {Rev. Mod. Phys.}\ }\textbf {\bibinfo {volume} {78}},\ \bibinfo
  {pages} {17} (\bibinfo {year} {2006})}\BibitemShut {NoStop}%
\bibitem [{\citenamefont {Broholm}\ \emph {et~al.}(2020)\citenamefont
  {Broholm}, \citenamefont {Cava}, \citenamefont {Kivelson}, \citenamefont
  {Nocera}, \citenamefont {Norman},\ and\ \citenamefont {Senthil}}]{QSL}%
  \BibitemOpen
  \bibfield  {author} {\bibinfo {author} {\bibfnamefont {C.}~\bibnamefont
  {Broholm}}, \bibinfo {author} {\bibfnamefont {R.~J.}\ \bibnamefont {Cava}},
  \bibinfo {author} {\bibfnamefont {S.~A.}\ \bibnamefont {Kivelson}}, \bibinfo
  {author} {\bibfnamefont {D.~G.}\ \bibnamefont {Nocera}}, \bibinfo {author}
  {\bibfnamefont {M.~R.}\ \bibnamefont {Norman}},\ and\ \bibinfo {author}
  {\bibfnamefont {T.}~\bibnamefont {Senthil}},\ }\bibfield  {title} {\bibinfo
  {title} {Quantum spin liquids},\ }\bibfield  {journal} {\bibinfo  {journal}
  {Science}\ }\textbf {\bibinfo {volume} {367}},\ \href
  {https://doi.org/10.1126/science.aay0668} {10.1126/science.aay0668} (\bibinfo
  {year} {2020})\BibitemShut {NoStop}%
\bibitem [{\citenamefont {Zhou}\ and\ \citenamefont {Lee}(2011)}]{PAL_sound}%
  \BibitemOpen
  \bibfield  {author} {\bibinfo {author} {\bibfnamefont {Y.}~\bibnamefont
  {Zhou}}\ and\ \bibinfo {author} {\bibfnamefont {P.~A.}\ \bibnamefont {Lee}},\
  }\bibfield  {title} {\bibinfo {title} {Spinon phonon interaction and
  ultrasonic attenuation in quantum spin liquids},\ }\href
  {https://doi.org/10.1103/PhysRevLett.106.056402} {\bibfield  {journal}
  {\bibinfo  {journal} {Phys. Rev. Lett.}\ }\textbf {\bibinfo {volume} {106}},\
  \bibinfo {pages} {056402} (\bibinfo {year} {2011})}\BibitemShut {NoStop}%
\bibitem [{Sup()}]{Supp}%
  \BibitemOpen
  \href@noop {} {}\bibinfo {note} {See Supplemental Material for more details,
  which includes references
  \cite{GStefanucci2013a,AKamenev2023,PBAllen1987,IEsterlis2021,HGuo2022a,CMVarma1989,AAPatel2023,CHerring1954,Pippard,Blount,ABBhatia1961,DChaudhuri2025,QiXu,Halperin76,MZacharias2015,IPaul2017,YWerman2017,HGuo2019,DVElse2021,HGuo2022a}.}\BibitemShut
  {Stop}%
\bibitem [{\citenamefont {Stefanucci}\ and\ \citenamefont {{van
  Leeuwen}}(2013)}]{GStefanucci2013a}%
  \BibitemOpen
  \bibfield  {author} {\bibinfo {author} {\bibfnamefont {G.}~\bibnamefont
  {Stefanucci}}\ and\ \bibinfo {author} {\bibfnamefont {R.}~\bibnamefont {{van
  Leeuwen}}},\ }\href {https://doi.org/10.1017/CBO9781139023979} {\emph
  {\bibinfo {title} {Nonequilibrium {{Many-Body Theory}} of {{Quantum
  Systems}}: {{A Modern Introduction}}}}}\ (\bibinfo  {publisher} {Cambridge
  University Press},\ \bibinfo {address} {Cambridge},\ \bibinfo {year}
  {2013})\BibitemShut {NoStop}%
\bibitem [{\citenamefont {Kamenev}(2023)}]{AKamenev2023}%
  \BibitemOpen
  \bibfield  {author} {\bibinfo {author} {\bibfnamefont {A.}~\bibnamefont
  {Kamenev}},\ }\href {https://doi.org/10.1017/9781108769266} {\emph {\bibinfo
  {title} {Field {{Theory}} of {{Non-Equilibrium Systems}}}}},\ \bibinfo
  {edition} {2nd}\ ed.\ (\bibinfo  {publisher} {Cambridge University Press},\
  \bibinfo {address} {Cambridge},\ \bibinfo {year} {2023})\BibitemShut
  {NoStop}%
\bibitem [{\citenamefont {Glorioso}\ and\ \citenamefont
  {Hartnoll}(2022)}]{PGlorioso2022a}%
  \BibitemOpen
  \bibfield  {author} {\bibinfo {author} {\bibfnamefont {P.}~\bibnamefont
  {Glorioso}}\ and\ \bibinfo {author} {\bibfnamefont {S.~A.}\ \bibnamefont
  {Hartnoll}},\ }\bibfield  {title} {\bibinfo {title} {Joule heating in bad and
  slow metals},\ }\href {https://doi.org/10.21468/SciPostPhys.13.4.095}
  {\bibfield  {journal} {\bibinfo  {journal} {SciPost Physics}\ }\textbf
  {\bibinfo {volume} {13}},\ \bibinfo {pages} {095} (\bibinfo {year}
  {2022})}\BibitemShut {NoStop}%
\bibitem [{\citenamefont {Chaudhuri}\ \emph {et~al.}(2025)\citenamefont
  {Chaudhuri}, \citenamefont {Barbalas}, \citenamefont {Mahmood}, \citenamefont
  {Liang}, \citenamefont {III}, \citenamefont {Legros}, \citenamefont {He},
  \citenamefont {Raffy}, \citenamefont {Bozovic},\ and\ \citenamefont
  {Armitage}}]{DChaudhuri2025}%
  \BibitemOpen
  \bibfield  {author} {\bibinfo {author} {\bibfnamefont {D.}~\bibnamefont
  {Chaudhuri}}, \bibinfo {author} {\bibfnamefont {D.}~\bibnamefont {Barbalas}},
  \bibinfo {author} {\bibfnamefont {F.}~\bibnamefont {Mahmood}}, \bibinfo
  {author} {\bibfnamefont {J.}~\bibnamefont {Liang}}, \bibinfo {author}
  {\bibfnamefont {R.~R.}\ \bibnamefont {III}}, \bibinfo {author} {\bibfnamefont
  {A.}~\bibnamefont {Legros}}, \bibinfo {author} {\bibfnamefont
  {X.}~\bibnamefont {He}}, \bibinfo {author} {\bibfnamefont {H.}~\bibnamefont
  {Raffy}}, \bibinfo {author} {\bibfnamefont {I.}~\bibnamefont {Bozovic}},\
  and\ \bibinfo {author} {\bibfnamefont {N.~P.}\ \bibnamefont {Armitage}},\
  }\href {https://doi.org/10.48550/arXiv.2503.15646} {\bibinfo {title}
  {Planckian dissipation, anomalous high temperature {{THz}} non-linear
  response and energy relaxation in the strange metal state of the cuprate
  superconductors}} (\bibinfo {year} {2025}),\ \Eprint
  {https://arxiv.org/abs/2503.15646} {arXiv:2503.15646 [cond-mat]} \BibitemShut
  {NoStop}%
\bibitem [{\citenamefont {Lee}(1989)}]{PALee1989}%
  \BibitemOpen
  \bibfield  {author} {\bibinfo {author} {\bibfnamefont {P.~A.}\ \bibnamefont
  {Lee}},\ }\bibfield  {title} {\bibinfo {title} {Gauge field,
  {{Aharonov-Bohm}} flux, and high-{${T}_{c}$} superconductivity},\ }\href
  {https://doi.org/10.1103/PhysRevLett.63.680} {\bibfield  {journal} {\bibinfo
  {journal} {Phys. Rev. Lett.}\ }\textbf {\bibinfo {volume} {63}},\ \bibinfo
  {pages} {680} (\bibinfo {year} {1989})}\BibitemShut {NoStop}%
\bibitem [{\citenamefont {Millis}(1993)}]{AJMillis1993}%
  \BibitemOpen
  \bibfield  {author} {\bibinfo {author} {\bibfnamefont {A.~J.}\ \bibnamefont
  {Millis}},\ }\bibfield  {title} {\bibinfo {title} {Effect of a nonzero
  temperature on quantum critical points in itinerant fermion systems},\ }\href
  {https://doi.org/10.1103/PhysRevB.48.7183} {\bibfield  {journal} {\bibinfo
  {journal} {Phys. Rev. B}\ }\textbf {\bibinfo {volume} {48}},\ \bibinfo
  {pages} {7183} (\bibinfo {year} {1993})}\BibitemShut {NoStop}%
\bibitem [{\citenamefont {Polchinski}(1994)}]{JPolchinski1994}%
  \BibitemOpen
  \bibfield  {author} {\bibinfo {author} {\bibfnamefont {J.}~\bibnamefont
  {Polchinski}},\ }\bibfield  {title} {\bibinfo {title} {Low-energy dynamics of
  the spinon-gauge system},\ }\href
  {https://doi.org/10.1016/0550-3213(94)90449-9} {\bibfield  {journal}
  {\bibinfo  {journal} {Nuclear Physics B}\ }\textbf {\bibinfo {volume}
  {422}},\ \bibinfo {pages} {617} (\bibinfo {year} {1994})}\BibitemShut
  {NoStop}%
\bibitem [{\citenamefont {Halperin}\ \emph {et~al.}(1993)\citenamefont
  {Halperin}, \citenamefont {Lee},\ and\ \citenamefont
  {Read}}]{BIHalperin1993}%
  \BibitemOpen
  \bibfield  {author} {\bibinfo {author} {\bibfnamefont {B.~I.}\ \bibnamefont
  {Halperin}}, \bibinfo {author} {\bibfnamefont {P.~A.}\ \bibnamefont {Lee}},\
  and\ \bibinfo {author} {\bibfnamefont {N.}~\bibnamefont {Read}},\ }\bibfield
  {title} {\bibinfo {title} {Theory of the half-filled {{Landau}} level},\
  }\href {https://doi.org/10.1103/PhysRevB.47.7312} {\bibfield  {journal}
  {\bibinfo  {journal} {Phys. Rev. B}\ }\textbf {\bibinfo {volume} {47}},\
  \bibinfo {pages} {7312} (\bibinfo {year} {1993})}\BibitemShut {NoStop}%
\bibitem [{\citenamefont {Kim}\ \emph {et~al.}(1994)\citenamefont {Kim},
  \citenamefont {Furusaki}, \citenamefont {Wen},\ and\ \citenamefont
  {Lee}}]{YBKim1994a}%
  \BibitemOpen
  \bibfield  {author} {\bibinfo {author} {\bibfnamefont {Y.~B.}\ \bibnamefont
  {Kim}}, \bibinfo {author} {\bibfnamefont {A.}~\bibnamefont {Furusaki}},
  \bibinfo {author} {\bibfnamefont {X.-G.}\ \bibnamefont {Wen}},\ and\ \bibinfo
  {author} {\bibfnamefont {P.~A.}\ \bibnamefont {Lee}},\ }\bibfield  {title}
  {\bibinfo {title} {Gauge-invariant response functions of fermions coupled to
  a gauge field},\ }\href {https://doi.org/10.1103/PhysRevB.50.17917}
  {\bibfield  {journal} {\bibinfo  {journal} {Phys. Rev. B}\ }\textbf {\bibinfo
  {volume} {50}},\ \bibinfo {pages} {17917} (\bibinfo {year}
  {1994})}\BibitemShut {NoStop}%
\bibitem [{\citenamefont {Nayak}\ and\ \citenamefont
  {Wilczek}(1994)}]{CNayak1994}%
  \BibitemOpen
  \bibfield  {author} {\bibinfo {author} {\bibfnamefont {C.}~\bibnamefont
  {Nayak}}\ and\ \bibinfo {author} {\bibfnamefont {F.}~\bibnamefont
  {Wilczek}},\ }\bibfield  {title} {\bibinfo {title} {Renormalization group
  approach to low temperature properties of a non-{{Fermi}} liquid metal},\
  }\href {https://doi.org/10.1016/0550-3213(94)90158-9} {\bibfield  {journal}
  {\bibinfo  {journal} {Nuclear Physics B}\ }\textbf {\bibinfo {volume}
  {430}},\ \bibinfo {pages} {534} (\bibinfo {year} {1994})}\BibitemShut
  {NoStop}%
\bibitem [{\citenamefont {Lee}(2009)}]{SSLee2009}%
  \BibitemOpen
  \bibfield  {author} {\bibinfo {author} {\bibfnamefont {S.-S.}\ \bibnamefont
  {Lee}},\ }\bibfield  {title} {\bibinfo {title} {Low-energy effective theory
  of {{Fermi}} surface coupled with {{U}}(1) gauge field in 2 + 1 dimensions},\
  }\href {https://doi.org/10.1103/PhysRevB.80.165102} {\bibfield  {journal}
  {\bibinfo  {journal} {Phys. Rev. B}\ }\textbf {\bibinfo {volume} {80}},\
  \bibinfo {pages} {165102} (\bibinfo {year} {2009})}\BibitemShut {NoStop}%
\bibitem [{\citenamefont {Metlitski}\ and\ \citenamefont
  {Sachdev}(2010)}]{MAMetlitski2010}%
  \BibitemOpen
  \bibfield  {author} {\bibinfo {author} {\bibfnamefont {M.~A.}\ \bibnamefont
  {Metlitski}}\ and\ \bibinfo {author} {\bibfnamefont {S.}~\bibnamefont
  {Sachdev}},\ }\bibfield  {title} {\bibinfo {title} {Quantum phase transitions
  of metals in two spatial dimensions. {{I}}. {{Ising-nematic}} order},\ }\href
  {https://doi.org/10.1103/PhysRevB.82.075127} {\bibfield  {journal} {\bibinfo
  {journal} {Phys. Rev. B}\ }\textbf {\bibinfo {volume} {82}},\ \bibinfo
  {pages} {075127} (\bibinfo {year} {2010})}\BibitemShut {NoStop}%
\bibitem [{\citenamefont {Mross}\ \emph {et~al.}(2010)\citenamefont {Mross},
  \citenamefont {McGreevy}, \citenamefont {Liu},\ and\ \citenamefont
  {Senthil}}]{DFMross2010}%
  \BibitemOpen
  \bibfield  {author} {\bibinfo {author} {\bibfnamefont {D.~F.}\ \bibnamefont
  {Mross}}, \bibinfo {author} {\bibfnamefont {J.}~\bibnamefont {McGreevy}},
  \bibinfo {author} {\bibfnamefont {H.}~\bibnamefont {Liu}},\ and\ \bibinfo
  {author} {\bibfnamefont {T.}~\bibnamefont {Senthil}},\ }\bibfield  {title}
  {\bibinfo {title} {Controlled expansion for certain non-{{Fermi-liquid}}
  metals},\ }\href {https://doi.org/10.1103/PhysRevB.82.045121} {\bibfield
  {journal} {\bibinfo  {journal} {Phys. Rev. B}\ }\textbf {\bibinfo {volume}
  {82}},\ \bibinfo {pages} {045121} (\bibinfo {year} {2010})}\BibitemShut
  {NoStop}%
\bibitem [{\citenamefont {Sur}\ and\ \citenamefont {Lee}(2014)}]{SSur2014}%
  \BibitemOpen
  \bibfield  {author} {\bibinfo {author} {\bibfnamefont {S.}~\bibnamefont
  {Sur}}\ and\ \bibinfo {author} {\bibfnamefont {S.-S.}\ \bibnamefont {Lee}},\
  }\bibfield  {title} {\bibinfo {title} {Chiral non-{{Fermi}} liquids},\ }\href
  {https://doi.org/10.1103/PhysRevB.90.045121} {\bibfield  {journal} {\bibinfo
  {journal} {Phys. Rev. B}\ }\textbf {\bibinfo {volume} {90}},\ \bibinfo
  {pages} {045121} (\bibinfo {year} {2014})}\BibitemShut {NoStop}%
\bibitem [{\citenamefont {Metlitski}\ \emph {et~al.}(2015)\citenamefont
  {Metlitski}, \citenamefont {Mross}, \citenamefont {Sachdev},\ and\
  \citenamefont {Senthil}}]{MAMetlitski2015}%
  \BibitemOpen
  \bibfield  {author} {\bibinfo {author} {\bibfnamefont {M.~A.}\ \bibnamefont
  {Metlitski}}, \bibinfo {author} {\bibfnamefont {D.~F.}\ \bibnamefont
  {Mross}}, \bibinfo {author} {\bibfnamefont {S.}~\bibnamefont {Sachdev}},\
  and\ \bibinfo {author} {\bibfnamefont {T.}~\bibnamefont {Senthil}},\
  }\bibfield  {title} {\bibinfo {title} {Cooper pairing in non-{{Fermi}}
  liquids},\ }\href {https://doi.org/10.1103/PhysRevB.91.115111} {\bibfield
  {journal} {\bibinfo  {journal} {Phys. Rev. B}\ }\textbf {\bibinfo {volume}
  {91}},\ \bibinfo {pages} {115111} (\bibinfo {year} {2015})},\ \Eprint
  {https://arxiv.org/abs/1403.3694} {arXiv:1403.3694 [cond-mat.str-el]}
  \BibitemShut {NoStop}%
\bibitem [{\citenamefont {Hartnoll}\ \emph {et~al.}(2014)\citenamefont
  {Hartnoll}, \citenamefont {Mahajan}, \citenamefont {Punk},\ and\
  \citenamefont {Sachdev}}]{SAHartnoll2014}%
  \BibitemOpen
  \bibfield  {author} {\bibinfo {author} {\bibfnamefont {S.~A.}\ \bibnamefont
  {Hartnoll}}, \bibinfo {author} {\bibfnamefont {R.}~\bibnamefont {Mahajan}},
  \bibinfo {author} {\bibfnamefont {M.}~\bibnamefont {Punk}},\ and\ \bibinfo
  {author} {\bibfnamefont {S.}~\bibnamefont {Sachdev}},\ }\bibfield  {title}
  {\bibinfo {title} {Transport near the {{Ising-nematic}} quantum critical
  point of metals in two dimensions},\ }\href
  {https://doi.org/10.1103/PhysRevB.89.155130} {\bibfield  {journal} {\bibinfo
  {journal} {Phys. Rev. B}\ }\textbf {\bibinfo {volume} {89}},\ \bibinfo
  {pages} {155130} (\bibinfo {year} {2014})}\BibitemShut {NoStop}%
\bibitem [{\citenamefont {Eberlein}\ \emph {et~al.}(2017)\citenamefont
  {Eberlein}, \citenamefont {Patel},\ and\ \citenamefont
  {Sachdev}}]{AEberlein2017}%
  \BibitemOpen
  \bibfield  {author} {\bibinfo {author} {\bibfnamefont {A.}~\bibnamefont
  {Eberlein}}, \bibinfo {author} {\bibfnamefont {A.~A.}\ \bibnamefont
  {Patel}},\ and\ \bibinfo {author} {\bibfnamefont {S.}~\bibnamefont
  {Sachdev}},\ }\bibfield  {title} {\bibinfo {title} {Shear viscosity at the
  {{Ising-nematic}} quantum critical point in two-dimensional metals},\ }\href
  {https://doi.org/10.1103/PhysRevB.95.075127} {\bibfield  {journal} {\bibinfo
  {journal} {Phys. Rev. B}\ }\textbf {\bibinfo {volume} {95}},\ \bibinfo
  {pages} {075127} (\bibinfo {year} {2017})}\BibitemShut {NoStop}%
\bibitem [{\citenamefont {Holder}\ and\ \citenamefont
  {Metzner}(2015{\natexlab{a}})}]{THolder2015}%
  \BibitemOpen
  \bibfield  {author} {\bibinfo {author} {\bibfnamefont {T.}~\bibnamefont
  {Holder}}\ and\ \bibinfo {author} {\bibfnamefont {W.}~\bibnamefont
  {Metzner}},\ }\bibfield  {title} {\bibinfo {title} {Anomalous dynamical
  scaling from nematic and {{U}}(1) gauge field fluctuations in two-dimensional
  metals},\ }\href {https://doi.org/10.1103/PhysRevB.92.041112} {\bibfield
  {journal} {\bibinfo  {journal} {Phys. Rev. B}\ }\textbf {\bibinfo {volume}
  {92}},\ \bibinfo {pages} {041112} (\bibinfo {year}
  {2015}{\natexlab{a}})}\BibitemShut {NoStop}%
\bibitem [{\citenamefont {Holder}\ and\ \citenamefont
  {Metzner}(2015{\natexlab{b}})}]{THolder2015a}%
  \BibitemOpen
  \bibfield  {author} {\bibinfo {author} {\bibfnamefont {T.}~\bibnamefont
  {Holder}}\ and\ \bibinfo {author} {\bibfnamefont {W.}~\bibnamefont
  {Metzner}},\ }\bibfield  {title} {\bibinfo {title} {Fermion loops and
  improved power-counting in two-dimensional critical metals with singular
  forward scattering},\ }\href {https://doi.org/10.1103/PhysRevB.92.245128}
  {\bibfield  {journal} {\bibinfo  {journal} {Phys. Rev. B}\ }\textbf {\bibinfo
  {volume} {92}},\ \bibinfo {pages} {245128} (\bibinfo {year}
  {2015}{\natexlab{b}})}\BibitemShut {NoStop}%
\bibitem [{\citenamefont {Fitzpatrick}\ \emph {et~al.}(2014)\citenamefont
  {Fitzpatrick}, \citenamefont {Kachru}, \citenamefont {Kaplan},\ and\
  \citenamefont {Raghu}}]{ALFitzpatrick2014}%
  \BibitemOpen
  \bibfield  {author} {\bibinfo {author} {\bibfnamefont {A.~L.}\ \bibnamefont
  {Fitzpatrick}}, \bibinfo {author} {\bibfnamefont {S.}~\bibnamefont {Kachru}},
  \bibinfo {author} {\bibfnamefont {J.}~\bibnamefont {Kaplan}},\ and\ \bibinfo
  {author} {\bibfnamefont {S.}~\bibnamefont {Raghu}},\ }\bibfield  {title}
  {\bibinfo {title} {Non-{{Fermi-liquid}} behavior of
  large-{{N}}{{{\textsubscript{B}}}} quantum critical metals},\ }\href
  {https://doi.org/10.1103/PhysRevB.89.165114} {\bibfield  {journal} {\bibinfo
  {journal} {Phys. Rev. B}\ }\textbf {\bibinfo {volume} {89}},\ \bibinfo
  {pages} {165114} (\bibinfo {year} {2014})},\ \Eprint
  {https://arxiv.org/abs/1312.3321} {arXiv:1312.3321 [cond-mat.str-el]}
  \BibitemShut {NoStop}%
\bibitem [{\citenamefont {Damia}\ \emph {et~al.}(2019)\citenamefont {Damia},
  \citenamefont {Kachru}, \citenamefont {Raghu},\ and\ \citenamefont
  {Torroba}}]{JADamia2019}%
  \BibitemOpen
  \bibfield  {author} {\bibinfo {author} {\bibfnamefont {J.~A.}\ \bibnamefont
  {Damia}}, \bibinfo {author} {\bibfnamefont {S.}~\bibnamefont {Kachru}},
  \bibinfo {author} {\bibfnamefont {S.}~\bibnamefont {Raghu}},\ and\ \bibinfo
  {author} {\bibfnamefont {G.}~\bibnamefont {Torroba}},\ }\bibfield  {title}
  {\bibinfo {title} {Two-{{Dimensional Non-Fermi-Liquid Metals}}: {{A Solvable
  Large- N Limit}}},\ }\href {https://doi.org/10.1103/PhysRevLett.123.096402}
  {\bibfield  {journal} {\bibinfo  {journal} {Phys. Rev. Lett.}\ }\textbf
  {\bibinfo {volume} {123}},\ \bibinfo {pages} {096402} (\bibinfo {year}
  {2019})}\BibitemShut {NoStop}%
\bibitem [{\citenamefont {Damia}\ \emph {et~al.}(2020)\citenamefont {Damia},
  \citenamefont {Sol{\'i}s},\ and\ \citenamefont {Torroba}}]{JADamia2020}%
  \BibitemOpen
  \bibfield  {author} {\bibinfo {author} {\bibfnamefont {J.~A.}\ \bibnamefont
  {Damia}}, \bibinfo {author} {\bibfnamefont {M.}~\bibnamefont {Sol{\'i}s}},\
  and\ \bibinfo {author} {\bibfnamefont {G.}~\bibnamefont {Torroba}},\
  }\bibfield  {title} {\bibinfo {title} {How non-{{Fermi}} liquids cure their
  infrared divergences},\ }\href {https://doi.org/10.1103/PhysRevB.102.045147}
  {\bibfield  {journal} {\bibinfo  {journal} {Phys. Rev. B}\ }\textbf {\bibinfo
  {volume} {102}},\ \bibinfo {pages} {045147} (\bibinfo {year}
  {2020})}\BibitemShut {NoStop}%
\bibitem [{\citenamefont {Damia}\ \emph {et~al.}(2021)\citenamefont {Damia},
  \citenamefont {Sol{\'i}s},\ and\ \citenamefont {Torroba}}]{JADamia2021}%
  \BibitemOpen
  \bibfield  {author} {\bibinfo {author} {\bibfnamefont {J.~A.}\ \bibnamefont
  {Damia}}, \bibinfo {author} {\bibfnamefont {M.}~\bibnamefont {Sol{\'i}s}},\
  and\ \bibinfo {author} {\bibfnamefont {G.}~\bibnamefont {Torroba}},\
  }\bibfield  {title} {\bibinfo {title} {Thermal effects in non-{{Fermi}}
  liquid superconductivity},\ }\href
  {https://doi.org/10.1103/PhysRevB.103.155161} {\bibfield  {journal} {\bibinfo
   {journal} {Phys. Rev. B}\ }\textbf {\bibinfo {volume} {103}},\ \bibinfo
  {pages} {155161} (\bibinfo {year} {2021})}\BibitemShut {NoStop}%
\bibitem [{\citenamefont {Ridgway}\ and\ \citenamefont
  {Hooley}(2015)}]{SPRidgway2015}%
  \BibitemOpen
  \bibfield  {author} {\bibinfo {author} {\bibfnamefont {S.~P.}\ \bibnamefont
  {Ridgway}}\ and\ \bibinfo {author} {\bibfnamefont {C.~A.}\ \bibnamefont
  {Hooley}},\ }\bibfield  {title} {\bibinfo {title} {Non-{{Fermi-Liquid
  Behavior}} and {{Anomalous Suppression}} of {{Landau Damping}} in {{Layered
  Metals Close}} to {{Ferromagnetism}}},\ }\href
  {https://doi.org/10.1103/PhysRevLett.114.226404} {\bibfield  {journal}
  {\bibinfo  {journal} {Phys. Rev. Lett.}\ }\textbf {\bibinfo {volume} {114}},\
  \bibinfo {pages} {226404} (\bibinfo {year} {2015})}\BibitemShut {NoStop}%
\bibitem [{\citenamefont {Abanov}\ and\ \citenamefont
  {Chubukov}(2020)}]{AAbanov2020}%
  \BibitemOpen
  \bibfield  {author} {\bibinfo {author} {\bibfnamefont {A.}~\bibnamefont
  {Abanov}}\ and\ \bibinfo {author} {\bibfnamefont {A.~V.}\ \bibnamefont
  {Chubukov}},\ }\bibfield  {title} {\bibinfo {title} {Interplay between
  superconductivity and non-{{Fermi}} liquid at a quantum critical point in a
  metal. {{I}}.},\ }\href {https://doi.org/10.1103/PhysRevB.102.024524}
  {\bibfield  {journal} {\bibinfo  {journal} {Phys. Rev. B}\ }\textbf {\bibinfo
  {volume} {102}},\ \bibinfo {pages} {024524} (\bibinfo {year}
  {2020})}\BibitemShut {NoStop}%
\bibitem [{\citenamefont {Wu}\ \emph {et~al.}(2020)\citenamefont {Wu},
  \citenamefont {Abanov}, \citenamefont {Wang},\ and\ \citenamefont
  {Chubukov}}]{YMWu2020}%
  \BibitemOpen
  \bibfield  {author} {\bibinfo {author} {\bibfnamefont {Y.-M.}\ \bibnamefont
  {Wu}}, \bibinfo {author} {\bibfnamefont {A.}~\bibnamefont {Abanov}}, \bibinfo
  {author} {\bibfnamefont {Y.}~\bibnamefont {Wang}},\ and\ \bibinfo {author}
  {\bibfnamefont {A.~V.}\ \bibnamefont {Chubukov}},\ }\bibfield  {title}
  {\bibinfo {title} {Interplay between superconductivity and non-{{Fermi}}
  liquid at a quantum critical point in a metal. {{II}}. {{The}} {{$\gamma$}}
  model at a finite {{T}} for {{0 $<$ $\gamma<$1}}},\ }\href
  {https://doi.org/10.1103/PhysRevB.102.024525} {\bibfield  {journal} {\bibinfo
   {journal} {Phys. Rev. B}\ }\textbf {\bibinfo {volume} {102}},\ \bibinfo
  {eid} {024525} (\bibinfo {year} {2020})},\ \Eprint
  {https://arxiv.org/abs/2006.02968} {arXiv:2006.02968 [cond-mat.supr-con]}
  \BibitemShut {NoStop}%
\bibitem [{\citenamefont {Wang}\ and\ \citenamefont {Berg}(2019)}]{XWang2019}%
  \BibitemOpen
  \bibfield  {author} {\bibinfo {author} {\bibfnamefont {X.}~\bibnamefont
  {Wang}}\ and\ \bibinfo {author} {\bibfnamefont {E.}~\bibnamefont {Berg}},\
  }\bibfield  {title} {\bibinfo {title} {Scattering mechanisms and electrical
  transport near an {{Ising}} nematic quantum critical point},\ }\href
  {https://doi.org/10.1103/PhysRevB.99.235136} {\bibfield  {journal} {\bibinfo
  {journal} {Phys. Rev. B}\ }\textbf {\bibinfo {volume} {99}},\ \bibinfo {eid}
  {235136} (\bibinfo {year} {2019})},\ \Eprint
  {https://arxiv.org/abs/1902.04590} {arXiv:1902.04590 [cond-mat.str-el]}
  \BibitemShut {NoStop}%
\bibitem [{\citenamefont {Klein}\ \emph {et~al.}(2020)\citenamefont {Klein},
  \citenamefont {Chubukov}, \citenamefont {Schattner},\ and\ \citenamefont
  {Berg}}]{AKlein2020}%
  \BibitemOpen
  \bibfield  {author} {\bibinfo {author} {\bibfnamefont {A.}~\bibnamefont
  {Klein}}, \bibinfo {author} {\bibfnamefont {A.~V.}\ \bibnamefont {Chubukov}},
  \bibinfo {author} {\bibfnamefont {Y.}~\bibnamefont {Schattner}},\ and\
  \bibinfo {author} {\bibfnamefont {E.}~\bibnamefont {Berg}},\ }\bibfield
  {title} {\bibinfo {title} {Normal {{State Properties}} of {{Quantum Critical
  Metals}} at {{Finite Temperature}}},\ }\href
  {https://doi.org/10.1103/PhysRevX.10.031053} {\bibfield  {journal} {\bibinfo
  {journal} {Phys. Rev. X}\ }\textbf {\bibinfo {volume} {10}},\ \bibinfo
  {pages} {031053} (\bibinfo {year} {2020})}\BibitemShut {NoStop}%
\bibitem [{\citenamefont {Grossman}\ \emph {et~al.}(2021)\citenamefont
  {Grossman}, \citenamefont {Hofmann}, \citenamefont {Holder},\ and\
  \citenamefont {Berg}}]{OGrossman2021}%
  \BibitemOpen
  \bibfield  {author} {\bibinfo {author} {\bibfnamefont {O.}~\bibnamefont
  {Grossman}}, \bibinfo {author} {\bibfnamefont {J.~S.}\ \bibnamefont
  {Hofmann}}, \bibinfo {author} {\bibfnamefont {T.}~\bibnamefont {Holder}},\
  and\ \bibinfo {author} {\bibfnamefont {E.}~\bibnamefont {Berg}},\ }\bibfield
  {title} {\bibinfo {title} {Specific heat of a quantum critical metal},\
  }\bibfield  {journal} {\bibinfo  {journal} {Physical Review Letters}\
  }\textbf {\bibinfo {volume} {127}},\ \href
  {https://doi.org/10.1103/physrevlett.127.017601}
  {10.1103/physrevlett.127.017601} (\bibinfo {year} {2021})\BibitemShut
  {NoStop}%
\bibitem [{\citenamefont {Chowdhury}\ and\ \citenamefont
  {Berg}(2020)}]{DChowdhury2020}%
  \BibitemOpen
  \bibfield  {author} {\bibinfo {author} {\bibfnamefont {D.}~\bibnamefont
  {Chowdhury}}\ and\ \bibinfo {author} {\bibfnamefont {E.}~\bibnamefont
  {Berg}},\ }\bibfield  {title} {\bibinfo {title} {The unreasonable
  effectiveness of {{Eliashberg}} theory for pairing of non-{{Fermi}}
  liquids},\ }\href {https://doi.org/10.1016/j.aop.2020.168125} {\bibfield
  {journal} {\bibinfo  {journal} {Annals of Physics}\ }\textbf {\bibinfo
  {volume} {417}},\ \bibinfo {eid} {168125} (\bibinfo {year} {2020})},\ \Eprint
  {https://arxiv.org/abs/1912.07646} {arXiv:1912.07646 [cond-mat.supr-con]}
  \BibitemShut {NoStop}%
\bibitem [{\citenamefont {Oganesyan}\ \emph {et~al.}(2001)\citenamefont
  {Oganesyan}, \citenamefont {Kivelson},\ and\ \citenamefont
  {Fradkin}}]{VOganesyan2001}%
  \BibitemOpen
  \bibfield  {author} {\bibinfo {author} {\bibfnamefont {V.}~\bibnamefont
  {Oganesyan}}, \bibinfo {author} {\bibfnamefont {S.~A.}\ \bibnamefont
  {Kivelson}},\ and\ \bibinfo {author} {\bibfnamefont {E.}~\bibnamefont
  {Fradkin}},\ }\bibfield  {title} {\bibinfo {title} {Quantum theory of a
  nematic {{Fermi}} fluid},\ }\href
  {https://doi.org/10.1103/PhysRevB.64.195109} {\bibfield  {journal} {\bibinfo
  {journal} {Phys. Rev. B}\ }\textbf {\bibinfo {volume} {64}},\ \bibinfo
  {pages} {195109} (\bibinfo {year} {2001})}\BibitemShut {NoStop}%
\bibitem [{\citenamefont {Chubukov}\ and\ \citenamefont
  {Maslov}(2017)}]{AVChubukov2017}%
  \BibitemOpen
  \bibfield  {author} {\bibinfo {author} {\bibfnamefont {A.~V.}\ \bibnamefont
  {Chubukov}}\ and\ \bibinfo {author} {\bibfnamefont {D.~L.}\ \bibnamefont
  {Maslov}},\ }\bibfield  {title} {\bibinfo {title} {Optical conductivity of a
  two-dimensional metal near a quantum critical point: {{The}} status of the
  extended {{Drude}} formula},\ }\href
  {https://doi.org/10.1103/PhysRevB.96.205136} {\bibfield  {journal} {\bibinfo
  {journal} {Phys. Rev. B}\ }\textbf {\bibinfo {volume} {96}},\ \bibinfo {eid}
  {205136} (\bibinfo {year} {2017})},\ \Eprint
  {https://arxiv.org/abs/1707.07352} {arXiv:1707.07352 [cond-mat.str-el]}
  \BibitemShut {NoStop}%
\bibitem [{\citenamefont {Maslov}\ and\ \citenamefont
  {Chubukov}(2017)}]{DLMaslov2017}%
  \BibitemOpen
  \bibfield  {author} {\bibinfo {author} {\bibfnamefont {D.~L.}\ \bibnamefont
  {Maslov}}\ and\ \bibinfo {author} {\bibfnamefont {A.~V.}\ \bibnamefont
  {Chubukov}},\ }\bibfield  {title} {\bibinfo {title} {Optical response of
  correlated electron systems},\ }\href
  {https://doi.org/10.1088/1361-6633/80/2/026503} {\bibfield  {journal}
  {\bibinfo  {journal} {Reports on Progress in Physics}\ }\textbf {\bibinfo
  {volume} {80}},\ \bibinfo {eid} {026503} (\bibinfo {year} {2017})},\ \Eprint
  {https://arxiv.org/abs/1608.02514} {arXiv:1608.02514 [cond-mat.str-el]}
  \BibitemShut {NoStop}%
\bibitem [{\citenamefont {Li}\ \emph {et~al.}(2023)\citenamefont {Li},
  \citenamefont {Sharma}, \citenamefont {Levchenko},\ and\ \citenamefont
  {Maslov}}]{SLi2023}%
  \BibitemOpen
  \bibfield  {author} {\bibinfo {author} {\bibfnamefont {S.}~\bibnamefont
  {Li}}, \bibinfo {author} {\bibfnamefont {P.}~\bibnamefont {Sharma}}, \bibinfo
  {author} {\bibfnamefont {A.}~\bibnamefont {Levchenko}},\ and\ \bibinfo
  {author} {\bibfnamefont {D.~L.}\ \bibnamefont {Maslov}},\ }\bibfield  {title}
  {\bibinfo {title} {Optical conductivity of a metal near an {{Ising-nematic}}
  quantum critical point},\ }\href
  {https://doi.org/10.1103/PhysRevB.108.235125} {\bibfield  {journal} {\bibinfo
   {journal} {Physical Review B}\ }\textbf {\bibinfo {volume} {108}},\ \bibinfo
  {pages} {235125} (\bibinfo {year} {2023})}\BibitemShut {NoStop}%
\bibitem [{\citenamefont {Aldape}\ \emph {et~al.}(2022)\citenamefont {Aldape},
  \citenamefont {Cookmeyer}, \citenamefont {Patel},\ and\ \citenamefont
  {Altman}}]{EEAldape2022}%
  \BibitemOpen
  \bibfield  {author} {\bibinfo {author} {\bibfnamefont {E.~E.}\ \bibnamefont
  {Aldape}}, \bibinfo {author} {\bibfnamefont {T.}~\bibnamefont {Cookmeyer}},
  \bibinfo {author} {\bibfnamefont {A.~A.}\ \bibnamefont {Patel}},\ and\
  \bibinfo {author} {\bibfnamefont {E.}~\bibnamefont {Altman}},\ }\bibfield
  {title} {\bibinfo {title} {Solvable theory of a strange metal at the
  breakdown of a heavy {{Fermi}} liquid},\ }\href
  {https://doi.org/10.1103/PhysRevB.105.235111} {\bibfield  {journal} {\bibinfo
   {journal} {Phys. Rev. B}\ }\textbf {\bibinfo {volume} {105}},\ \bibinfo
  {eid} {235111} (\bibinfo {year} {2022})},\ \Eprint
  {https://arxiv.org/abs/2012.00763} {arXiv:2012.00763 [cond-mat.str-el]}
  \BibitemShut {NoStop}%
\bibitem [{\citenamefont {Esterlis}\ \emph {et~al.}(2021)\citenamefont
  {Esterlis}, \citenamefont {Guo}, \citenamefont {Patel},\ and\ \citenamefont
  {Sachdev}}]{IEsterlis2021}%
  \BibitemOpen
  \bibfield  {author} {\bibinfo {author} {\bibfnamefont {I.}~\bibnamefont
  {Esterlis}}, \bibinfo {author} {\bibfnamefont {H.}~\bibnamefont {Guo}},
  \bibinfo {author} {\bibfnamefont {A.~A.}\ \bibnamefont {Patel}},\ and\
  \bibinfo {author} {\bibfnamefont {S.}~\bibnamefont {Sachdev}},\ }\bibfield
  {title} {\bibinfo {title} {Large {{N}} theory of critical {{Fermi}}
  surfaces},\ }\href {https://doi.org/10.1103/PhysRevB.103.235129} {\bibfield
  {journal} {\bibinfo  {journal} {Phys. Rev. B}\ }\textbf {\bibinfo {volume}
  {103}},\ \bibinfo {pages} {235129} (\bibinfo {year} {2021})},\ \Eprint
  {https://arxiv.org/abs/2103.08615} {arXiv:2103.08615 [cond-mat.str-el]}
  \BibitemShut {NoStop}%
\bibitem [{\citenamefont {Kim}\ \emph {et~al.}(1995)\citenamefont {Kim},
  \citenamefont {Lee},\ and\ \citenamefont {Wen}}]{YBKim1995a}%
  \BibitemOpen
  \bibfield  {author} {\bibinfo {author} {\bibfnamefont {Y.~B.}\ \bibnamefont
  {Kim}}, \bibinfo {author} {\bibfnamefont {P.~A.}\ \bibnamefont {Lee}},\ and\
  \bibinfo {author} {\bibfnamefont {X.-G.}\ \bibnamefont {Wen}},\ }\bibfield
  {title} {\bibinfo {title} {Quantum {{Boltzmann}} equation of composite
  fermions interacting with a gauge field},\ }\href
  {https://doi.org/10.1103/PhysRevB.52.17275} {\bibfield  {journal} {\bibinfo
  {journal} {Phys. Rev. B}\ }\textbf {\bibinfo {volume} {52}},\ \bibinfo
  {pages} {17275} (\bibinfo {year} {1995})}\BibitemShut {NoStop}%
\bibitem [{\citenamefont {Delacr{\'e}taz}\ \emph {et~al.}(2022)\citenamefont
  {Delacr{\'e}taz}, \citenamefont {Du}, \citenamefont {Mehta},\ and\
  \citenamefont {Son}}]{LVDelacretaz2022a}%
  \BibitemOpen
  \bibfield  {author} {\bibinfo {author} {\bibfnamefont {L.~V.}\ \bibnamefont
  {Delacr{\'e}taz}}, \bibinfo {author} {\bibfnamefont {Y.-H.}\ \bibnamefont
  {Du}}, \bibinfo {author} {\bibfnamefont {U.}~\bibnamefont {Mehta}},\ and\
  \bibinfo {author} {\bibfnamefont {D.~T.}\ \bibnamefont {Son}},\ }\bibfield
  {title} {\bibinfo {title} {Nonlinear bosonization of {{Fermi}} surfaces:
  {{The}} method of coadjoint orbits},\ }\href
  {https://doi.org/10.1103/PhysRevResearch.4.033131} {\bibfield  {journal}
  {\bibinfo  {journal} {Phys. Rev. Res.}\ }\textbf {\bibinfo {volume} {4}},\
  \bibinfo {pages} {033131} (\bibinfo {year} {2022})}\BibitemShut {NoStop}%
\bibitem [{\citenamefont {Han}\ \emph {et~al.}(2023)\citenamefont {Han},
  \citenamefont {Desrochers},\ and\ \citenamefont {Kim}}]{SEHan2023}%
  \BibitemOpen
  \bibfield  {author} {\bibinfo {author} {\bibfnamefont {S.}~\bibnamefont
  {Han}}, \bibinfo {author} {\bibfnamefont {F.}~\bibnamefont {Desrochers}},\
  and\ \bibinfo {author} {\bibfnamefont {Y.~B.}\ \bibnamefont {Kim}},\ }\href
  {https://doi.org/10.48550/arXiv.2306.14955} {\bibinfo {title} {Bosonization
  of {{Non-Fermi Liquids}}}} (\bibinfo {year} {2023}),\ \Eprint
  {https://arxiv.org/abs/2306.14955} {arXiv:2306.14955 [cond-mat,
  physics:hep-th]} \BibitemShut {NoStop}%
\bibitem [{\citenamefont {Else}\ \emph {et~al.}(2021)\citenamefont {Else},
  \citenamefont {Thorngren},\ and\ \citenamefont {Senthil}}]{DVElse2021a}%
  \BibitemOpen
  \bibfield  {author} {\bibinfo {author} {\bibfnamefont {D.~V.}\ \bibnamefont
  {Else}}, \bibinfo {author} {\bibfnamefont {R.}~\bibnamefont {Thorngren}},\
  and\ \bibinfo {author} {\bibfnamefont {T.}~\bibnamefont {Senthil}},\
  }\bibfield  {title} {\bibinfo {title} {Non-{{Fermi Liquids}} as {{Ersatz
  Fermi Liquids}}: {{General Constraints}} on {{Compressible Metals}}},\ }\href
  {https://doi.org/10.1103/PhysRevX.11.021005} {\bibfield  {journal} {\bibinfo
  {journal} {Phys. Rev. X}\ }\textbf {\bibinfo {volume} {11}},\ \bibinfo
  {pages} {021005} (\bibinfo {year} {2021})}\BibitemShut {NoStop}%
\bibitem [{\citenamefont {Shi}\ \emph {et~al.}(2022)\citenamefont {Shi},
  \citenamefont {Goldman}, \citenamefont {Else},\ and\ \citenamefont
  {Senthil}}]{ZDShi2022}%
  \BibitemOpen
  \bibfield  {author} {\bibinfo {author} {\bibfnamefont {Z.~D.}\ \bibnamefont
  {Shi}}, \bibinfo {author} {\bibfnamefont {H.}~\bibnamefont {Goldman}},
  \bibinfo {author} {\bibfnamefont {D.~V.}\ \bibnamefont {Else}},\ and\
  \bibinfo {author} {\bibfnamefont {T.}~\bibnamefont {Senthil}},\ }\bibfield
  {title} {\bibinfo {title} {Gifts from anomalies: {{Exact}} results for
  {{Landau}} phase transitions in metals},\ }\href
  {https://doi.org/10.21468/SciPostPhys.13.5.102} {\bibfield  {journal}
  {\bibinfo  {journal} {SciPost Physics}\ }\textbf {\bibinfo {volume} {13}},\
  \bibinfo {pages} {102} (\bibinfo {year} {2022})}\BibitemShut {NoStop}%
\bibitem [{\citenamefont {Qi}\ and\ \citenamefont {Xu}(2009)}]{QiXu}%
  \BibitemOpen
  \bibfield  {author} {\bibinfo {author} {\bibfnamefont {Y.}~\bibnamefont
  {Qi}}\ and\ \bibinfo {author} {\bibfnamefont {C.}~\bibnamefont {Xu}},\
  }\bibfield  {title} {\bibinfo {title} {Global phase diagram for magnetism and
  lattice distortion of iron-pnictide materials},\ }\href
  {https://doi.org/10.1103/PhysRevB.80.094402} {\bibfield  {journal} {\bibinfo
  {journal} {Phys. Rev. B}\ }\textbf {\bibinfo {volume} {80}},\ \bibinfo
  {pages} {094402} (\bibinfo {year} {2009})}\BibitemShut {NoStop}%
\bibitem [{\citenamefont {Bergman}\ and\ \citenamefont
  {Halperin}(1976)}]{Halperin76}%
  \BibitemOpen
  \bibfield  {author} {\bibinfo {author} {\bibfnamefont {D.~J.}\ \bibnamefont
  {Bergman}}\ and\ \bibinfo {author} {\bibfnamefont {B.~I.}\ \bibnamefont
  {Halperin}},\ }\bibfield  {title} {\bibinfo {title} {Critical behavior of an
  ising model on a cubic compressible lattice},\ }\href
  {https://doi.org/10.1103/PhysRevB.13.2145} {\bibfield  {journal} {\bibinfo
  {journal} {Phys. Rev. B}\ }\textbf {\bibinfo {volume} {13}},\ \bibinfo
  {pages} {2145} (\bibinfo {year} {1976})}\BibitemShut {NoStop}%
\bibitem [{\citenamefont {Guo}\ \emph {et~al.}(2024)\citenamefont {Guo},
  \citenamefont {Valentinis}, \citenamefont {Schmalian}, \citenamefont
  {Sachdev},\ and\ \citenamefont {Patel}}]{HGuo2024}%
  \BibitemOpen
  \bibfield  {author} {\bibinfo {author} {\bibfnamefont {H.}~\bibnamefont
  {Guo}}, \bibinfo {author} {\bibfnamefont {D.}~\bibnamefont {Valentinis}},
  \bibinfo {author} {\bibfnamefont {J.}~\bibnamefont {Schmalian}}, \bibinfo
  {author} {\bibfnamefont {S.}~\bibnamefont {Sachdev}},\ and\ \bibinfo {author}
  {\bibfnamefont {A.~A.}\ \bibnamefont {Patel}},\ }\bibfield  {title} {\bibinfo
  {title} {Cyclotron resonance and quantum oscillations of critical {{Fermi}}
  surfaces},\ }\href {https://doi.org/10.1103/PhysRevB.109.075162} {\bibfield
  {journal} {\bibinfo  {journal} {Phys. Rev. B}\ }\textbf {\bibinfo {volume}
  {109}},\ \bibinfo {pages} {075162} (\bibinfo {year} {2024})}\BibitemShut
  {NoStop}%
\bibitem [{\citenamefont {Patel}\ and\ \citenamefont
  {Sachdev}(2014)}]{AAPatel2014}%
  \BibitemOpen
  \bibfield  {author} {\bibinfo {author} {\bibfnamefont {A.~A.}\ \bibnamefont
  {Patel}}\ and\ \bibinfo {author} {\bibfnamefont {S.}~\bibnamefont
  {Sachdev}},\ }\bibfield  {title} {\bibinfo {title} {Dc resistivity at the
  onset of spin density wave order in two-dimensional metals},\ }\href
  {https://doi.org/10.1103/PhysRevB.90.165146} {\bibfield  {journal} {\bibinfo
  {journal} {Phys. Rev. B}\ }\textbf {\bibinfo {volume} {90}},\ \bibinfo
  {pages} {165146} (\bibinfo {year} {2014})}\BibitemShut {NoStop}%
\bibitem [{\citenamefont {Lu}\ \emph {et~al.}(2016)\citenamefont {Lu},
  \citenamefont {Zhang}, \citenamefont {Hwang}, \citenamefont {Ofori-Okai},
  \citenamefont {Fleischer},\ and\ \citenamefont {Nelson}}]{keithnelson}%
  \BibitemOpen
  \bibfield  {author} {\bibinfo {author} {\bibfnamefont {J.}~\bibnamefont
  {Lu}}, \bibinfo {author} {\bibfnamefont {Y.}~\bibnamefont {Zhang}}, \bibinfo
  {author} {\bibfnamefont {H.~Y.}\ \bibnamefont {Hwang}}, \bibinfo {author}
  {\bibfnamefont {B.~K.}\ \bibnamefont {Ofori-Okai}}, \bibinfo {author}
  {\bibfnamefont {S.}~\bibnamefont {Fleischer}},\ and\ \bibinfo {author}
  {\bibfnamefont {K.~A.}\ \bibnamefont {Nelson}},\ }\bibfield  {title}
  {\bibinfo {title} {Nonlinear two-dimensional terahertz photon echo and
  rotational spectroscopy in the gas phase},\ }\href
  {https://doi.org/10.1073/pnas.1609558113} {\bibfield  {journal} {\bibinfo
  {journal} {Proceedings of the National Academy of Sciences}\ }\textbf
  {\bibinfo {volume} {113}},\ \bibinfo {pages} {11800} (\bibinfo {year}
  {2016})},\ \Eprint
  {https://arxiv.org/abs/https://www.pnas.org/doi/pdf/10.1073/pnas.1609558113}
  {https://www.pnas.org/doi/pdf/10.1073/pnas.1609558113} \BibitemShut {NoStop}%
\bibitem [{\citenamefont {Wan}\ and\ \citenamefont {Armitage}(2019)}]{2DCS}%
  \BibitemOpen
  \bibfield  {author} {\bibinfo {author} {\bibfnamefont {Y.}~\bibnamefont
  {Wan}}\ and\ \bibinfo {author} {\bibfnamefont {N.~P.}\ \bibnamefont
  {Armitage}},\ }\bibfield  {title} {\bibinfo {title} {Resolving continua of
  fractional excitations by spinon echo in thz 2d coherent spectroscopy},\
  }\href {https://doi.org/10.1103/PhysRevLett.122.257401} {\bibfield  {journal}
  {\bibinfo  {journal} {Phys. Rev. Lett.}\ }\textbf {\bibinfo {volume} {122}},\
  \bibinfo {pages} {257401} (\bibinfo {year} {2019})}\BibitemShut {NoStop}%
\bibitem [{\citenamefont {Mahmood}\ \emph {et~al.}(2021)\citenamefont
  {Mahmood}, \citenamefont {Chaudhuri}, \citenamefont {Gopalakrishnan},
  \citenamefont {Nandkishore},\ and\ \citenamefont {Armitage}}]{Mahmood2021}%
  \BibitemOpen
  \bibfield  {author} {\bibinfo {author} {\bibfnamefont {F.}~\bibnamefont
  {Mahmood}}, \bibinfo {author} {\bibfnamefont {D.}~\bibnamefont {Chaudhuri}},
  \bibinfo {author} {\bibfnamefont {S.}~\bibnamefont {Gopalakrishnan}},
  \bibinfo {author} {\bibfnamefont {R.}~\bibnamefont {Nandkishore}},\ and\
  \bibinfo {author} {\bibfnamefont {N.~P.}\ \bibnamefont {Armitage}},\
  }\bibfield  {title} {\bibinfo {title} {Observation of a marginal fermi
  glass},\ }\href {https://doi.org/10.1038/s41567-020-01149-0} {\bibfield
  {journal} {\bibinfo  {journal} {Nature Physics}\ }\textbf {\bibinfo {volume}
  {17}},\ \bibinfo {pages} {627} (\bibinfo {year} {2021})}\BibitemShut
  {NoStop}%
\bibitem [{\citenamefont {Barbalas}\ \emph {et~al.}(2023)\citenamefont
  {Barbalas}, \citenamefont {III}, \citenamefont {Chaudhuri}, \citenamefont
  {Mahmood}, \citenamefont {Nair}, \citenamefont {Schreiber}, \citenamefont
  {Schlom}, \citenamefont {Shen},\ and\ \citenamefont
  {Armitage}}]{DBarbalas2023}%
  \BibitemOpen
  \bibfield  {author} {\bibinfo {author} {\bibfnamefont {D.}~\bibnamefont
  {Barbalas}}, \bibinfo {author} {\bibfnamefont {R.~R.}\ \bibnamefont {III}},
  \bibinfo {author} {\bibfnamefont {D.}~\bibnamefont {Chaudhuri}}, \bibinfo
  {author} {\bibfnamefont {F.}~\bibnamefont {Mahmood}}, \bibinfo {author}
  {\bibfnamefont {H.~P.}\ \bibnamefont {Nair}}, \bibinfo {author}
  {\bibfnamefont {N.~J.}\ \bibnamefont {Schreiber}}, \bibinfo {author}
  {\bibfnamefont {D.~G.}\ \bibnamefont {Schlom}}, \bibinfo {author}
  {\bibfnamefont {K.~M.}\ \bibnamefont {Shen}},\ and\ \bibinfo {author}
  {\bibfnamefont {N.~P.}\ \bibnamefont {Armitage}},\ }\href
  {https://doi.org/10.48550/arXiv.2312.13502} {\bibinfo {title} {Energy
  {{Relaxation}} and dynamics in the correlated metal {{Sr}}{$_2$}{{RuO}}{$_4$}
  via {{THz}} two-dimensional coherent spectroscopy}} (\bibinfo {year}
  {2023}),\ \Eprint {https://arxiv.org/abs/2312.13502} {arXiv:2312.13502
  [cond-mat]} \BibitemShut {NoStop}%
\bibitem [{\citenamefont {Bo{\v z}ovi{\'c}}\ \emph {et~al.}(2016)\citenamefont
  {Bo{\v z}ovi{\'c}}, \citenamefont {He}, \citenamefont {Wu},\ and\
  \citenamefont {Bollinger}}]{IBozovic2016}%
  \BibitemOpen
  \bibfield  {author} {\bibinfo {author} {\bibfnamefont {I.}~\bibnamefont
  {Bo{\v z}ovi{\'c}}}, \bibinfo {author} {\bibfnamefont {X.}~\bibnamefont
  {He}}, \bibinfo {author} {\bibfnamefont {J.}~\bibnamefont {Wu}},\ and\
  \bibinfo {author} {\bibfnamefont {A.~T.}\ \bibnamefont {Bollinger}},\
  }\bibfield  {title} {\bibinfo {title} {Dependence of the critical temperature
  in overdoped copper oxides on superfluid density},\ }\href
  {https://doi.org/10.1038/nature19061} {\bibfield  {journal} {\bibinfo
  {journal} {Nature}\ }\textbf {\bibinfo {volume} {536}},\ \bibinfo {pages}
  {309} (\bibinfo {year} {2016})}\BibitemShut {NoStop}%
\bibitem [{\citenamefont {Mahmood}\ \emph {et~al.}(2018)\citenamefont
  {Mahmood}, \citenamefont {He}, \citenamefont {Bozovic},\ and\ \citenamefont
  {Armitage}}]{FMahmood2018}%
  \BibitemOpen
  \bibfield  {author} {\bibinfo {author} {\bibfnamefont {F.}~\bibnamefont
  {Mahmood}}, \bibinfo {author} {\bibfnamefont {X.}~\bibnamefont {He}},
  \bibinfo {author} {\bibfnamefont {I.}~\bibnamefont {Bozovic}},\ and\ \bibinfo
  {author} {\bibfnamefont {N.~P.}\ \bibnamefont {Armitage}},\ }\bibfield
  {title} {\bibinfo {title} {Locating the missing superconducting electrons in
  overdoped cuprates},\ }\href@noop {} {\bibfield  {journal} {\bibinfo
  {journal} {ArXiv e-prints}\ } (\bibinfo {year} {2018})},\ \Eprint
  {https://arxiv.org/abs/1802.02101} {arXiv:1802.02101 [cond-mat.supr-con]}
  \BibitemShut {NoStop}%
\bibitem [{\citenamefont {{Giraldo-Gallo}}\ \emph {et~al.}(2018)\citenamefont
  {{Giraldo-Gallo}}, \citenamefont {Galvis}, \citenamefont {Stegen},
  \citenamefont {Modic}, \citenamefont {Balakirev}, \citenamefont {Betts},
  \citenamefont {Lian}, \citenamefont {Moir}, \citenamefont {Riggs},
  \citenamefont {Wu}, \citenamefont {Bollinger}, \citenamefont {He},
  \citenamefont {Bo{\v z}ovi{\'c}}, \citenamefont {Ramshaw}, \citenamefont
  {McDonald}, \citenamefont {Boebinger},\ and\ \citenamefont
  {Shekhter}}]{PGiraldo-Gallo2018}%
  \BibitemOpen
  \bibfield  {author} {\bibinfo {author} {\bibfnamefont {P.}~\bibnamefont
  {{Giraldo-Gallo}}}, \bibinfo {author} {\bibfnamefont {J.~A.}\ \bibnamefont
  {Galvis}}, \bibinfo {author} {\bibfnamefont {Z.}~\bibnamefont {Stegen}},
  \bibinfo {author} {\bibfnamefont {K.~A.}\ \bibnamefont {Modic}}, \bibinfo
  {author} {\bibfnamefont {F.~F.}\ \bibnamefont {Balakirev}}, \bibinfo {author}
  {\bibfnamefont {J.~B.}\ \bibnamefont {Betts}}, \bibinfo {author}
  {\bibfnamefont {X.}~\bibnamefont {Lian}}, \bibinfo {author} {\bibfnamefont
  {C.}~\bibnamefont {Moir}}, \bibinfo {author} {\bibfnamefont {S.~C.}\
  \bibnamefont {Riggs}}, \bibinfo {author} {\bibfnamefont {J.}~\bibnamefont
  {Wu}}, \bibinfo {author} {\bibfnamefont {A.~T.}\ \bibnamefont {Bollinger}},
  \bibinfo {author} {\bibfnamefont {X.}~\bibnamefont {He}}, \bibinfo {author}
  {\bibfnamefont {I.}~\bibnamefont {Bo{\v z}ovi{\'c}}}, \bibinfo {author}
  {\bibfnamefont {B.~J.}\ \bibnamefont {Ramshaw}}, \bibinfo {author}
  {\bibfnamefont {R.~D.}\ \bibnamefont {McDonald}}, \bibinfo {author}
  {\bibfnamefont {G.~S.}\ \bibnamefont {Boebinger}},\ and\ \bibinfo {author}
  {\bibfnamefont {A.}~\bibnamefont {Shekhter}},\ }\bibfield  {title} {\bibinfo
  {title} {Scale-invariant magnetoresistance in a cuprate superconductor},\
  }\href {https://doi.org/10.1126/science.aan3178} {\bibfield  {journal}
  {\bibinfo  {journal} {Science}\ }\textbf {\bibinfo {volume} {361}},\ \bibinfo
  {pages} {479} (\bibinfo {year} {2018})}\BibitemShut {NoStop}%
\bibitem [{\citenamefont {Zacharias}\ \emph {et~al.}(2015)\citenamefont
  {Zacharias}, \citenamefont {Paul},\ and\ \citenamefont
  {Garst}}]{MZacharias2015}%
  \BibitemOpen
  \bibfield  {author} {\bibinfo {author} {\bibfnamefont {M.}~\bibnamefont
  {Zacharias}}, \bibinfo {author} {\bibfnamefont {I.}~\bibnamefont {Paul}},\
  and\ \bibinfo {author} {\bibfnamefont {M.}~\bibnamefont {Garst}},\ }\bibfield
   {title} {\bibinfo {title} {Quantum {{Critical Elasticity}}},\ }\href
  {https://doi.org/10.1103/PhysRevLett.115.025703} {\bibfield  {journal}
  {\bibinfo  {journal} {Physical Review Letters}\ }\textbf {\bibinfo {volume}
  {115}},\ \bibinfo {pages} {025703} (\bibinfo {year} {2015})}\BibitemShut
  {NoStop}%
\bibitem [{\citenamefont {Paul}\ and\ \citenamefont {Garst}(2017)}]{IPaul2017}%
  \BibitemOpen
  \bibfield  {author} {\bibinfo {author} {\bibfnamefont {I.}~\bibnamefont
  {Paul}}\ and\ \bibinfo {author} {\bibfnamefont {M.}~\bibnamefont {Garst}},\
  }\bibfield  {title} {\bibinfo {title} {Lattice {{Effects}} on {{Nematic
  Quantum Criticality}} in {{Metals}}},\ }\href
  {https://doi.org/10.1103/PhysRevLett.118.227601} {\bibfield  {journal}
  {\bibinfo  {journal} {Physical Review Letters}\ }\textbf {\bibinfo {volume}
  {118}},\ \bibinfo {pages} {227601} (\bibinfo {year} {2017})}\BibitemShut
  {NoStop}%
\bibitem [{\citenamefont {Werman}\ \emph
  {et~al.}(2017{\natexlab{a}})\citenamefont {Werman}, \citenamefont
  {Kivelson},\ and\ \citenamefont {Berg}}]{YWerman2017}%
  \BibitemOpen
  \bibfield  {author} {\bibinfo {author} {\bibfnamefont {Y.}~\bibnamefont
  {Werman}}, \bibinfo {author} {\bibfnamefont {S.~A.}\ \bibnamefont
  {Kivelson}},\ and\ \bibinfo {author} {\bibfnamefont {E.}~\bibnamefont
  {Berg}},\ }\bibfield  {title} {\bibinfo {title} {Quantum chaos in an
  electron-phonon bad metal},\ }\href@noop {} {\bibfield  {journal} {\bibinfo
  {journal} {arXiv e-prints}\ ,\ \bibinfo {eid} {arXiv:1705.07895}} (\bibinfo
  {year} {2017}{\natexlab{a}})},\ \Eprint {https://arxiv.org/abs/1705.07895}
  {arXiv:1705.07895 [cond-mat.str-el]} \BibitemShut {NoStop}%
\bibitem [{\citenamefont {Guo}\ \emph {et~al.}(2019)\citenamefont {Guo},
  \citenamefont {Gu},\ and\ \citenamefont {Sachdev}}]{HGuo2019}%
  \BibitemOpen
  \bibfield  {author} {\bibinfo {author} {\bibfnamefont {H.}~\bibnamefont
  {Guo}}, \bibinfo {author} {\bibfnamefont {Y.}~\bibnamefont {Gu}},\ and\
  \bibinfo {author} {\bibfnamefont {S.}~\bibnamefont {Sachdev}},\ }\bibfield
  {title} {\bibinfo {title} {Transport and chaos in lattice
  {{Sachdev-Ye-Kitaev}} models},\ }\href
  {https://doi.org/10.1103/PhysRevB.100.045140} {\bibfield  {journal} {\bibinfo
   {journal} {Phys. Rev. B}\ }\textbf {\bibinfo {volume} {100}},\ \bibinfo
  {pages} {045140} (\bibinfo {year} {2019})},\ \Eprint
  {https://arxiv.org/abs/1904.02174} {arXiv:1904.02174 [cond-mat.str-el]}
  \BibitemShut {NoStop}%
\bibitem [{\citenamefont {Zhang}\ \emph {et~al.}(2017)\citenamefont {Zhang},
  \citenamefont {{Levenson-Falk}}, \citenamefont {Ramshaw}, \citenamefont
  {Bonn}, \citenamefont {Liang}, \citenamefont {Hardy}, \citenamefont
  {Hartnoll},\ and\ \citenamefont {Kapitulnik}}]{JZhang2017a}%
  \BibitemOpen
  \bibfield  {author} {\bibinfo {author} {\bibfnamefont {J.}~\bibnamefont
  {Zhang}}, \bibinfo {author} {\bibfnamefont {E.~M.}\ \bibnamefont
  {{Levenson-Falk}}}, \bibinfo {author} {\bibfnamefont {B.~J.}\ \bibnamefont
  {Ramshaw}}, \bibinfo {author} {\bibfnamefont {D.~A.}\ \bibnamefont {Bonn}},
  \bibinfo {author} {\bibfnamefont {R.}~\bibnamefont {Liang}}, \bibinfo
  {author} {\bibfnamefont {W.~N.}\ \bibnamefont {Hardy}}, \bibinfo {author}
  {\bibfnamefont {S.~A.}\ \bibnamefont {Hartnoll}},\ and\ \bibinfo {author}
  {\bibfnamefont {A.}~\bibnamefont {Kapitulnik}},\ }\bibfield  {title}
  {\bibinfo {title} {Anomalous thermal diffusivity in underdoped
  {{YBa2Cu3O6}}+x},\ }\href {https://doi.org/10.1073/pnas.1703416114}
  {\bibfield  {journal} {\bibinfo  {journal} {Proceedings of the National
  Academy of Sciences}\ }\textbf {\bibinfo {volume} {114}},\ \bibinfo {pages}
  {5378} (\bibinfo {year} {2017})}\BibitemShut {NoStop}%
\bibitem [{\citenamefont {Zhang}\ \emph
  {et~al.}(2019{\natexlab{a}})\citenamefont {Zhang}, \citenamefont {Kountz},
  \citenamefont {{Levenson-Falk}}, \citenamefont {Song}, \citenamefont
  {Greene},\ and\ \citenamefont {Kapitulnik}}]{JZhang2019}%
  \BibitemOpen
  \bibfield  {author} {\bibinfo {author} {\bibfnamefont {J.}~\bibnamefont
  {Zhang}}, \bibinfo {author} {\bibfnamefont {E.~D.}\ \bibnamefont {Kountz}},
  \bibinfo {author} {\bibfnamefont {E.~M.}\ \bibnamefont {{Levenson-Falk}}},
  \bibinfo {author} {\bibfnamefont {D.}~\bibnamefont {Song}}, \bibinfo {author}
  {\bibfnamefont {R.~L.}\ \bibnamefont {Greene}},\ and\ \bibinfo {author}
  {\bibfnamefont {A.}~\bibnamefont {Kapitulnik}},\ }\bibfield  {title}
  {\bibinfo {title} {Thermal {{Diffusivity Above Mott-Ioffe-Regel Limit}}},\
  }\href {https://doi.org/10.1103/PhysRevB.100.241114} {\bibfield  {journal}
  {\bibinfo  {journal} {Physical Review B}\ }\textbf {\bibinfo {volume}
  {100}},\ \bibinfo {pages} {241114} (\bibinfo {year} {2019}{\natexlab{a}})},\
  \Eprint {https://arxiv.org/abs/1808.07564} {arXiv:1808.07564 [cond-mat]}
  \BibitemShut {NoStop}%
\bibitem [{\citenamefont {Zhang}\ \emph
  {et~al.}(2019{\natexlab{b}})\citenamefont {Zhang}, \citenamefont {Kountz},
  \citenamefont {Behnia},\ and\ \citenamefont {Kapitulnik}}]{JZhang2019a}%
  \BibitemOpen
  \bibfield  {author} {\bibinfo {author} {\bibfnamefont {J.}~\bibnamefont
  {Zhang}}, \bibinfo {author} {\bibfnamefont {E.~D.}\ \bibnamefont {Kountz}},
  \bibinfo {author} {\bibfnamefont {K.}~\bibnamefont {Behnia}},\ and\ \bibinfo
  {author} {\bibfnamefont {A.}~\bibnamefont {Kapitulnik}},\ }\bibfield  {title}
  {\bibinfo {title} {Thermalization and possible signatures of quantum chaos in
  complex crystalline materials},\ }\href
  {https://doi.org/10.1073/pnas.1910131116} {\bibfield  {journal} {\bibinfo
  {journal} {Proceedings of the National Academy of Sciences}\ }\textbf
  {\bibinfo {volume} {116}},\ \bibinfo {pages} {19869} (\bibinfo {year}
  {2019}{\natexlab{b}})}\BibitemShut {NoStop}%
\bibitem [{\citenamefont {Werman}\ \emph
  {et~al.}(2017{\natexlab{b}})\citenamefont {Werman}, \citenamefont
  {Kivelson},\ and\ \citenamefont {Berg}}]{YWerman2017a}%
  \BibitemOpen
  \bibfield  {author} {\bibinfo {author} {\bibfnamefont {Y.}~\bibnamefont
  {Werman}}, \bibinfo {author} {\bibfnamefont {S.~A.}\ \bibnamefont
  {Kivelson}},\ and\ \bibinfo {author} {\bibfnamefont {E.}~\bibnamefont
  {Berg}},\ }\bibfield  {title} {\bibinfo {title} {Non-quasiparticle transport
  and resistivity saturation: A view from the large-{{N}} limit},\ }\href
  {https://doi.org/10.1038/s41535-017-0009-8} {\bibfield  {journal} {\bibinfo
  {journal} {npj Quantum Materials}\ }\textbf {\bibinfo {volume} {2}},\
  \bibinfo {eid} {7} (\bibinfo {year} {2017}{\natexlab{b}})},\ \Eprint
  {https://arxiv.org/abs/1607.05725} {arXiv:1607.05725 [cond-mat.str-el]}
  \BibitemShut {NoStop}%
\bibitem [{\citenamefont {Tulipman}\ and\ \citenamefont
  {Berg}(2020)}]{ETulipman2020a}%
  \BibitemOpen
  \bibfield  {author} {\bibinfo {author} {\bibfnamefont {E.}~\bibnamefont
  {Tulipman}}\ and\ \bibinfo {author} {\bibfnamefont {E.}~\bibnamefont
  {Berg}},\ }\bibfield  {title} {\bibinfo {title} {Strongly coupled quantum
  phonon fluid in a solvable model},\ }\href
  {https://doi.org/10.1103/PhysRevResearch.2.033431} {\bibfield  {journal}
  {\bibinfo  {journal} {Physical Review Research}\ }\textbf {\bibinfo {volume}
  {2}},\ \bibinfo {pages} {033431} (\bibinfo {year} {2020})}\BibitemShut
  {NoStop}%
\bibitem [{\citenamefont {Tulipman}\ and\ \citenamefont
  {Berg}(2021)}]{ETulipman2021}%
  \BibitemOpen
  \bibfield  {author} {\bibinfo {author} {\bibfnamefont {E.}~\bibnamefont
  {Tulipman}}\ and\ \bibinfo {author} {\bibfnamefont {E.}~\bibnamefont
  {Berg}},\ }\bibfield  {title} {\bibinfo {title} {Strongly coupled phonon
  fluid and {{Goldstone}} modes in an anharmonic quantum solid: {{Transport}}
  and chaos},\ }\href {https://doi.org/10.1103/PhysRevB.104.195113} {\bibfield
  {journal} {\bibinfo  {journal} {Physical Review B}\ }\textbf {\bibinfo
  {volume} {104}},\ \bibinfo {pages} {195113} (\bibinfo {year}
  {2021})}\BibitemShut {NoStop}%
\bibitem [{\citenamefont {Xu}\ \emph {et~al.}(2020)\citenamefont {Xu},
  \citenamefont {Klein}, \citenamefont {Sun}, \citenamefont {Chubukov},\ and\
  \citenamefont {Meng}}]{XYXu2020}%
  \BibitemOpen
  \bibfield  {author} {\bibinfo {author} {\bibfnamefont {X.~Y.}\ \bibnamefont
  {Xu}}, \bibinfo {author} {\bibfnamefont {A.}~\bibnamefont {Klein}}, \bibinfo
  {author} {\bibfnamefont {K.}~\bibnamefont {Sun}}, \bibinfo {author}
  {\bibfnamefont {A.~V.}\ \bibnamefont {Chubukov}},\ and\ \bibinfo {author}
  {\bibfnamefont {Z.~Y.}\ \bibnamefont {Meng}},\ }\bibfield  {title} {\bibinfo
  {title} {Identification of non-{{Fermi}} liquid fermionic self-energy from
  quantum {{Monte Carlo}} data},\ }\href
  {https://doi.org/10.1038/s41535-020-00266-6} {\bibfield  {journal} {\bibinfo
  {journal} {npj Quantum Mater.}\ }\textbf {\bibinfo {volume} {5}},\ \bibinfo
  {pages} {65} (\bibinfo {year} {2020})}\BibitemShut {NoStop}%
\bibitem [{\citenamefont {Xu}\ \emph {et~al.}(2017)\citenamefont {Xu},
  \citenamefont {Sun}, \citenamefont {Schattner}, \citenamefont {Berg},\ and\
  \citenamefont {Meng}}]{XYXu2017}%
  \BibitemOpen
  \bibfield  {author} {\bibinfo {author} {\bibfnamefont {X.~Y.}\ \bibnamefont
  {Xu}}, \bibinfo {author} {\bibfnamefont {K.}~\bibnamefont {Sun}}, \bibinfo
  {author} {\bibfnamefont {Y.}~\bibnamefont {Schattner}}, \bibinfo {author}
  {\bibfnamefont {E.}~\bibnamefont {Berg}},\ and\ \bibinfo {author}
  {\bibfnamefont {Z.~Y.}\ \bibnamefont {Meng}},\ }\bibfield  {title} {\bibinfo
  {title} {Non-{{Fermi Liquid}} at ( 2 + 1 ) {{D Ferromagnetic Quantum Critical
  Point}}},\ }\href {https://doi.org/10.1103/PhysRevX.7.031058} {\bibfield
  {journal} {\bibinfo  {journal} {Phys. Rev. X}\ }\textbf {\bibinfo {volume}
  {7}},\ \bibinfo {pages} {031058} (\bibinfo {year} {2017})}\BibitemShut
  {NoStop}%
\bibitem [{\citenamefont {Rothwarf}\ and\ \citenamefont
  {Taylor}(1967)}]{ARothwarf1967}%
  \BibitemOpen
  \bibfield  {author} {\bibinfo {author} {\bibfnamefont {A.}~\bibnamefont
  {Rothwarf}}\ and\ \bibinfo {author} {\bibfnamefont {B.~N.}\ \bibnamefont
  {Taylor}},\ }\bibfield  {title} {\bibinfo {title} {Measurement of
  {{Recombination Lifetimes}} in {{Superconductors}}},\ }\href
  {https://doi.org/10.1103/PhysRevLett.19.27} {\bibfield  {journal} {\bibinfo
  {journal} {Physical Review Letters}\ }\textbf {\bibinfo {volume} {19}},\
  \bibinfo {pages} {27} (\bibinfo {year} {1967})}\BibitemShut {NoStop}%
\bibitem [{\citenamefont {Mason}(1955)}]{mason}%
  \BibitemOpen
  \bibfield  {author} {\bibinfo {author} {\bibfnamefont {W.~P.}\ \bibnamefont
  {Mason}},\ }\bibfield  {title} {\bibinfo {title} {Ultrasonic attenuation due
  to lattice-electron interaction in normal conducting metals},\ }\href
  {https://doi.org/10.1103/PhysRev.97.557} {\bibfield  {journal} {\bibinfo
  {journal} {Phys. Rev.}\ }\textbf {\bibinfo {volume} {97}},\ \bibinfo {pages}
  {557} (\bibinfo {year} {1955})}\BibitemShut {NoStop}%
\bibitem [{\citenamefont {Morse}(1955)}]{morse}%
  \BibitemOpen
  \bibfield  {author} {\bibinfo {author} {\bibfnamefont {R.~W.}\ \bibnamefont
  {Morse}},\ }\bibfield  {title} {\bibinfo {title} {Ultrasonic attenuation in
  metals by electron relaxation},\ }\href
  {https://doi.org/10.1103/PhysRev.97.1716} {\bibfield  {journal} {\bibinfo
  {journal} {Phys. Rev.}\ }\textbf {\bibinfo {volume} {97}},\ \bibinfo {pages}
  {1716} (\bibinfo {year} {1955})}\BibitemShut {NoStop}%
\bibitem [{\citenamefont {Pippard}(1955)}]{Pippard}%
  \BibitemOpen
  \bibfield  {author} {\bibinfo {author} {\bibfnamefont {A.}~\bibnamefont
  {Pippard}},\ }\bibfield  {title} {\bibinfo {title} {Cxxii. ultrasonic
  attenuation in metals},\ }\href {https://doi.org/10.1080/14786441008521122}
  {\bibfield  {journal} {\bibinfo  {journal} {The London, Edinburgh, and Dublin
  Philosophical Magazine and Journal of Science}\ }\textbf {\bibinfo {volume}
  {46}},\ \bibinfo {pages} {1104} (\bibinfo {year} {1955})},\ \Eprint
  {https://arxiv.org/abs/https://doi.org/10.1080/14786441008521122}
  {https://doi.org/10.1080/14786441008521122} \BibitemShut {NoStop}%
\bibitem [{\citenamefont {Blount}(1959)}]{Blount}%
  \BibitemOpen
  \bibfield  {author} {\bibinfo {author} {\bibfnamefont {E.~I.}\ \bibnamefont
  {Blount}},\ }\bibfield  {title} {\bibinfo {title} {Ultrasonic attenuation by
  electrons in metals},\ }\href {https://doi.org/10.1103/PhysRev.114.418}
  {\bibfield  {journal} {\bibinfo  {journal} {Phys. Rev.}\ }\textbf {\bibinfo
  {volume} {114}},\ \bibinfo {pages} {418} (\bibinfo {year}
  {1959})}\BibitemShut {NoStop}%
\bibitem [{\citenamefont {Tsuneto}(1961)}]{tsuneto}%
  \BibitemOpen
  \bibfield  {author} {\bibinfo {author} {\bibfnamefont {T.}~\bibnamefont
  {Tsuneto}},\ }\bibfield  {title} {\bibinfo {title} {Ultrasonic attenuation in
  superconductors},\ }\href {https://doi.org/10.1103/PhysRev.121.402}
  {\bibfield  {journal} {\bibinfo  {journal} {Phys. Rev.}\ }\textbf {\bibinfo
  {volume} {121}},\ \bibinfo {pages} {402} (\bibinfo {year}
  {1961})}\BibitemShut {NoStop}%
\bibitem [{\citenamefont {Khan}\ and\ \citenamefont {Allen}(1987)}]{Allen_US}%
  \BibitemOpen
  \bibfield  {author} {\bibinfo {author} {\bibfnamefont {F.~S.}\ \bibnamefont
  {Khan}}\ and\ \bibinfo {author} {\bibfnamefont {P.~B.}\ \bibnamefont
  {Allen}},\ }\bibfield  {title} {\bibinfo {title} {Sound attenuation by
  electrons in metals},\ }\href {https://doi.org/10.1103/PhysRevB.35.1002}
  {\bibfield  {journal} {\bibinfo  {journal} {Phys. Rev. B}\ }\textbf {\bibinfo
  {volume} {35}},\ \bibinfo {pages} {1002} (\bibinfo {year}
  {1987})}\BibitemShut {NoStop}%
\bibitem [{\citenamefont {Bhatia}\ and\ \citenamefont
  {Moore}(1961)}]{ABBhatia1961}%
  \BibitemOpen
  \bibfield  {author} {\bibinfo {author} {\bibfnamefont {A.~B.}\ \bibnamefont
  {Bhatia}}\ and\ \bibinfo {author} {\bibfnamefont {R.~A.}\ \bibnamefont
  {Moore}},\ }\bibfield  {title} {\bibinfo {title} {Ultrasonic {{Attenuation}}
  in {{Normal Metals}} at {{Low Temperatures}}},\ }\href
  {https://doi.org/10.1103/PhysRev.121.1075} {\bibfield  {journal} {\bibinfo
  {journal} {Physical Review}\ }\textbf {\bibinfo {volume} {121}},\ \bibinfo
  {pages} {1075} (\bibinfo {year} {1961})}\BibitemShut {NoStop}%
\bibitem [{\citenamefont {Herring}(1954)}]{CHerring1954}%
  \BibitemOpen
  \bibfield  {author} {\bibinfo {author} {\bibfnamefont {C.}~\bibnamefont
  {Herring}},\ }\bibfield  {title} {\bibinfo {title} {Role of low-energy
  phonons in thermal conduction},\ }\href
  {https://doi.org/10.1103/PhysRev.95.954} {\bibfield  {journal} {\bibinfo
  {journal} {Physical Review}\ }\textbf {\bibinfo {volume} {95}},\ \bibinfo
  {pages} {954} (\bibinfo {year} {1954})}\BibitemShut {NoStop}%
\bibitem [{\citenamefont {Mahan}(2000)}]{GDMahan2000}%
  \BibitemOpen
  \bibfield  {author} {\bibinfo {author} {\bibfnamefont {G.~D.}\ \bibnamefont
  {Mahan}},\ }\href {https://doi.org/10.1007/978-1-4757-5714-9} {\emph
  {\bibinfo {title} {Many-Particle Physics}}},\ \bibinfo {edition} {3rd}\ ed.,\
  Physics of Solids and Liquids\ (\bibinfo  {publisher} {Springer Science \&
  Business Media},\ \bibinfo {address} {New York},\ \bibinfo {year}
  {2000})\BibitemShut {NoStop}%
\end{thebibliography}%

\clearpage
\newpage

\begin{center}
    {\large \textbf{End Matter}}
\end{center}

{\it Asymptotic large-$N$ limit.-} The theory studied in the main text can be formally controlled using the following large-$N$ extension of the Yukawa-SYK model. We assign additional flavor indices to the fields labeled by $a,b,c,\dots=1,\dots,N$.  The action then reads 
\begin{subequations}
\beq\label{eq:LargeN}
   \calL_e & =& \int \rd\tau \Big[\sum_{\vec{k}} c_{\vec{k}, a}^\dagger \left(\partial_\tau+\varepsilon_{\vec{k}}\right)c_{\vec{k}, a} + \frac{v}{\sqrt{N}} \epsilon^{(3)}_{ab} \sum_{\vec{x}} c^\dagger_{\vec{x},a} c_{\vec{x},b}    \nn\\
   &&+\sum_{\vec{q}} \varphi_{-\vec{q} ,a}\left(\partial_\tau^2+v_{\vp}^2\vec{q}^2+ r\right)\varphi_{\vec{q}, a} \nn \\
     && + \frac{g}{N} \epsilon^{(1)}_{abc} \sum_{\vec{k},\vec{q}} f_{\vec{k},\vec{{q}}} c_{\vec{k+q/2},a}^\dagger c_{\vec{k-q/2},b}\varphi_{\vec{q},c} \nn   \\ 
     && + \frac{g'}{N} \sum_{\vec{k},\vec{q}} \epsilon^{(2)}_{\vec{q},abc} f_{\vec{k},\vec{{q}}} c_{\vec{k+q/2},a}^\dagger c_{\vec{k-q/2},b}\varphi_{\vec{q},c} \nn     \Big], \,\\
     %\label{eq:Lp} 
  \calL_{p} &=& \int\rd \tau \sum_i\sum_{\vn{q}<\omd/c}X_{i,-\vec{q},a} \left(\partial_{\tau}^2+c^2\vec{q}^2\right) X_{i,\vec{q},a}\,, \\
  \calL_{ep}^{(1)} &=& \int\rd \tau \sum_{\vec{k},\vec{q}} \epsilon^{(4)}_{abc}\frac{M^i_{\vec{k},\vec{q}}}{N} c^\dagger_{\vec{k+q/2},a}c_{\vec{k-q/2},b} X_{i,\vec{q},c}\,,\\
  %\label{eq:coupling2}
  \calL_{ep}^{(2)} &=&\int \rd \tau \sum_{\vec{q}} \frac{1}{\sqrt{N}}\epsilon_{ab}^{(5)} N_{\vec{q}}^i X_{i,-\vec{q},a} \varphi_{\vec{q},b}\,,\\
  %\label{eq:coupling3}
  \calL_{ep}^{(3)} &=& \frac{1}{2}\int \rd \tau \sum_{\vec{k},\vec{{q}}} \epsilon_{abc}^{(6)} \frac{L^{i}_{\vec{k},\vec{q}}}{N} \varphi_{-\vec{k-q/2},a} \varphi_{\vec{k+q/2},b} X_{i,\vec{q},c}\,.
\eeq
\end{subequations} Here the repeated indices $a,b,c$ are summed over by Einstein convention. The couplings are now dressed by a tensor factor $\epsilon^{(1)-(6)}$ which are drawn from independent, Gaussian ensembles with zero mean. Due to hermiticity of the action $\epsilon^{(5)}_{ab}$ and $\epsilon^{(6)}_{abc}$ are real and the others are complex. They satisfy the condition $\epsilon^{(1)}_{abc}={\epsilon^{(1)}_{bac}}^*$, $\epsilon^{(2)}_{ab}={\epsilon^{(2)}_{ba}}^*$, $\epsilon^{(3)}_{\vec{q},abc}={\epsilon^{(3)}_{-\vec{q},bac}}^*$, $\epsilon^{(4)}_{abc}={\epsilon^{(4)}_{bac}}^*$, $\epsilon^{6}_{abc}=\epsilon^{(6)}_{bac}$, and otherwise they are independent, and have unit variance. In Eq.\eqref{eq:LargeN} we have explicitly included two types of disorder, i.e. the disorder in the Yukawa coupling $c^\dagger c\varphi$ and the potential disorder for the electrons. Both of these are shown to lead to dynamical exponent $z_\varphi=2$ and a marginal Fermi liquid self-energy for the electrons \cite{AAPatel2023, HGuo2022a}, and the former also leads to linear-in-$T$ resistivity of the electrons \cite{AAPatel2023}. The large-$N$ saddle point of the above action leads to the self-energies \change{described by Eliashberg equations}, which are used in the rest of the work.

{\it Energy-relaxation due to deformation potential coupling.-} 
The leading contribution due to $\calL_{ep}^{(1)}$ in Eq.~\eqref{eq:coupling1} leads to the phonon self-energy in Fig.~\ref{fig:fig1}(b). When $T_e\approx T_p\approx T$, we obtain $\partial_t E=\kappa^{(1)}(T_e-T_p)$, where the general expression for $\kappa^{(1)}$ is presented below in Eq.\eqref{eq:kappa12}  and derived in the supplement \cite{Supp}. However, neglecting the electron self-energies and assuming $c\ll v_F$, we obtain 
\begin{equation}\label{eq:kappa_free}
  \kappa_\text{free}^{(1)}=\frac{\calN}{4 v_F a_z}\int\frac{\rd^3\vec{q}}{(2\pi)^3}\overline{|M(\vn{q})|^2}\frac{c^2\vn{q}^2}{\vn{q_{\text{2D}}}}\frac{1}{T^2 \sinh^2\frac{c\vn{q}}{2T}}\,,
\end{equation} 
where $\vec{q}$ is the momentum transfer between the phonons and the particle-hole excitations, $a_z$ is the lattice constant in the $z-$direction and $\calN$ is the electronic density of states. The $1/\vn{q_\text{2D}}$ factor arises from projecting onto the 2D Fermi surface, which however does not lead to any IR divergence. We have also replaced $\sum_{i} |M^i_{\vec{k}+\vec{q/2},\vec{q}}|^2$ by its average over the Fermi surface $\overline{|M(\vec{q})|^2}$. When $T\ll \omd$ (Debye frequency), each factor of $\vn{q}$ and $\vn{q_\text{2D}}$ is replaced by $T$ (recall that $|M(\vn{q})|^2\propto\vn{q}^2$), and we obtain $\kappa^{(1)}_\text{free}\propto T^4$. This result can also be alternatively derived using Fermi's Golden rule, as follows:
\begin{equation}
\begin{split}
   & \kappa^{(1)}\sim \overbrace{\calN}^\text{Fermi surface DOS}\times \overbrace{T^3}^{\text{phonon density}} \times  
\overbrace{(\sqrt{T})^2}^{\text{Goldstone coupling}} \\
&\times \underbrace{T}_{\text{Energy transfer per particle}}
\times \underbrace{\frac{1}{T}}_{\text{Expand around equilibrium}}\,.
\end{split}
\end{equation}
Here a typical phonon has energy $T$ and the coupling enters as $\sqrt{T}$ because the displacement field $X$ is related to the creation and annihilation operators via $X_{\vec{q}}=(1/\sqrt{2c\vn{q}})(a_{\vec{q}}+a_{-\vec{q}}^\dagger)$ \cite{GDMahan2000}. {Additionally, we note that this scaling is not sensitive to the Fermi surface geometry, because energy relaxation is not constrained by small-angle scattering, unlike momentum relaxation.}
On the other hand, when $T\gg \omd$, Eq.~\eqref{eq:kappa_free} becomes $T$-independent, which is the classic expectation that in the equipartition regime of the phonons, it becomes difficult for them to absorb energy from the electrons. 
%The numerical result of Eq.~\eqref{eq:kappa_free} is plotted in Fig.~\ref{fig:results}(a), which reproduces Allen's result \cite{PBAllen1987}; 

Next, we include the marginal Fermi liquid self-energy. The general expression for $\kappa^{(1)}$ derived from the Keldysh formalism \cite{Supp} reads 
\begin{equation}\label{eq:kappa12}
\begin{split}
  &\kappa^{(1)}=\frac{1}{8 a_z} \int\frac{\rd^3\vec{q}}{(2\pi)^3}\int\frac{\rd \nu}{2\pi} \calN^2 \frac{4\pi^2}{k_F} \overline{|M(\vec{q})|^2} \\ &\times
  \left[\frac{1}{\sqrt{\vn{q_\text{2D}}^2+(A+iB)^2}}+\frac{1}{\sqrt{\vn{q_\text{2D}}^2+(A-iB)^2}}\right]\\
  &\times \frac{c\vn{q}}{2T^2 \sinh^2\frac{c\vn{q}}{2T}}\left[\tanh\bigg(\frac{\nu+c\vn{q}}{2T}\bigg)-\tanh\bigg(\frac{\nu}{2T}\bigg)\right]\,,
\end{split}
\end{equation} 
where $A,B$ are given by 
\begin{subequations}
\beq\label{eq:facA}
  A&=&-\frac{\Sigma''(\nu)+\Sigma''(\nu+c\vn{q})}{v_F}\,\\
  \label{eq:facB}
  B&=&\frac{c\vn{q}+\Sigma'(\nu)-\Sigma'(\nu+c\vn{q})}{v_F}\,.
\eeq
\end{subequations}
%In Eq.~\eqref{eq:kappa12}, $\vec{q}$ denotes the momentum transfer between the phonons and the electronic particle-hole excitations, $\nu$ denotes the typical frequency of the fermion, $\vec{q}_{2D}$ is the projection of $\vec{q}$ onto the $xy$-plane, $a_z$ is the lattice constant in the $z-$direction and $\calN$ is the electron density of states on the Fermi surface. 
Here, $\nu$ is the typical frequency of the fermion and the retarded self-energy of the fermion is $\Sigma_R(\omega)=\Sigma'(\omega)+i\Sigma''(\omega)$. In a marginal Fermi liquid, we take $\Sigma''(\omega)=-\Gamma/2-\alpha|\omega|$, where $\Gamma$ is the elastic scattering rate due to disorders, and $\alpha$ is a dimensionless coefficient related to fermion-boson Yukawa coupling and disorder strength \cite{HGuo2022a,AAPatel2023}. The real part $\Sigma'(\omega)$ is related via the Kramers-Kr\"onig relation. In the free limit where self-energy becomes negligible and $c\ll v_F$, Eq.~\eqref{eq:kappa12} reduces to Eq.~\eqref{eq:kappa_free}.

%As discussed in the main text, Eq.~\eqref{eq:kappa12} reproduces Allen's result \cite{PBAllen1987} in the limit where we drop the electron self-energy and for $c\ll v_F$, which allows us to drop the $A,~B$ terms above. 

Next, let us discuss the effects of including the fermion self-energies. 

(a) We first focus on the elastic scattering processes whose effects are primarily encoded in $A$. The elastic scattering rate $\Gamma$ introduces a new energy scale $T_{\rm{elastic}}=c \Gamma/v_F$, such that for $T<T_{\rm{elastic}}$, $\kappa$ crosses over from $T^4$ to $T^5$. However, this energy scale may not be relevant to actual experiments, because typical values of $\Gamma$ are of order $10\rm{K}$ and $c/v_F \sim 10^{-3}$ in cuprates. However, in systems where the Fermi velocity is small such as disordered, flat-band systems, this energy scale might be experimentally accessible, which we leave for future investigation. 

(b) The second process is the inelastic scattering process of marginal Fermi liquid. Because the $\tanh(...)$ factors in Eq.~\eqref{eq:kappa12} imply that the fermion frequency $\nu$ is comparable to phonon energy $c\vn{q}$, it has a weak effect in the $A$ factor which only leads to a small decrease of the integral at the order of $c/v_F$. The real part of the fermion self-energy (encoded in the factor $B$), on the other hand, can lead to a potential enhancement due to the divergence of the integrand when $B\sim \vn{q}$. Rewriting Eq.~\eqref{eq:facB} as $B\sim c\vn{q}/Z(c\vn{q}) v_F$, where $Z(c\vn{q})\sim 1/\ln(E_F/c\vn{q})$ is the energy-dependent quasiparticle residue of the marginal Fermi liquid, we find that $B$ only becomes comparable to $\vn{q}$ at exponentially low-energy scales, $E_*\sim E_F \exp(-v_F/c)$. Therefore, for the typical range of temperatures in the normal state that are experimentally relevant between the superconducting $T_c$ and $O(100~\rm{K})$, the energy relaxation rate due to the above coupling in a MFL yields results that are nearly identical to the Fermi liquid.

{\it Computation of $\kappa^{(2,3)}$ in regime (iii) and effects of coupling to out-of-plane strain.-}
 In this part we discuss the computations of $\kappa^{(2,3)}$ in the regime (iii) shown in Fig.~\ref{fig:results}. We start from a modified version of Eq.~\eqref{eq:kappa21}, which includes effects of coupling to out-of-plane strain:
 \begin{equation}\label{eq:kappa21_app}
\kappa^{(2)} \approx \frac{A_N}{c^5} \int_0^\pi d\theta\, \frac{\gamma T^5 \sin\theta(\sin^2\theta+\lambda\cos^2\theta)}{\gamma^2 T^2 + \left[m^2(T) + T^2 v_\varphi^2 \sin^2\theta / c^2\right]^2},
\end{equation} Here, the coupling to out-of-plane strain enters through replacing $\vn{q_\text{2D}}^2\propto \sin^2 \theta$ by $\lambda q_z^2\propto \lambda \cos^2 \theta$. In the limit where $T\gg \gamma c^2/v_\varphi^2$, the $\sin^2\theta$ term dominates the denominator. But retaining only this term in the denominator leads to an IR divergence, which is cutoff at $\theta=\theta_\text{min}$.  $\theta_\text{min}$ can be determined by balancing the two competing terms in the denominator, i.e. $(T^4 v_\varphi^4/c^4)\sin^4 \theta\sim \Delta(T)^4$, yielding $\theta_\text{min}\sim (c/v_\varphi)\Delta(T)/T$. Recall that $\Delta(T)=\max(\sqrt{\gamma T},m(T))$.  Therefore, the scaling of $\kappa^{(2)}$ can be obtained by 
\begin{equation}\label{}
\begin{split}
  \kappa^{(2)}&\sim \frac{A_N}{c} \int_{\theta_\text{min}}^{\infty} \rd \theta \frac{\gamma T^5 (\theta^3+\lambda\theta)}{T^4 v_\varphi^4 \theta^4}\\
  &\sim \frac{A_N \gamma T}{c v_\varphi^4}\left(\ln\frac{1}{\theta_\text{min}}+\frac{\lambda}{\theta_\text{min}^2}\right)\,,
\end{split} 
\end{equation} which yields the result reported in main text. In regime (iii), $\theta_\text{min}\ll 1$ signaling enhanced contributions from phonons moving in the $z$-direction.

This estimate can be generalized to $\kappa^{(3)}$, where 
\begin{equation}\label{}
\begin{split}
  \kappa^{(3)} &\approx\frac{A_L}{a_zc^5} T^5 \int_0^\pi \rd \theta \sin \theta(\sin^2\theta+\lambda\cos^2\theta) \int \rd^2\vec{k}_\text{2D} \\
  &\times A_\varphi(T,\vec{k}_\text{2D}) A_\varphi(T,\vec{k}_\text{2D}+\vec{q}_\text{2D}(\theta))\,.
\end{split}
\end{equation} Here we have generalized Eq.\eqref{eq:kappa32}.  Here the spectral function $A_\varphi$ is given in Eq.~\eqref{eq:Aphi}. Here the $\vec{k}_\text{2D}$ integral involves two peaks centered at $\vec{k}_\text{2D}=0$ and $\vec{k}_\text{2D}=-\vec{q}_\text{2D}(\theta)$, but due to symmetry they contribute equally and we consider the peak at  $\vec{k}_\text{2D}=0$. In regime (iii), the $\vec{k}_\text{2D}$ integral is controlled by the first $A_\varphi(T,\vec{k}_\text{2D})$ factor, and therefore $\int \rd^2\vec{k}_\text{2D}$ is replaced by $k_\text{typ}^2$  where $k_\text{typ}\sim \Delta(T)/v_\varphi$. The remaining $A_\varphi(T,\vec{k}_\text{2D}+\vec{q}_\text{2D}(\theta))$ factor has a similar IR divergence to the case of $\kappa^{(2)}$, and is therefore handled similarly, with the result 
\begin{equation}\label{}
  \kappa^{(3)}\sim \frac{A_L}{a_z c v_\varphi^6} \frac{\gamma^2 T^3}{\Delta(T)^2}\left(\ln\frac{1}{\theta_\text{min}}+\frac{\lambda}{\theta_\text{min}^2}\right).
\end{equation}

\newpage

\clearpage

\foreach \x in {1,...,5}
{
\clearpage
\includepdf[pages={\x},angle=0]{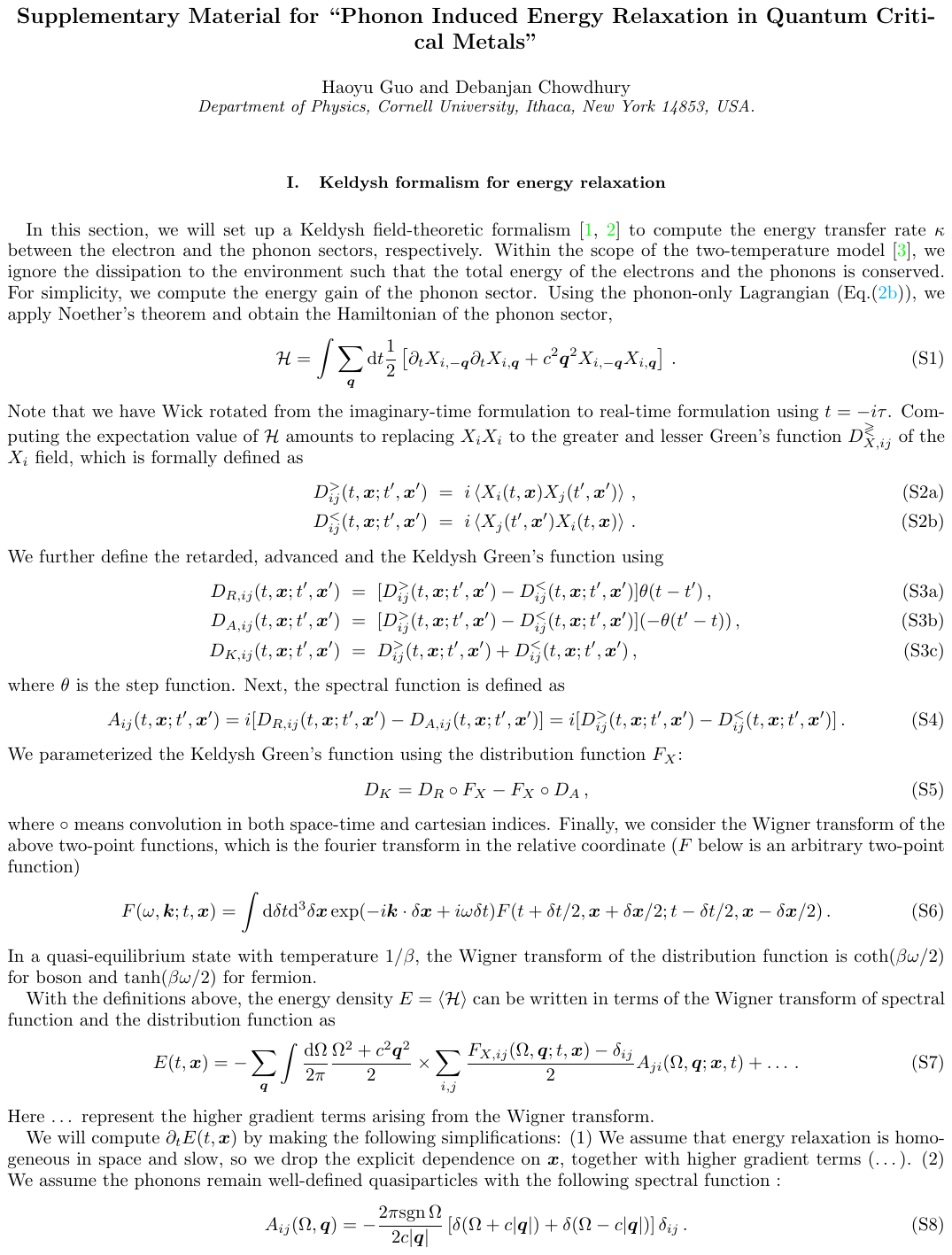}
}

\end{document}